
\input harvmac.tex

\def\frac#1#2{{#1\over#2}}
\def\coeff#1#2{{\textstyle{#1\over #2}}}
\def\hf{\coeff12}

\def\exp{{\rm exp}}

\def\slash#1{\mathord{\mathpalette\c@ncel{#1}}}
\overfullrule=0pt

\def\steepslash{\c@ncel}
\def\frac#1#2{{#1\over #2}}

\def\p {\partial}

\def\C{{\bf C}}
\def\inbar{\,\vrule height1.5ex width.4pt depth0pt}
\def\IB{\relax{\rm I\kern-.18em B}}
\def\IC{\relax\hbox{$\inbar\kern-.3em{\rm C}$}}
\def\IP{\relax{\rm I\kern-.18em P}}
\def\IR{\relax{\rm I\kern-.18em R}}
\def\IZ{\relax\ifmmode\mathchoice
{\hbox{Z\kern-.4em Z}}{\hbox{Z\kern-.4em Z}}
{\lower.9pt\hbox{Z\kern-.4em Z}}
{\lower1.2pt\hbox{Z\kern-.4em Z}}\else{Z\kern-.4em Z}\fi}

\def\pr{\Pi_\rho}

\def\pol{{1\over 2}}

\catcode`\@=12

\def\pol{{1\over 2}}
\input epsf
\noblackbox
\def\inbar{\,\vrule height1.5ex width.4pt depth0pt}
\def\IC{\relax\hbox{$\inbar\kern-.3em{\rm C}$}}
\def\IR{\relax{\rm I\kern-.18em R}}
\font\cmss=cmss10 \font\cmsss=cmss10 at 7pt
\def\IZ{\relax\ifmmode\mathchoice
{\hbox{\cmss Z\kern-.4em Z}}{\hbox{\cmss Z\kern-.4em Z}}
{\lower.9pt\hbox{\cmsss Z\kern-.4em Z}}
{\lower1.2pt\hbox{\cmsss Z\kern-.4em Z}}\else{\cmss Z\kern-.4em Z}\fi}

\font\manual=manfnt \def\dbend{\lower3.5pt\hbox{\manual\char127}}

\def\figin{\epsfcheck\figin}\def\figins{\epsfcheck\figins}
\def\epsfcheck{\ifx\epsfbox\UnDeFiNeD
\message{(NO epsf.tex, FIGURES WILL BE IGNORED)}
\gdef\figin##1{\vskip2in}\gdef\figins##1{\hskip.5in}
\else\message{(FIGURES WILL BE INCLUDED)}%
\gdef\figin##1{##1}\gdef\figins##1{##1}\fi}
\def\DefWarn#1{}
\def\figinsert{\goodbreak\midinsert}
\def\ifig#1#2#3{\DefWarn#1\xdef#1{fig.~\the\figno}
\writedef{#1\leftbracket fig.\noexpand~\the\figno}%
\figinsert\figin{\centerline{#3}}\medskip\centerline{\vbox{\baselineskip12pt
\advance\hsize by -1truein\noindent\footnotefont{\bf Fig.~\the\figno:} #2}}
\bigskip\endinsert\global\advance\figno by1}

\catcode`\@=11   
\def\xeqn{\expandafter\xe@n}\def\xe@n(#1){#1}
\def\xeqna#1{\expandafter\xe@na#1}\def\xe@na\hbox#1{\xe@nap #1}
\def\xe@nap$(#1)${\hbox{$#1$}}
\def\eqns#1{(\e@ns #1{\hbox{}})}
\def\e@ns#1{\ifx\und@fined#1\message{eqnlabel \string#1 is undefined.}%
\xdef#1{(?.?)}\fi \edef\next{#1}\ifx\next\em@rk\def\next{}%
\else\ifx\next#1\xeqn#1\else\def\n@xt{#1}\ifx\n@xt\next#1\else\xeqna#1\fi
\fi\let\next=\e@ns\fi\next}
\catcode`\@=12


 \def\cmp{Comm. Math. Phys. }
\def\np{Nucl. Phys. } \def\pl{Phys. Lett. }
\def\pr{Phys. Rev. } \def\prl{Phys. Rev. Lett. }
\lref\Bax{Baxter, R.J.: Partition function of the Eight-vertex lattice model.
Ann. Phys. {\bf 70}, 193--228 (1972) }
\lref\Lu{Rosso, M.: Finite-dimensional representations of the quantum analog of
the enveloping algebra of a complex simple Lie algebra. \cmp {\bf 117},
581-593 (1988)\semi
Lusztig, G.: Quantum deformations of certain simple modules over enveloping
algebras. Adv. in Math. {\bf 70},237-249 (1988)}
\lref\YBE{Yang-Baxter equation in integrable systems, ed. Jimbo, M.
Singapore: World Scientific (1989)}
\lref\bernard{Bernard, D., Felder, J.: Fock representation and BRST
cohmology in SL(2) current algebra. \cmp {\bf 127}, 145-168 (1990)}
\lref\yang{Yang, C.N.: Some results for the many-body problem in
one dimension with repulsive delta-function interaction. \prl {\bf 19} 1312
(1967)\semi
Yang, C.N.: S matrix for the one-dimensional N-body problem with repulsive or
attractive \ $\delta$-function interaction. \pr {\bf 168 }
1920-1923 (1968)}
\lref\BPZ{Belavin, A.A., Polyakov, A.M.,  Zamolodchikov, A.B.:
Infinite conformal symmetries in two-dimensional quantum field theory.
Nucl. Phys. B{\bf 241}, 333-380 (1984) }
\lref\DotsFat{ Dotsenko, Vl. S. and Fateev, V. A.: Conformal algebra and
multipoint correlation functions in 2d statistical models. Nucl. Phys.
B{\bf 240}[FS{\bf 12}], 312 (1984) \semi Dotsenko, Vl. S. and  Fateev, V. A.:
Four-point
correlation functions and the operator algebra in 2d conformal invariant
theories with central charge $c\le1$. Nucl. Phys. B{\bf 251}[FS{\bf 13}] 691
(1985) .}
\lref\abf{Andrews, G., Baxter, R., Forrester, J.: Eight-vertex SOS model
and generalized Rogers-Ramanujan identities. J. Stat. Phys. {\bf 35},
193-266 (1984).}
\lref\japAn{Date E., Jimbo M., Kuniba A., Miwa T., Okado M.: Exactly
solvable  SOS models: Local height probabilities and theta functions
identities. \np B{\bf 290}, 231, (1987)\semi
Date E., Jimbo M., Kuniba A., Miwa T., Okado M.: Exactly
solvable  SOS models 2: Proof of the star-triangle relation and combinatorial
identities. Adv. stud. Pure Math. {\bf 16}, 17-122, Tokyo: Kinokuniya (1988)}
\lref\huse{Huse, D.A.: Exact exponents for infinitely many new multicritical
points. \pr B{\bf 30}, 3908-3915 (1984)}
\lref\pasqu{Pasquier, V.: Etiology of IRF models. \cmp {\bf 118}, 355-364
(1988)}
\lref\LF{Fateev, V.A., Zamolodchikov, A.B.:
Conformal quantum field theory models in
two dimensions having\ $ {\bf Z}_3$\  symmetry. Nucl.Phys. B{\bf 280} [FS 18],
644-660 (1987) \semi
Fateev, V.A., Lukyanov, S.L.:The models of two-dimensional quantum field theory
with \ ${\bf Z}_n$\ symmetry. Int.J.Mod.Phys. A {\bf 3},  507-520 (1988)}
\lref\KZ{Knizhnik V.G., Zamolodchikov A.B.: Current algebra and Wess-Zumino
models in two dimensions. \np B{\bf 247} 83 (1984) }
\lref\rehren{Tsuchiya, A., Kanie, Y.: Vertex operators in conformal fiels
theory
on \ $CP^1$\ and monodromy representations of braid group.
Adv.Stud.Pure Math. {\bf 16} 297, Kinokunia, Tokyo (1989)\semi
Rehren, K.-H.: Locality of conformal fields in two dimensions:
Exchange algebra on the light cone \cmp {\bf 116}, 675-685 (1988)\semi
Froehlich, J.: Statistics of fields, the Yang-Baxter equation, and the
theory of Knots and links. In:"Non-perturbative quantum field theory",
G. `t Hooft et.al. (eds.), New York: Plenum (1988)
}
\lref\gerv{Gervais, J.-L. and Neveu, A.: Novel triangle relation and absense of
tachyons in Liouvill string theory.
\np B {\bf 238}, 125-141 (1984)
\semi Gervais, J.-L. and Neveu, A.: New quantum treatment of Liouville field
theory B {\bf 224}, 329 (1983).}
\lref\Yang{Yang, C.N., Yang, C.P.:Phys Rev. {\bf 150}, 321-327 (1966)}
\lref\bet{Bethe, H.A.: Zur Theorie der Metalle: I. Eigenwerte
und Eigenfunktionen der linearen Atomkette. Z. Phys. {\bf 71},
205-226 (1931) }
\lref\ber{Berezin, F.A., Pokhil, G.P., Finkelberg, V.M.:
Schr$\ddot {\rm o}$dinger's equation  for
systems of   one-dimensional particles
with pointwise interactions.
Vestn. Mosk. Gos. Univ., Ser. Mat. Mekh. {\bf 1}, 21-28 (1964) }
\lref\gui{McGuire, J.B.:
Study of exactly soluble one-dimensional N-body
problems.
Math. Phys. {\bf 5}, 622-629 (1964)}
\lref\gard{Gardner, C.S., Greene, J.M, Kruskal, M.D.,
Miura, R.H.: Method for solving the Korteveg-de Vries equation.
Phys. Rev. Lett. {\bf 19}, 1095-1097 (1967)}
\lref\zakh{Zakharov, V.E., Shabat, A.B.: Exact theory of
two-dimensional self-focusing and one-dimensional
self-modulation of waves in nonlinear media. Zh. Eksp. Teor. Fiz.
{\bf 61}, 118-134 (1971)}
\lref\kul{Kulish, P.P.:
Factorization of the classical and quantum S-matrix
and conservation laws. Theor. Math. Phys.
{\bf 26}, 198-205 (1976)}
\lref\z{Zamolodchikov, A.B.:
Exact two-particle S-matrix of quantum sine-Gordon solitons.
Commun. Math. Phys. {\bf 55}, 183-186 (1977) }
\lref\shr{Karowski, M., Thun, H.J., Truong, T.T.:
On the uniqueness of a purely elastic S-matrix in (1+1)
dimensions.
Phys. Lett.
B{\bf 67}, 321-322 (1977)}
\lref\Zkvadrat{Zamolodchikov, A.B., Zamolodchikov, Al.B.:
Factorized S-matrices in two dimensions as the exact
solutions of certain relativistic quantum field theory models.
Ann. Phys. (N.Y.) {\bf 120}, 253-291 (1979) }
\lref\og{Ogievetski, E.I., Reshetikhin, N.Yu., Wiegmann, P.B.:
The principal chiral field in two dimensions on classical Lie
algebras: The Bethe-ansatz solution and factorized theory
of scattering.
Nucl. Phys. B{\bf 280} [FS 18], 45-96 (1987)}
\lref\fat{Fateev, V.A., Zamolodchikov, A.B.:
Conformal field theory and purely elastic S-matrices. In:
Brink, L., Friedan, D., Polyakov, A.M. (eds)
Physics and Mathematics of Strings.
Memorial volume for Vadim Knizhnik,
pp. 245-270. Singapore: World Scientific 1989}
\lref\fre{Freund, P.G.O., Classen, T.R.,  Melzer, E.:
S-matrices for perturbations of certain conformal field theories.
Phys. Lett. B{\bf 229}, 243-247 (1989)}
\lref\sot{Sotkov, G., Zhu, G.-J.:
Bootstrap fusions and tricritical Potts model away
from criticality.
Phys. Lett. B{\bf 229}, 391-397 (1989)}
\lref\mar{Christe, J.L., Mussardo, G.:
Integrable systems away from criticality: the
Toda field theory and S-matrix of the tricritical Ising model.
Nucl. Phys. B{\bf 330}, 465-487 (1990)}
\lref\muss{Mussardo, G.: Off-critical statistical
models factorized scattering theories and
bootstrap program. Phys. Rep. {\bf 218}, 215-379 (1992)}
\lref\kar{Karowski,  M., Weisz, P.:
Exact form factors in (1+1)-dimensional
field theoretic models with solution behavior.
Nucl. Phys. B{\bf 139}, 455-476 (1978)}
\lref\berg{Berg, B., Karowski, M., Weisz, P.:
Construction of Green's functions from an
exact S-matrix.
Phys. Rev. D{\bf 19}, 2477-2479 (1979)}
\lref\smi{Smirnov, F.A.:
Quantum Gelfand-Levitan-Marchenko equations and form-factors
in the sine-Gordon model. J. Phys. A{\bf 17}, L1873-L1891 (1984)}
\lref\sm{Smirnov, F.A.: Solitons form-factors in the
sine-Gordon model. J. Phys. A{\bf 19}, L575-L578 (1986)}
\lref\ki{Kirillov, A.N., Smirnov, F.A.:
A representation of current algebra connected with SU(2)
invariant Thirring model. Phys. Lett. B{\bf 198}, 506-510 (1987)}
\lref\smirn{Smirnov, F.A.: Form-factors in completely
integrable models of
quantum field theory. Singapore: World Scientific 1992}
\lref\yur{Yurov, V.P., Zamolodchikov, Al.B.:
Correlations functions of integrable 2D models of the
relativistic field theory; Ising model.
Int. J. Mod. Phys. A{\bf 6}, 3419-3440 (1991) }
\lref\fad{Faddeev, L.D.: Quantum completely integrable models
in field theory.
Sov. Sci. Rev. Math. Phys. C{\bf 1}, 107-155 (1980)}
\lref\pea{Mitter, P.K.,  Weisz, P.H.:
Asymptotic scale invariance in a massive
Thirring model with U(n) symmetry.
Phys. Rev. D{\bf 8}, 4410-4429 (1973) }
\lref\bank{Banks, T., Horn, D.,  Neuberger, H.:
Bosonization of the SU(N) Thirring models.
Nucl. Phys. B{\bf 108}, 119-129 (1976) }
\lref\hal{Halpern, M.B.:
Quantum "solitons" which are SU(N) fermions.
Phys. Rev. D{\bf 12}, 1684-1699 (1976)}
\lref\col{Ablowitz, M.J.,  Kaup, D.J.,
Newell, A.C., Segur, H.:
Method for Solving the Sine-Gordon Equation.
Phys. Rev. Lett. {\bf 30}, 1262-1264 (1973)}
\lref\ztt{Takhtadzhian, L.A.,
Faddeev, L.D.:
Essentially nonlinear one-dimensional model of
classical field theory. Theor. Math. Phys.
{\bf 21}, 160-174 (1974)}
\lref\c{Zamolodchikov, A.B.: unpublished}
\lref\fr{Frenkel, I.B., Reshetikhin, N.Yu.: Quantum affine algebras
and holonomic difference equations. Commun. Math.
Phys.\ {\bf 146}, 1-60 (1992)}
\lref\sm{Smirnov, F.A.: Dynamical symmetries of
massive integrable models I, II.
Int. J. Mod. Phys. A{\bf 7}, Suppl. 1B, 813-837, 839-858 (1992)}
\lref\ztt{V.E. Zakharov, L.A. Takhtadjyan and
L.D. Faddeev, Sov. Phys. Dokl. 19
(1974/75).}
\lref\japone{Foda, O., Miwa, T.: Corner transfer matrices and quantum
affine algebras. Int. J. Mod. Phys. A{\bf 7}, Suppl.
1A, 279-302 (1992) }
\lref\japtwo{Davies, B., Foda, O., Jimbo, M., Miwa, T.,
Nakayashiki, A.:
Diagonalization of the XXZ Hamiltonian by
Vertex Operators. Commun. Math. Phys. {\bf 151}, 89-153  (1993)}
\lref\japthree{Jimbo, M., Miki, K., Miwa, T., Nakayashiki, A.:
Correlation functions of the XXZ model for $\Delta<-1$.
Phys. Lett. A{\bf 168}, 256-263 (1992)}
\lref\newjap{Jimbo, M., Miwa, T., Nakayashiki, A.:
Difference equations for the correlation functions
of the eight-vertex model. J. Phys. A: Math. Gen. {\bf 26},
2199-2209 (1993)}
\lref\japellip{Foda, O., Iohara, K.,
Jimbo, M., Kedem, R., Miwa, T. and Yan, H.:
An elliptic quantum algebra for $\widehat{sl}_2$. Lett. Math. Phys.
{\bf 32}, 259-268 (1994) \semi
Foda, O., Iohara, K.,
Jimbo, M., Kedem, R., Miwa, T. and Yan, H.:
Notes on highest weight modules of the elliptic algebra ${\cal
A}_{q,p}\left(\widehat{sl}_2\right)$. (hep-th/9405058)}
\lref\japising{Foda, O., Jimbo, M., Miwa, T., Miki, K., Nakayashiki, A.:
Vertex operators in solvable lattice models. J. math. Phys.
{\bf 35}, 13-46 (1994)}
\lref\fren{Frenkel, I.B., Jing, N.: Vertex
representations of quantum affine algebras.
Proc. Natl. Acad. Sci. USA
{\bf 85}, 9373-9377 (1988)}
\lref\cls{Clavelli, L., Shapiro, J.A.: Pomeron Factorization in
General Dual Models. Nucl. Phys. B{\bf 57}, 490-535 (1973)}
\lref\shat{Lukyanov, S., Shatashvili, S.:
Free field representation for the classical limit
of the quantum affine algebra.
Phys. Lett. B{\bf 298}, 111-115 (1993)}
\lref\fei{Feigin, B.L.,  Fuchs, D.B.: Representations
of the Virasoro algebra. In: Topology, Proceedings, Leningrad 1982.
Faddeev, L.D., Mal'cev, A.A. (eds.). Lecture Notes in Mathematics,
vol. {\bf 1060.} Berlin Heidelberg, New York: Springer 1984.
}
\lref\bour{Abada, A.,  Bougourzi, A.H., El Gradechi, M.A.:
Deformation of the Wakimoto construction.
Mod. Phys. Lett. A{\bf 8}, 715-724 (1993)}
\lref\berg{Berg, B., Karowski, M., Kurak, V., Weisz, P.:
Factorized U(n) symmetric S-matrices in two dimensions.
Nucl. Phys. B{\bf 134}, 125-132 (1978)}
\lref\berg{B. Berg, M. Karowski, V. Kurak and P. Weisz,
Nucl. Phys. B134 (1978) 125.}
\lref\lus{L$\ddot {\rm u}$scher, M.:
Quantum non-local charges and absence of particle production
in the two-dimensional non-linear $\sigma$ -model. Nucl. Phys.
B{\bf 135}, 1-19 (1978) }
\lref\lec{Bernard, D., LeClair, A.: Quantum group
symmetries and non-local currents in 2D
QFT. Commun. Math. Phys. {\bf 142}, 99-138 (1991)}
\lref\lecl{LeClair, A.:
Restricted Sine-Gordon theory and the minimal conformal
series. Phys. Lett. B{\bf 230}, 103-107 (1989)}
\lref\resm{Reshetikhin, N.Yu., Smirnov, F.A.:
Hidden quantum group symmetry and integrable perturbations
of conformal field theories. Commun. Math. Phys. {\bf 131},
157-177 (1990)}
\lref\rks{Kulish, P.P., Reshetikhin, N.Yu., Sklyanin, E.K.:
Yang-Baxter equation and representation theory. Lett.Math.
Phys. {\bf 5}, 393-403 (1981) }
\lref\dr{Drinfel'd, V.G.: Quantum Groups. In: Proceedings of
the International Congress of Mathematics. Berkeley 1986,
{\bf 1}, pp. 798-820. California: Acad. Press 1987\semi
Jimbo, M.: A q-difference analogue of \ $U{\cal G}$\ and Yang-Baxter
equation. Lett. Math. Phys. {\bf 10}, 63-68 (1985)
}
\lref\sklyanin{Sklyanin, E.K.: Some algebraic structures connected with the
Yang-Baxter
equation. Funct. Anal. and Appl.
{\bf 16} 263-270 (1983)\semi
Sklyanin, E.K.: Some algebraic structures connected with the Yang-Baxter
equation. Representations of quantum algebras. Funct. Anal. and Appl.
{\bf 17} 273-284 (1984)}
\lref\sk{Sklyanin, E.K., Faddeev, L.D.:
Quantum mechanical approach to completely integrable
field theory models. Sov. Phys. Dokl. {\bf 23}, 902-904 (1978)}
\lref\lecla{A. LeClair, Spectrum Generating Affine Lie Algebras
and Correlation Functions in
Massive Field Theory, Cornell preprint (1993)
CLNS 93/1220.}
\lref\kirre{A.N. Kirillov and N.Yu. Reshetikhin,
ed. V.G. Kac, World Scientific, Singapore (1989).}
\lref\crtthrn{T.L. Curtright and C.B. Thorn, \prl {\bf 48} (1982) 1309;
E. Braaten, T. Curtright and C. Thorn, \pl {\bf 118B}
(1982) 115; Ann. Phys. {\bf 147} (1983) 365;
E. Braaten, T. Curtright, G. Ghandour and C. Thorn, \prl {\bf 51}
(1983) 19; Ann. Phys. {\bf 153} (1984) 147.}
\lref\gradsh{I.S. Gradshteyn and I.M. Ryzhik, {\it Tables of Integrals,
Series, and Products\/}, Academic Press (1980).}
\lref\zkvadratdva{A.B. Zamolodchikov and Al.B. Zamolodchikov,
``Massless factorized scattering and sigma models with topological terms,''
Nucl. Phys. B379 (1992) 602.}
\lref\FLM{Frenkel, I., Lepowsky, J., Meurman, A.:
Vertex Operator Algebras and the Monster Group,
Academic Press (1988).}
\lref\Kor{V. Korepin, Theor. and Math. Physics 41 (1979) 953.}
\lref\ChTh{A. Chodos and C. Thorn, ``Making the massless string massive,''
Nucl. Phys. B72 (1974) 509.}
\lref\bax{Baxter, R.J.:
Exactly Solved Models in Statistical Mechanics,
Academic Press, London (1982) }
\lref\singordon{Lukyanov, S.: Free Field Representation for Massive
Integrable Models,
Rutgers preprint RU-93-30 (1993) (hep-th/9307196)}
\lref\fa{L.D. Faddeev and L.A. Takhtajan, Hamiltonian Method in the
Theory of Solitons, Springer, N.Y. (1987)}
\lref\josti{Lukyanov, S.: Correlators of the Jost Functions in
the Sine-Gordon Model. \pl {\bf B325}, 409-417 (1994) }
\lref\lecla{LeClair, A.:
Particle-Field Duality and Form Factors from
Vertex Operators.
CLNS 93/1263 Cornell preprint (1993)}
\lref\kirre{Kirillov, A.N., Reshetikhin, N.Yu.:
Representations of the algebra\  $U_q(sl(2))$, q-orthogonal
polynomials and invariants of links. In: Kac, V.G. (ed.)
Infinite-dimensional
Lie algebras and groups.
Singapore: World Scientific 1989}
\lref\ffk{Felder, G., Froehlich, J., Keller, G.:
Braid matrices and structure constants for minimal conformal models.
Commun. Math. Phys. {\bf 124}, 647-664 (1989)}
\lref\alg{Alvarez-Gaume, L., Gomez, G.,  Sierra, C.:
Quantum group
interpretation of some conformal field theories. \pl B {\bf 220},
142-152 (1989)}
\lref\mor{Moore, G., Seiberg, N.:
Classical and quantum conformal field theory.
Commun. Math. Phys. {\bf 123}, 177-254 (1989)}
\lref\sau{Moore G., Reshetikhin, N.Yu.:
A comment on quantum group symmetry in conformal field theory.
Nucl. Phys. B{\bf 328}, 557-574 (1989) }
\lref\felders{ Felder, G.: BRST approach to minimal models.
Nucl. Phys. B{\bf 317}, 215-236 (1989)}
\lref\fatee{Fateev, V.A., Zamolodchikov, A.B.:
A model factorized S-matrix and an integrable spin-1 Heisenberg chain.
Sov. J. Nucl. Phys. {\bf 32}, 298-303 (1980)}
\lref\TK{Tsuchiya A., Kanie Y.: Vertex operators in conformal fiels theory
on \ $CP^1$\ and monodromy representations of braid group.
Adv.Stud.Pure Math. {\bf 16} 297, Kinokunia, Tokyo (1989)}
\lref\kitaitzi{Chaichian, M., Presnajder, P.: Sugawara construction
and the q-deformation of Virasoro (super)algebra. Phys. Lett. B{\bf 277}
109-118 (1992)}
\lref\resh{Reshetikhin N.Yu., Semenov-Tyan-Shansky M.A.: Central extension
of quantum current groups. Lett. Math. Phys. {\bf 19} 133-142 (1990)}
\lref\japnew{Matsuo, A.: Free Field Representation of
Quantum Affine Algebra \ $U_q(\widehat{sl(2)})$,
Nagoya University preprint, August 1992
\semi Shiraishi, J.:
Free Boson Representation of $U_q(\widehat{sl(2)})$.
UT-617 Tokyo preprint (1992) \semi
Kato, A., Quano, Y., Shiraishi, J.: Free Boson Representation
of q-Vertex Operators and Their
Correlation Functions. UT-618 Tokyo preprint  (1992)\semi
Abada, A.,  Bougourzi, A.H., El Gradechi, M.A.:
Deformation of the Wakimoto construction.
Mod. Phys. Lett. A{\bf 8}, 715-724 (1993)}
\lref\gelfand{Gelfand I.M. Fuks B.B. : Funct. anal. and its appl.}
\lref\belavin{Belavin, A.A. : KdV and W . Adv.Stud.Pure Math.{\bf 19}, 117
(1989)}
\lref\exton{Gasper, G., Rahman, M., Basic hypergeometric series,
Cambridge University Press (1990)}

\Title{\vbox{\baselineskip12pt\hbox{RU-94-41}
                               \hbox{hep-th/9412128}}}
{\vbox{\centerline{
Bosonization of ZF Algebras: Direction Toward}
\vskip6pt\centerline{Deformed Virasoro Algebra}}}

\centerline{Sergei Lukyanov\footnote{$\dagger$}
{Current address after September 1, 1994: Newman Laboratory,
Cornell University, Ithaca, NY 14853-5001 }\ and Yaroslav Pugai}
\bigskip\centerline{Department of Physics and Astronomy}
\centerline{Rutgers University, Piscataway, NJ 08855-049, USA}
\centerline{and}
\centerline{L.D. Landau Institute for Theoretical
Physics}
\centerline{Chernogolovka, 142432, Russia}
\bigskip\bigskip\bigskip
\centerline{\bf{Abstract}}
These lectures were prepared to be presented at A.A. Belavin seminar
on CFT at Landau Institute for Theoretical Physics. We review
bosonization of CFT and show how it can be
applied to the studying of representations of Zamolodchikov-Faddeev (ZF)
algebras.
In the bosonic construction we obtain explicit
realization of chiral vertex operators interpolating between
irreducible representations of
the deformed Virasoro algebra. The
commutation relations of these operators
are determined by the elliptic matrix
of IRF type and their matrix elements are given in the form
of the contour integrals
of some meromorphic functions.
\Date{May, 94}

\newsec{Introduction.}

The development of CFT was initiated
in the fundamental work \refs{\BPZ} by
Belavin, Polyakov and Zamolodchikov where
the system of axioms describing CFT was proposed.
The main idea of BPZ is that
fields in CFT are classified by
irreducible representations of the Virasoro algebra.
{}From mathematical point of view, the
studying
of the CFT is equivalent
to the description of representations of the Virasoro
algebra and deriving the matrix
elements of vertex operators
interpolating between different irreducible
representations of the
Virasoro algebra \ ${\cal L}_{\lambda}$\ specified by highest weight
\ $\Delta_\lambda$.
Among vertex operators there is
a set of basic ones, called primary operators
$$
\Phi^{\lambda\ \mu}_{\Delta}(\zeta)\ :
\ {\cal L}_{\mu}\rightarrow
{\cal L}_{\lambda}\otimes \C(\zeta)\ \zeta^{\Delta_\lambda-\Delta_\mu}\ .
$$
Matrix elements of these operators ("conformal
blocks") are multivalued
analytical
functions. Knowing conformal
blocks one can reconstruct physical correlation functions
which satisfy the requirement of locality.
The analytical properties
of conformal blocks \refs{\DotsFat} are dictated by the
commutation relations in the algebra of
chiral vertex operators \ $\Phi^{\lambda\ \mu}_{\Delta}(\zeta)$:
\eqn\tegdfrdr{\Phi^{\lambda_{3}\ \lambda_4}_{\Delta_1}(\zeta_1)
\Phi^{\lambda_{4}\ \lambda_1}_{\Delta_2}(\zeta_2)
|_{{\cal L}_{\lambda_1}}
={\sum_{\lambda_2}}{\bf W}_{\Delta_1 \Delta_2}\biggl[\matrix{\lambda_3&
\lambda_2\cr
\lambda_4&\lambda_1}\
\biggr]\
\Phi^{\lambda_{3}\ \lambda_2}_{\Delta_2}(\zeta_2)
\Phi^{\lambda_2\ \lambda_1}_{\Delta_1}(\zeta_1)|_{{\cal L}_{\lambda_1}}\ .}
This quadratic algebra turns out to be associative,
which follows from the fact that
matrix \ ${\bf W}_{\Delta_1 \Delta_2}$\ satisfies to the
Yang-Baxter equation (YBE)
in the IRF form  \refs{\bax}.

It is important that CFT can be
alternatively considered
as a representation theory
of this algebra \refs{\gerv},\
\refs{\rehren}
which will be called further as Zamolodchikov-Faddeev (ZF)
algebra of IRF type.
In this way Virasoro algebra can be considered
as an algebra of transformations preserving the
commutation relations. Under the appropriate
choice of the set of irreducible representations of Virasoro algebra
\ ${\cal L}_{\lambda}$\ the
representations of the
chiral vertex algebra \tegdfrdr\
possess the realization in the direct sum
\ $\oplus_{\lambda}{\cal L}_{\lambda}$.
ZF algebra of IRF type is deeply
connected with another
associative quadratic algebra
\eqn\FZzzz{Z_a(\zeta_1)Z_b(\zeta_2)={\bf R}_{ab}^{cd}\ Z_d(\zeta_2)Z_c(\zeta_1)
\ ,}
where matrix \ ${\bf R}^{ab}_{cd}$\ is a solution of the
Yang-Baxter equation.
We call \FZzzz\ by Zamolodchikov-Faddeev algebra
\refs{\Zkvadrat},\ \refs{\fad}
of vertex type. It can be realized in
the extension of the space   \ $\oplus_{\lambda}{\cal L}_{\lambda}$\
by taking irreducible representations
of Virasoro algebra with proper multiplicities:
\eqn\tryue{\pi_Z=\oplus_{\lambda} \ {\cal L}_{\lambda}
\otimes {\cal V}_{\lambda}\    ,}
where \ ${\cal V}_{\lambda}$\ are some
finite-dimensional vector spaces
\refs{\sau}.

Thus, CFT can be described basing on the notions of two
algebras which have different forms
and require, at the first glance, different approaches of the investigation.
The first one is infinite-dimensional Virasoro algebra,
while the second is associative quadratic algebra
which is determined by some finite-dimensional matrices. However,
both algebras
are deeply connected to each other and can be represented in the same
space.
The initial success of CFT was based on well-developed
representation theory of the Virasoro algebra \refs{\fei}.
At the same time, ZF algebra approach seems to be
more general since
algebraic structures like \tegdfrdr - \FZzzz\ with
matrices \ ${\bf R}$\ and \ ${\bf W}$\ depending on
\ $\zeta_1{\zeta_2}^{-1}$, play crucial role in the
two-dimensional Integrable models of both Statistical Mechanics
(\refs{\fr},\ \refs{\japone},\ \refs{\japtwo},\ \refs{\japthree},\
\refs{\japising}) and Quantum Field Theory (see e.g. \refs{\smirn})
In the hierarchy of the solutions
of YBE
the constant solutions corresponding to algebras \tegdfrdr - \FZzzz\
are the simplest ones. They might be obtained
from trigonometric and elliptic \ ${\bf R}$\ and \ ${\bf W}$\
matrices as a result of the well-known limiting procedure. Conversely, one can
consider
more complicated solutions of YBE as parametric deformations
of the constant ones. It is reasonable to expect that
ZF algebras
corresponding to the trigonometric and elliptic
matrices \ ${\bf R}$\ ,\ ${\bf W}$\
are deformations of the algebras of conformal vertex
operators.
In this way we run into the following questions:
\it
\par\noindent
1.\ Is it possible to describe the representations
of elliptic and trigonometric ZF algebras using the
methods of CFT, or,
more explicitly, can ZF algebras be realized in the
direct sum of irreducible
representations of some infinite-dimensional
algebras generalizing Virasoro algebra?

\par\noindent
2.\ What is the exact form of the commutation relations of
these deformed
Virasoro algebras and their
geometrical and physical meaning?
\rm

The present work is mainly devoted to
studying of the first problem.
We construct the representation of
the elliptic deformation
of the conformal vertex operator algebra \tegdfrdr.
Our main idea is to deform in the appropriate manner
the bosonization procedure developed in the works  \refs{\DotsFat},\
\refs{\fei},\
\refs{\felders} for
CFT. Let us recall that
the central objects in the bosonization
are screening operators. Indeed, to
realize the irreducible
representations
of the Virasoro algebra in the Fock space,
one needs to know only explicit bosonic
realization of intertwining (screening) operators.
It is remarkable that the explicit form of
commutation relations of the Virasoro algebra
is not really used.
We describe a generalization of the conformal bosonization
basing on deformed screening operators which was proposed
in the work \refs{\singordon}.
It allows one to get the
explicit bosonic realization
of the chiral vertex operators which satisfy the commutation relations
of the form  \tegdfrdr\ with elliptic \ ${\bf W}$\ matrices.
Quadratic associative algebras obtained in this
approach seem to be deeply connected with algebras
of vertex operators found in the works \refs{\japellip}.
In particular, we expect that deformed Virasoro
algebra is some reduction of the
elliptic \ $sl(2)$\ algebra as well as ordinary
Virasoro algebra is a result of
quantum Hamiltonian reduction of affine Kac-Moody algebra \ ${\hat sl}(2)$
\refs{\belavin}.

The present work is prepared to
be presented at the A.A.Belavin's
seminar on CFT and Integrable Models at Chernogolovka.
Its largest part is devoted to bosonization of CFT.
In the simplest cases
we re-examine why and when bosonization is still working.
The material  from the first and the third sections seems
to be known to experts and it is contained (but sometimes
in the hidden form)
in the works \refs{\DotsFat}, \ \refs{\fei}, \refs{\felders}.
Our aim here
is just to emphasize those subtle moments
which appeared to be crucial in the
generalization of bosonization \refs{\singordon}.
As soon as we will understand and
correctly formulate conformal case, we will be
able to construct the representation of the
elliptic ZF algebra.
In particular,
we obtain integral representation for matrix elements
of  vertex operators
generalizing Dotsenko-Fateev formulas
for conformal blocks.
For instance, in the simplest case
of four-point function it is equivalent to the
following integral representation of q-hypergeometrical function:
\eqn\jdhgleorww{\eqalign{
\int_C&
\frac{d z}{2\pi i} z^{c-1}
\frac{(q^{\frac{1+a}{2}} z^{-1};q)_{\infty}}
{(q^{\frac{1-a}{2}} z^{-1};q)_{\infty}}\
\frac{(q^{\frac{1+b}{2}}\zeta z;q)_{\infty}}
{(q^{\frac{1-b}{2}}\zeta z;q)_{\infty}}=
\cr
&=\
q^{\frac{c(1-a)}{2}}\  \frac{\Gamma_q(c+a)}{\Gamma_q(c+1)\ \Gamma_q(a)}
\ F_q(a+c,b,c+1;  q^{1-\frac{a+b}{2}} \zeta)\ .}}
Notice, that matrix elements of
deformed vertex operators are written in the form of
ordinary contour integrals
rather than Jackson's integrals in the bosonization scheme
of the works \refs{\japnew}.
We also consider trigonometric
limit of elliptic
construction which corresponds to Sin-Gordon model \ \refs{\singordon},
\refs{\josti},\ \refs{\shat}.
In this limit we show how to reconstruct
the ZF algebra of vertex type
from IRF ZF algebra \refs{\pasqu}.

Let us make some notational conventions.
\par\noindent
i) Throughout this
work we will denote objects which have similar
meaning by the same letter, distinguishing it by "prime" symbol '.
It should not be confused with derivation symbol \ $\partial$.
\par\noindent
ii) For the technical reason it is convenient for
us to make ordering procedure in the exponential operators
like \ $e^{i \phi}$\ in the final step.
There is no difference in this prescription with
ordinary one in the conformal case, and it is not principal
but useful in the deformed case. It will be explained in the
section 4.

\newsec{Preliminaries}

{\it 2.1}\ Let \ $R_{ab}^{cd}(t)$\ be numerical matrix
depending on complex parameter\ $t$. The indexes \ $a,b,c,d$\
take value in the set of integer numbers. One can consider this
matrix as an operator acting in
tensor product \ ${\cal V}\otimes{\cal V}$,
where \ ${\cal V}$ is finite dimensional
linear space, which vectors are specified by indexes \ $a$.
We will call by Yang-Baxter equation of vertex type
the following algebraic equation
\refs{\bax}\ :
\eqn\yb{\eqalign{R^{c_1c_2}_{a_1a_2}(t_1t_2^{-1})
R^{b_1c_3}_{c_1a_3}&
(t_1t_3^{-1})
R^{b_2b_3}_{c_2c_3}
(t_2t_3^{-1})
=\cr
&=R^{c_1b_3}_{a_1c_3}
(t_1t_3^{-1})
R^{c_2c_3}_{a_2a_3}
(t_2t_3^{-1})
R^{b_1b_2}_{c_1c_2}
(t_1t_2^{-1})\ .}}
This equation play fundamental role in the
2 D Integrable Models of  both Quantum Field
Theory and Statistical Mechanics.
At present a lot of its nontrivial solutions
are found (see e.g. \refs{\YBE}). In this work we
will be  interested in the  simplest one when the
space \ ${\cal V}$\ has the dimension equal 2.
We will enumerate basic vectors of \ ${\cal V}$ by\ $\pm$.
In 1972 Baxter found the following remarkable
solution of\ \yb \ \refs{\Bax}:
\eqn\ldk{\eqalign{&R_{--}^{--}(t,p,q)=
\frac{\Theta_q(p^2q^{\frac{1}{2}})\
\Theta_q(t q^{\frac{1}{2}})}{\Theta_q(q^{\frac{1}{2}})\
\Theta_q(p^2 t q^{\frac{1}{2}})}\ ,\cr
&R_{+-}^{+-}(t,p,q)= R^{-+}_{-+} (t,p,q)=
p\ \frac{\Theta_q(p^2q^{\frac{1}{2}})\
\Theta_q(t )}{\Theta_q(q^{\frac{1}{2}})\
\Theta_q(p^2 t )}\ ,\cr
& R_{+-}^{-+}(t,p,q)=
R^{+-}_{-+}(t,p,q)=
t^{\frac{1}{2}}\ \frac{\Theta_q(p^2)\
\Theta_q(t q^{\frac{1}{2}})}{\Theta_q(q^{\frac{1}{2}})\
\Theta_q(p^2 t )}\ ,\cr
& R_{++}^{--}(t,p,q)=
R^{--}_{++}(t,p,q)=
p^{-1}t^{-\frac{1}{2}} q^{\frac{1}{4}}\ \frac{\Theta_q(p^2)\
\Theta_q(t )}{\Theta_q(q^{\frac{1}{2}})\
\Theta_q(p^2 t q^{\frac{1}{2}} )}\ ,}}
where

$$\Theta_q(s)=(q;q)_{\infty}\ (s;q)_{\infty}\ (qs^{-1};q)_{\infty}\ ,$$
is the Jacobi elliptic function and we use the
standard notation:
$$(z;   q_1,...,q_k)_{\infty}=\prod_{n_1,...n_k=0}^{\infty}
(1-z q_1^{n_1}...q_k^{n_k})\ .$$
The  \ $R$-matrix depends on two additional parameters\ $p,q$. In this
work we  suppose that parameter\ $q$ is real
and\ $0\leq q<1$. At the same time  \ $p$\ will
be complex number such that\ $|p|^2=1$. In this case
the matrix elements are real  numbers
if  \ $|t|^2=1$.
Together with the Yang-Baxter equation
the matrix\ $R_{ab}^{cd}$\ satisfies also to so-called unitarity condition:
\eqn\uni{R^{b_1b_2}_{a_1a_2}(t)R^{c_1c_2}_{b_1b_2}(t^{-1})
=\delta^{c_1}_{a_1}\delta^{c_2}_{a_2}.}
As it was pointed out  by A.B. Zamolodchikov \refs{\Zkvadrat}
the equations\ \yb ,\ \uni \ can be treated as correspondingly
self consistency and associativity condition
in the formal algebra:
\eqn\fz{Z_{a_1}(t_1)Z_{a_2}(t_2)
=S^{b_1b_2}_{a_1a_2}(t_1 t_2^{-1})Z_{b_2}(t_2)Z_{b_1}(t_1)\ .}
This algebra  was also considered in the works of L.D.
Faddeev \refs{\fad} and we will call it as ZF algebra.

If the parameter\ $q\to 0$ the matrix
extremely simplifies since\ $\Theta_q(t)\to 1-t$
and nontrivial elements of the matrix \ $R_{ab}^{cd}(t,p)$\
read:
\eqn\fsdq{\eqalign{& R_{++}^{++}(t,p)=
R_{--}^{--}(t,p)=1,\cr
&R_{+-}^{+-}(t,p)=R^{-+}_{-+} (t,p)
=p\ \frac{1-t}{1-p^2 t}\ ,\cr
&R_{+-}^{-+}(t,p)=
R^{+-}_{-+}(t,p)=t^{\frac{1}{2}}\ \frac{1-p^2}{1-p^2 t}\ .}}
We   can also get a solution of Yang-Baxter equation which
does not depend on spectral parameter \ $t$  by
the following limiting procedure:
\eqn\gdtrs{ R_{ab}^{cd}(\sigma,p)= p^{\frac{\sigma}{2}}
\lim_{L\to +\infty}e^{\frac{(a-c)\sigma L}{4}}
R_{ab}^{cd}(e^{\sigma L},p)\  ,}
here\ $\sigma=\pm$.
Non zero elements of matrix\ $R_{ab}^{cd}(+,p)$ are defined by
the relations
\eqn\fsdq{\eqalign{& R_{++}^{++}(+,p)=
R_{--}^{--}(+,p)=p^{\frac{1}{2}},\cr
&R_{+-}^{+-}(+,p)=R^{-+}_{-+} (+,p)
=p^{-\frac{1}{2}} ,\cr
&R_{+-}^{-+}(+,p)=p^{-\frac{1}{2}}(p-p^{-1})\ .}}
Note that matrices \ $R(+,p)$\ and\ $R(-,p)$\ are connected
by the relation
\eqn\osnm{R_{ab}^{cd}(-,p)= [R^{-1}]^{dc}_{ba}(+,p)\ ,}
which can be considered as analogue of unitarity condition
for constant  solution of Yang-Baxter equation.
The matrix \ \fsdq\
coincides with  the fundamental R-matrix for
quantum algebra\ $U_p(sl(2))$ \refs{\dr} .
We will need
some facts concerning this algebra, so let
us recall them here.

{\it 2.2.}
The quantum universal enveloping algebra \ $U_p(sl(2))$\
is an algebra on generators
$X^{\pm}, T$\ subject to the following relations:
\eqn\kdjh{\eqalign{&X^+X^--p^2X^-X^+=\frac{1-T^2}{1-p^{-2}}\,\ \cr
&T X^{\pm}=p^{\mp 2}X^{\pm} T\ .}}
It is well known that \ \kdjh\  has  Hopf algebra  structure.
In particular,
the comultiplication \ $\Delta$\ is defined by
\eqn\ksjhhd{\Delta(X^{\pm})=T\otimes X^{\pm}+X^{\pm}
\otimes 1\ ,\Delta(T)=T\otimes T\ .}
The parameter  \ $p$\ is complex number. The  most relevant cases
for us will be those with
$p=e^{i\pi\frac{\xi+1}{\xi}}$\ or\ $e^{i\pi\frac{\xi}{\xi+1}}$\
where \ $\xi$\ is a real irrational number greater then 1.
In these cases the algebra \ $U_p(sl(2))$\ admits the
irreducible representations in the finite dimensional
spaces  \ ${\cal V}_l$\ with the \ $dim\ {\cal V}_l=l$.
Construction of such representations is quite
similar to the construction of
irreducible representations of
\ $sl(2)$\ Lie algebra with the spin \ $j=\frac{l-1}{2}$\  \refs{\kirre},
\refs{\Lu} .
The basic vectors ${\bf e}_l^m\in {\cal V}_l$\ of representation
are specified
by number \ $m=-j,-j+1,...,j$ and conditions
\eqn\hsgd{\eqalign{&T{\bf e}_{l}^{m}=
p^{- 2m}{\bf e}_{l}^{m}\ ,\cr
&X^{-}{\bf e}_{l}^{m}={\bf e}_{l}^{m-1}\  .}}
The existence of
Hopf structure means that tensor product
${\cal V}_{l_1}\otimes{\cal V}_{l_2}$
of two representations of \ $U_p(sl(2))$\ would
also carry the structure of representation of this algebra.
Moreover,
if \ $\xi$\ is an irrational number, then
this representation turns to be completely reducible.
Since the tensor product
of two finite-dimensional irreducible
representations can be represented as direct sum of
irreducible representations, then
one can write down the Clebsh-Gordon decomposition of the form:
\eqn\mcnv{\eqalign{{\bf e}_{2j_1+1}^{m_1}\otimes
{\bf e}_{2j_2+1}^{m_2}=\sum_l
\left(\matrix{j_1&j_2&j\cr
m_1&m_2&m_1+m_2\cr}\right)_p\
{\bf e}_{2j+1}^{m}\ .}}
In present work we will need the explicit form of
the following Clebsh-Gordon coefficients:
\eqn\mdsj{\eqalign{&\left(\matrix{\frac{1}{2}&j&j+\frac{1}{2}\cr
\frac{1}{2}&m&m+\frac{1}{2}\cr}\right)_p=p^{-2j} \frac{[j+m+1]_p}{[2j+1]_p}
\ ,\cr
&\left(\matrix{\frac{1}{2}&j&j+\frac{1}{2}\cr
-\frac{1}{2}&m&m-\frac{1}{2}\cr}\right)_p=
\frac{p^{-j+m}}{ [2j+1]_p}\ ,\cr
&\left(\matrix{\frac{1}{2}&j&j-\frac{1}{2}\cr
\frac{1}{2}&m&m+\frac{1}{2}\cr}\right)_p=
(-1)^{2j}\ \frac{[j-m]_p}{[2j+1]_p} \ ,\cr
&\left(\matrix{\frac{1}{2}&j&j-\frac{1}{2}\cr
-\frac{1}{2}&m&m-\frac{1}{2}\cr}\right)_p=
(-1)^{2j+1}\ \frac{p^{-j+m-1}}{ [2j+1]_p}
\ ,}}
here we use the notation \ $[x]_p=\frac{p^x-p^{-x}}{p-p^{-1}}$.

%



\newsec{Free Fermions ZF algebra and \ $c=-2$\ Virasoro algebra}
In this section we will  discuss
ZF algebra generated by operators \ $Z_a(\zeta)\ (a=\pm ,\zeta\in\C $),
satisfying the simple commutation relations
\eqn\ki{Z_{a}(\zeta_1)Z_{b}(\zeta_2)=-Z_{b}(\zeta_2)Z_{a}(\zeta_1),
\ \ \zeta_1\not=\zeta_2\ .}
We consider its irreducible representations
characterized by   operator product expansions:
\eqn\lik{\eqalign{&Z_{\pm}(\zeta_2)Z_{\mp}(\zeta_1)=
\pm \frac{2\zeta_1\zeta_2}{(\zeta_2-\zeta_1)^2}+O(1)\ ,\cr
&Z_{\pm}(\zeta_2)Z_{\pm}(\zeta_1)=O(1),\  \zeta_1\to\zeta_2\  .}}
This algebra is well known  in the physical literature as
so called\ $b-c$\ system\ $(Z_+=b,\   Z_-=\p c)$. Its
representations admit  decompositions into  direct sum of irreducible
representations of the Virasoro algebra.
Let us note now that the commutation relations \ \ki\ and
operator product expansions\ \lik\ are invariant
with respect to a linear transformation
$$Z_a[r]\to {\bf G}^{ab} Z_b[r]\ ,$$
if the  determinant of \ ${\bf G}$\  is equal to one.
We will concentrate on the case when this symmetry
is not broken and the representation space of
the ZF algebra is classified
by the \ $SL(2)$\ symmetry together with Virasoro one.
In this case Virasoro algebra has the central charge
equal to  -2. We will try to analyze these well-known results in the form
which leaves room for generalization.

\subsec{Irreducible representations of the free fermions algebra}
The  algebra\ \ki\  possesses
two different  types of representation
denoted by\  ${\pi}_Z^R$\  (Ramond)
and\  ${\pi}_Z^{NS}$\   (Neveu-Schwartz) according to type
of boundary conditions imposed on the generators\ $Z_a(\zeta)$:
\eqn\loi{\eqalign{R\ &:\ \ Z_a(e^{2\pi i} \zeta)=Z_a(\zeta)\ ,\cr
NS\ &:\ \ Z_a(e^{2\pi i} \zeta)=-Z_a(\zeta)\ .}}
For this reason, \ $Z_a$\ have
the following decomposition in
Laurent series:
\eqn\jdgfbv{\ Z_a(\zeta)=\sum_{
r\not=
0}Z_a[r]\ \zeta^{-r}\ ,}
where \ $r\in {\bf Z}$\ for R sector and \ $r\in {\bf Z}+\frac{1}{2}$ for NS
one.
It follows from\  \ki - \lik\ ,
that modes \ $Z_a[r]
$\ obey the anticommutation
relations:
\eqn\jshgd{\{Z_a[r],Z_b[m]\}_{+}
=2\ r\ a\  \delta_{a+b,0}\delta_{r+m,0}\ .}
The spaces ${\pi}_Z^R$ and ${\pi}_Z^{NS}$\ are defined as
Fock modules of fermionic algebra
\  \jshgd \ created by the operators \ $Z_a[r]\ ,
r<0$\
under the action on the corresponding R- and NS-vacuum states.
The vacuum states
are specified  by the condition that they
are annihilated by any operator
$Z_a[r]$\ with \ $ r>0. $\
We assume that R- and NS- vacuum states are scalars
with respect to the \ $SL(2)$\ transformations.
Then \ $SL(2)$\ structure of the spaces\  $\pi_Z^{R,NS}$\
is uniquely determined by the condition that ZF operators are arranged
in the fundamental \ $SL(2)$\  doublet.

The next fact which we will need further is that
the dual space \ ${\pi}_Z^{R,NS\star}$\ also  admits the structure
of representation of the algebra\ \jshgd . It goes as follows. Given a linear
space \ ${\pi}_Z$\ its dual \ ${\pi}_Z^{\star}$\ is the set of linear maps
\ ${\pi}^{\star}_Z:
\ \pi_Z\to\C$\ determined by action \ ${\bf u}\matrix{ {\bf u}^{\star}\cr
\longrightarrow\cr{}}$$<{\bf u}^{\star},{\bf u}>.$
Choose the dual basis by the canonical pairing
\ $<{\bf u}^{\star}_j,{\bf u}_k>
=\delta_{jk}$. Since algebra \jshgd \
admits the  anti-involution
\eqn\hg{{\cal A}
\ \big\{Z_a[r]\big\}=Z_{-a}[-r]\ ,    }
then the spaces\ ${\pi}^{\star}_Z={\pi}_Z^{R,NS\star}$\
can be endowed by the structure of
representation of ZF algebra
through the formula
\eqn\mj{
<{\bf u}^{\star}_1 Z^{\star}_a[n],{\bf u}_2>=\ <{\bf u}^{\star}_1,{\cal A}
\ \big\{Z_{a}[n]\big\}\ {\bf u}_2>\ . }

\subsec{Bosonization of fermionic ZF algebra}
Now   we want to show how
the representations of the ZF algebra \ \ki -\lik\
can be  realized in
direct sums of
boson  Fock
modules.
We start  with  the  space ${\pi}_Z^R$.
Let \ $\{b_n, P, Q\ |n\in{\bf Z}/ \{ 0 \} \}$\  be a set of operators
satisfying the  commutation relations:
\eqn\lnsgd{\eqalign{&[b_m,b_n]=m\ \delta_{m+n,0}\ ,\cr
&[P,Q]=-i,\ \ [P,b_n]=[Q,b_n]=0\ .}}
The bosonic Fock module\ ${\cal F}_p$\ is generated by acting of
creation
operators\ $b_{-n}, n>0$ \ on the
highest vector\ ${\bf f}_p:\
b_{n}\ {\bf f}_p=0,\ \ n>0\ ; \  P\ {\bf f}_p=p\ {\bf f}_p\ .$
In the  direct sum of the Fock  modules
$\oplus_{k\in {\bf Z}}
{\cal F}_{k-\frac{1}{2}} $
action of the following operators is well-defined:
\eqn\idreg{\eqalign{
&X=\oint\frac{d z}{2\pi i z}\  e^{-2 i \phi(z)},\cr
&X'=\oint\frac{d z}{2\pi i z}\  e^{ i \phi(z)},}}
where the integration contours are around the zero and
$$\phi(z)=Q-i\ P\ln z+\sum_{\scriptstyle m \in {\bf Z}\scriptstyle\atop
m\not=0}\ \frac{b_m}{i\  m} z^{-m}\ .$$
It is easy to see that operator \ $X'$\ is nilpotent one,
$X'^2=0\ .$
Let us define the following spaces:
\eqn\gstaqfd{\eqalign{
&{\cal F}^+_{k-\frac{1}{2}}=Ker_{{\cal F}_{k-\frac{1}{2}}}[ X']
\ ,\cr
& {\cal F}^{-}_{k-\frac{1}{2}}=
{\cal F}_{k-\frac{1}{2}}/{\cal F}^{+}_{k-\frac{1}{2}} .}}
Since the   operators
\ $X,X', P$\  obey the commutation relations:
\eqn\kdhfg{[X,X']=0,\ \   [P,X]=-2 X\ ,
\ \ [P,X']=X'\ ,}
then they act as
\eqn\kdfjfhj{\eqalign{
&{\cal F}^{\pm}_{k+\frac{1}{2}}\matrix{X\cr\longrightarrow\cr{}}
{\cal F}^{\pm}_{k-\frac{3}{2}}\ ,\cr
&{\cal F}^{-}_{k-\frac{1}{2}}\matrix{X'\cr\longrightarrow
\cr{}}  {\cal F}^{+}_{k+\frac{1}{2}}\matrix{X'\cr\longrightarrow
\cr{}}0\ .}}
In the simple case we are considering now, it is not
hard to guess the expressions of \ $Z_a(\zeta)$\ through the
field \ $\phi$.
Finally, the bosonization of \ ${\pi}_Z^R$\
is described by the

\noindent
{\bf Proposition 3.1}
\noindent
\it
The map \ $\pi_Z^R\to \oplus_{k\in Z}^{\infty}
{\cal F}^+_{k-\frac{1}{2}}$ given by the identification of the R-vacuum
state of \ ${\pi}_Z^R$\ with
vector \ ${\bf f}_{\frac{1}{2}}$ and  bosonization rules
\eqn\lshh{\eqalign{
&Z_+(\zeta)= \ e^{i\phi(\zeta)}\ ,\cr
&Z_-(\zeta)= \ [X,Z_+(\zeta)]=2  \zeta\p_{\zeta}
e^{-i\phi(\zeta)}}}
is an isomorphism of modules.

\rm
\noindent
Note that it  follows from the formulae \kdhfg\ that operators\
$X$\ and $H=P-\frac{1}{2}$\ can be considered as  generators of
the  Borel subalgebra of \ $sl(2)$\ algebra  acting in
the space\ $\pi^{R}_Z$.

Basing on the proposition 3.1, one can treat
\ ${\pi}_Z^R$\ as a direct sum of Virasoro algebra modules.
The crucial observation here
is that the subspaces \ $Ker_{{\cal F}^+_{l-\frac{1}{2}}}[X^l], l>0\ $
can be endowed by the
structure of irreducible representation of the Virasoro
algebra. To explain this important property
let us introduce \ $SL(2)$\ scalar  field\ $T(\zeta)$\ by the formula:
\eqn\bxvc{- 4\ \zeta^2T(\zeta)=Z_+(\zeta)Z_-(\zeta)-Z_-(\zeta)Z_+(\zeta)\ .}
In terms of bosonic field\ $\phi$\ it reads
\eqn\kdhfg{   2\  T(\zeta)=
f^2-\p_{\zeta} f\ ,}
where \ $f=i\p_{\zeta}\phi(\zeta)-\frac{1}{2\zeta}$.
It is easy to check that \ $L_n$\ :\ $
T(\zeta)=\sum_{n\in{\bf Z}} \ L_n \zeta^{-n-2}\ $
generate the Virasoro algebra \ $Vir_c$\
\eqn\gsfd{[L_n,L_m]=(n-m)L_{n+m}+\frac{c}{12}(n^3-n)\ }
with
$\ c=-2$.
Let us denote by \ ${\cal L}_{l-\frac{1}{2}}, \ l>0$\ the
irreducible Verma
module with
highest weight
\eqn\peiorut{\Delta_{l-\frac{1}{2}}=\frac{l(l-1)}{2}}
built upon
highest vector \ ${\bf v}_{l-\frac{1}{2}}$. If generators of the Virasoro
algebra are realized as \kdhfg \ then any
highest weight vector \ ${\bf f}_{l-\frac{1}{2}}$\ of
Fock space turns to be the highest weight vector of Verma module
of Virasoro algebra with
highest weight \peiorut .
Moreover, consider in bosonic space \ ${\pi}_Z^R$\
non-zero vectors which can be obtained from \ ${\bf f}_{l-\frac{1}{2}}$\
by the action of
operator \ $X^k,\ k=0,1,...$\ . We claim that generators
\ $L_n$\ with \ $n>0$\ annihilate any such vector
while \ $L_0$\
acts by multiplication on constant \ $\Delta_{l-\frac{1}{2}}$.
It
follows from the fact that
generators \kdhfg \ commute with operator \ $X$ as it can be
easily checked.

Let us
consider formal tensor product of irreducible
Virasoro algebra module \ ${\cal L}_{l-\frac{1}{2}}$\
and \ $l$-dimensional \ $sl(2)$\   irreducible representation
\ ${\cal V}_{l}^{+}$\ with basic vectors
\ ${\bf e}^{m+}_{l}$\ ,
where \ $m=-j,...,j$ and \ $l=2j+1$,
\ $j=0, \frac{1}{2}, 1,...$\  . We will
denote it as \ ${\cal L}_{l-\frac{1}{2}}\otimes
{\cal V}^{+}_{l}$.
The meaning of the additional index "+"  will be clarified later.
Standard arguments \refs{\fei}
possess to prove the following proposition describing Virasoro
structure of the space \ ${\pi}_Z^R$ ( fig. 1):

\par
\noindent
{\bf Proposition 3.2}

\noindent
\it
The map \ ${\pi}_Z^R\rightarrow
\oplus_{l=1}^{\infty}{\cal L}_{l-\frac{1}{2}}\otimes
{\cal V}^{+}_l$\ given by correspondence
\ $X^{j-m}\ {\bf f}_{l-\frac{1}{2}}
\rightarrow {\bf v}_{l-\frac{1}{2}}\otimes {\bf e}^{m+}_{l}\ (l=2j+1)$
and bosonization rule \kdhfg\ is an isomorphism of
modules of the  Virasoro algebra  with
\par
\noindent
i)
\eqn\teryru{Ker_{{\cal F}^{+}_{l-\frac{1}{2}}}[X^{l}]
\cong {\cal L}_{l-\frac{1}{2}}\ , l>0;}
\par
\noindent
ii)
\eqn\hdgff{{\cal F}^+_{2m+\frac{1}{2}}\cong \oplus_{k=0}^{\infty}\
{\cal L}_{2|m|+2k+\frac{1}{2}}
\otimes {\bf e}^{m+}_{2|m|+2k+1}\ , \ m\in \frac{1}{2} {\bf Z} \ . }

\rm
\ifig\ffplus{The Virasoro structure of the  space \ $\oplus_{k\in Z}
{\cal F}^+_{k-\frac{1}{2}}$\ }
{\epsfxsize4.5in\epsfbox{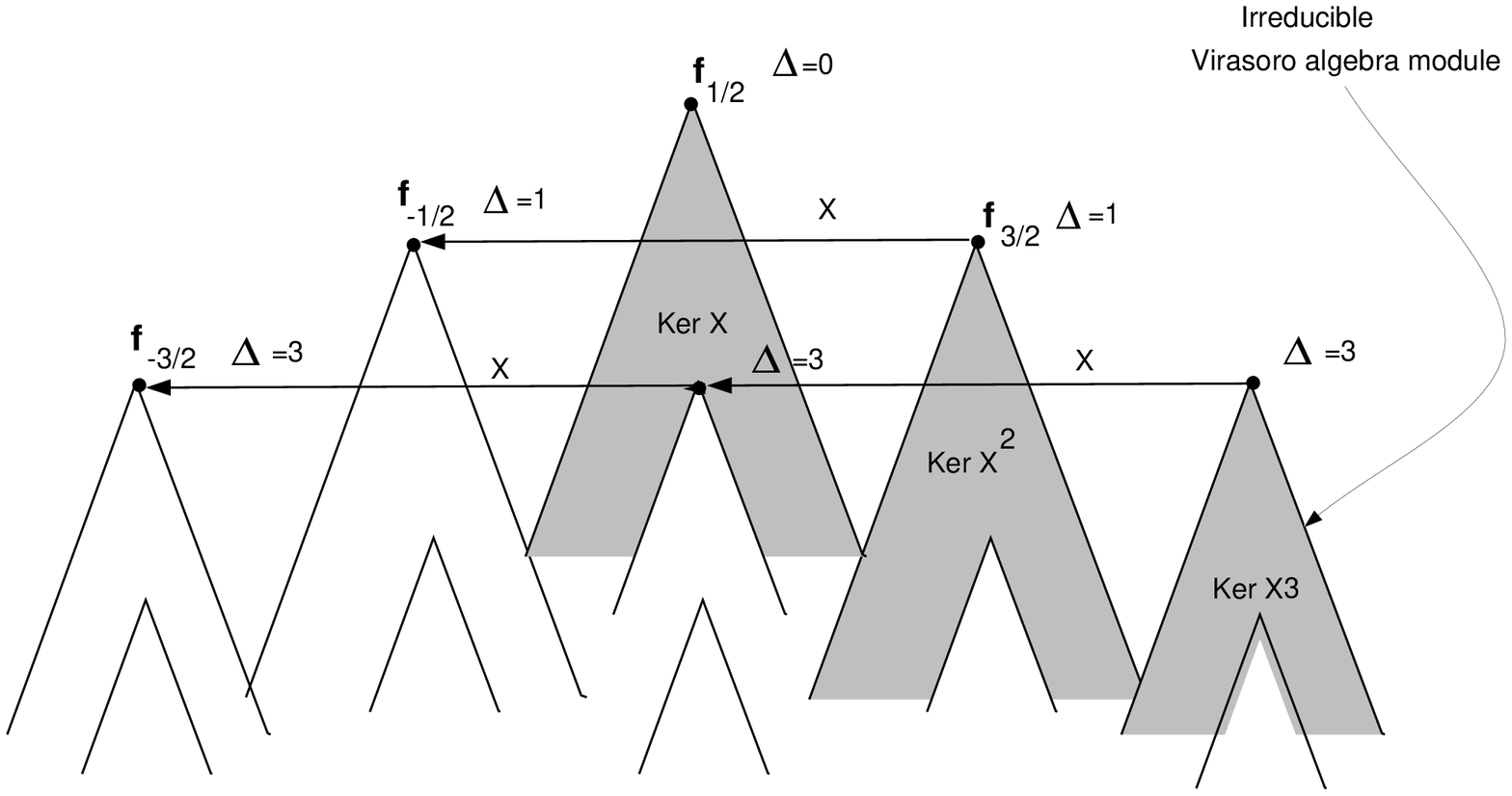}}


\par
\noindent
Notice, that such as operators\ $L_n$\ and $X'$\ commute, then
the space\ ${\cal F}^{-}_{k-\frac{1}{2}}$
has the similar Virasoro structure as \ ${\cal F}^{+}_{k+\frac{1}{2}}$:
\eqn\lakjsk{{\cal F}^{-}_{2m-\frac{1}{2}}\cong \oplus_{k=0}\
{\cal L}_{2|m|+2k+\frac{1}{2}}\otimes {\bf e}^{m-}_{2|m|+2k+1}\ ,
\ m\in \frac{1}{2}{\bf Z}, }
where we denoted by  \ ${\bf e}^{m-}_{l}$\ the basis vectors
in \ $l$-dimensional space\ ${\cal V}^{-}_{l}$. Note
that vector\ ${\bf v}_{l-\frac{1}{2}}\otimes {\bf e}_l^{-m,-}$\
where \ ${\bf v}_{l-\frac{1}{2}}$\  is highest vectors in
Virasoro module is
identified with the state
from \ ${\cal F}^-_{-2m-\frac{1}{2}}$\ which is the proimage
of
\  $\ X^{ j+m}\
{\bf f}_{-l+\frac{1}{2}}\ $\ with respect to the action
of operator \ $X'$. In particular
\ ${\bf v}_{l-\frac{1}{2}}\otimes{\bf e}^{-j, -}_{l}
\equiv {\bf f}_{-l+\frac{1}{2}}\ mod\  Ker X'.$
The difference which appears in this case
is in the fact that irreducible Virasoro
algebra modules are identified with
factor spaces rather than subspaces of
Fock space.
\noindent

\subsec{Scalar product in Fock space.}
Now we turn to the bosonization of dual representation
\ ${\pi}_Z^{R\star}$\ of ZF algebra. This problem is closely related with
the proper choice of the scalar product in the direct sum of
Fock spaces
\ $\oplus_{k\in Z}{\cal F}_{k-\frac{1}{2}}$.
The obvious
guess is
\eqn\vcsda{({\bf u}_1b_n,{\bf u}_2)_{0}=-({\bf u}_1,b_{-n}{\bf u}_2)_{0},
\  ({\bf u}_1P,{\bf u}_2)_{0}=-({\bf u}_1,P{\bf u}_2)_{0}\ . }
However, it does not conform with  conjugation condition
\mj\  for operators \ $Z_a$.
In thinking about this problem, it is rather natural
to introduce independent bosonic representation  for \ ${\pi}_Z^{R\star}$\
and then try to identify it with some subspaces
in the Fock space
\ $\oplus_{k\in Z}{\cal F}_{k-\frac{1}{2}}$. To do this, we  define
new set of generators\ $\{b^{\star}_n,P^{\star},
Q^{\star}|n=\pm1,\pm2,...\}$\
with the commutation relations as in\ \lnsgd . Let analogues for
integral operators \idreg \ be
\eqn\shtrih{\eqalign{
&X^{'\star}=
\oint\frac{d z}{2\pi i z} e^{ i \phi^{\star}(z)}\ ,\cr
& X^{\star}=
\oint\frac{d z}{2\pi i z} e^{-2 i  \phi^{\star}(z)}\ .}}
Repeating step by step the analysis above, one can describe the
bosonization of the representation  \ ${\pi}_Z^{R\star}$\ and
introduce Virasoro algebra structure in it. The formulae for this
case are rather evident.
In particular \ $\pi_Z^{R\star}\cong \oplus_{k\in Z} \
{\cal F}^{\star +}_{k-\frac{1}{2}}
$, where we denote by \ ${\cal F}^{\star +}_{k-\frac{1}{2}}$\
the kernel
of the operator
\ $X^{'\star}:\ {\cal F}^{\star}_{k-\frac{1}{2}}
\rightarrow {\cal F}^{\star}_{k+\frac{1}{2}}$. And any space
\ ${\cal F}^{\star +}_{k+\frac{1}{2}}$\
turns to be isomorphic to direct sum of irreducible
Virasoro algebra modules :
\eqn\hdgff{ {\cal F}^{\star +}_{2m+\frac{1}{2}}\cong
\oplus_{k=0}^{\infty}
{\cal L}^{\star}_{2|m|+2k+\frac{1}{2}}
\otimes {\bf e}^{\star m}_{2|m| +2k+1} \ .}
Here the states \ $X^{\star j-m}{\bf f}^{\star}_{l-\frac{1}{2}} \ (l=2j+1)$\
correspond to states
\ ${\bf v}_{l-\frac{1}{2}}^\star \otimes {\bf e}^{\star m}_{l}$.
Notice, that the symbol \ $\star$\ in the \ ${\cal L}^\star_{l-\frac{1}{2}}$
is used in order to emphasize the fact that generators
of Virasoro algebra are built from fields \ $\phi^\star$\
rather than from \ $\phi$'s. To proceed further, let us
note, that the decompositions \lakjsk \ , \hdgff \ of spaces
\ ${\pi}_Z^{R\star}$\ and \ $\oplus_{k\in{\bf Z}}F^{-}_{k-\frac{1}{2}}$\
include
the identical sets of irreducible representations of the Virasoro
algebra. Indeed, any two representations \ ${\cal L}_{l-\frac{1}{2}}$\
and \ ${\cal L}_{l-\frac{1}{2}}^\star$\ are isomorphic as\ $ c=-2$\  Virasoro
algebra representations since they have the same highest weight
\ $\Delta_{l-\frac{1}{2}}$. In the bosonic realization
this means that one can identify the Virasoro algebra generators.
\foot{Because of this identification   one
can consider
fields \ $i\p_{\zeta}\phi-\frac{1}{2\zeta}$\
and \ $i\p_{\zeta} \phi^{\star}-\frac{1}{2\zeta}$\
as different solutions
of   quantum version of the Riccati equation\ \kdhfg \ \refs{\gerv}\ .}
To extend isomorphism of Virasoro algebra modules
on the isomorphism of \ ${\pi}_Z^{R\star}$\
and \ $\oplus_{k\in{\bf Z}}{\cal F}^{-}_{k-\frac{1}{2}}$\
one needs establish the correspondence between
vectors in \ ${\cal V}_{l}^\star$\ and \ ${\cal V}^{-}_{l}$\ .
Let us identify
vectors \ ${\bf e}^{m\star}_{l}\in {\cal V}_{l}^\star$\
and \ $c^j_m\ {\bf e}^{-m,-}_{l}\in {\cal V}^{-}_{l}.$
The constants \ $c_m^j$\ can be fixed by the condition
that the operators \ $X,  X^{\star}$\ and \ $H=-P-\frac{1}{2}$\
acting in the space\ $\oplus_{k\in Z}{\cal F}^-_{k-\frac{1}{2}}
\equiv\oplus_{k\in Z}{\cal F}^{\star +}_{k-\frac{1}{2}}$
generate the \ $sl(2)$\ Lie algebra.
\eqn\gsfddw{[X,  X^{\star}]=H,
\ \  [ H,X]=2 X,\
[ H, X^{\star}]=-2 X^{\star}\ .}
Then we will have\ $c_m^j=\frac{(j-m)!(2j)!}{(j+m)!}$.
Hence, we conclude that the representation
\ ${\pi}_Z^{R\star}$\
can be realized as following:
\eqn\vcxasdge{
{\pi}_Z^{R\star}=\oplus_{k\in{\bf Z}}{\cal F}_{k-\frac{1}{2}}^-\ .}
Note that due to  this identification, an action
of the operator
\ $X^{\star}$\ in the space\
$\oplus_{k\in{\bf Z}}{\cal F}_{k-\frac{1}{2}}^-$
can be
uniquely specified by the condition that the
following diagram is
commutative:
\eqn\sgdft{\eqalign{\matrix{&{\ } & {\ }    &X' &{\ }\cr
&{\ } &{\cal F}^-_{k-\frac{3}{2}}
&\longrightarrow &{\cal F}^+_{k-\frac{1}{2}} &{\ }\cr
& X^{\star}&\biggl\downarrow & X'&\biggl\downarrow & X^{\star}\cr
&{\ } &{\cal F}^-_{k+\frac{1}{2}}
&\longrightarrow &{\cal F}^+_{k+\frac{3}{2}}&{\ }\cr}\ .}}

Now we are able to describe total symmetry algebra\ $Symm$\  which acts
in the space \ ${\pi}_Z^R\oplus {\pi}_Z^{R\star}$. First, it includes
infinite dimensional Virasoro algebra
with central charge \ $c=-2$.
Second, it contains \ $sl(2)$\  subalgebra.
Finally,  the operators\ $X'$ and\ $X'^{\star}$ \
generate the Clifford  subalgebra:
\eqn\jshdgp{\{X',X^{'\star}\}_+=1,\ \
X'^2= \big[X^{'\star}\big]^2=0\ .}
Note that the operators \ $X', X'^{\star}$\  commute with
Virasoro generators and\ $X,X^{\star}$ but do not commute
with \ $H$.
The decomposition of the space\ ${\pi}_Z^R\oplus
{\pi}_Z^{R\star}$\
into direct sum of irreducible representations of the algebra \ $Symm$\ has the
form:
\eqn\jshdg{{\pi}_Z^R\oplus {\pi}_Z^{\star R}
=\oplus_{l=0}^{\infty}\
{\cal L}_{l-\frac{1}{2}}
\otimes {\cal V}_{l}\otimes{\cal V}'_{2}\ ,}
where we denoted the \ $l$-dimensional irreducible representation
of algebra\ $su(2)$\ as \ ${\cal V}_{l}$\ and \ ${\cal V}'_{2}$\
is
two dimensional representation of
Clifford algebra\ \jshdgp .
Decomposition \jshdg \  makes clear the symbolic notations
introduced earlier. Namely, vectors \ ${\bf e}^{m\pm}_{l}$\
are basic vectors in the
irreducible \ $l$-dimensional
representations
of \ $sl(2)$\ algebra
with spin \ $j=\frac{l-1}{2}$\ and momentum projection \ $m$.
Any pair of vectors \ ${\bf e}^{m-}_{l}$\
and \ ${\bf e}^{m+}_{l}$\ is arranged in doublet
of the Clifford algebra.

Let us discuss now the Hilbert structure of the space
\ ${\pi}_Z^R\oplus {\pi}_Z^{R\star}$.
This space has a canonical
scalar product   \  $({\bf u}_1 ,{\bf u}_2 )$\  induced by condition:\
$({\bf u}_1,{\bf u}_2)=({\bf u}^{\star}_1, {\bf u}^{\star}_2)=0,\
({\bf u}_1^{\star},{\bf u}_2)=
({\bf u}_1,{\bf u}_2^{\star})^*=<{\bf u}_1^{\star},{\bf u}_2>
{\rm for\  any}\
{\bf u}_{1,2}\in {\pi}_Z^R,\ {\bf u}_{1,2}^{\star}
\in {\pi}_Z^{R\star}$.
{}From the other side such scalar product is equivalently
described as following:
\eqn\gfor{({\bf e}^{n\  b}_{2j'+1},{\bf e}^{m\  a}_{2j+1})=
(-1)^{j+m}\frac{2 ^{2j}(2j)!}{2j+1}\
\delta_{j,j'}\ \delta_{m+n,0}\ \delta_{a+b,0}\ }
and the scalar product in the irreducible representations
of Virasoro
algebra  is defined by the conjugation conditions
\eqn\hsfdd{({\bf v}_1L_n,{\bf v}_2)=({\bf v}_1,L_{-n}{\bf v}_2)\ .}
We will also  assume that the scalar
product of  the highest Virasoro vectors  have the form
\eqn\jdhg{({\bf v}_{l_1-\frac{1}{2}}^{\star}, {\bf v}_{l_2-\frac{1}{2}})=
l_1\ \delta_{l_1,l_2}\ .}
Notice that  the scalar product\ \gfor -
\hsfdd\   is drastically different from\ \vcsda  .

\subsec{NS sector.}
Now let us shortly describe the bosonization
of NS sector of representation of ZF algebra. The consideration
here is quite similar as it have been done for R-sector and we
present only results. The space\ ${\pi}_Z^{NS}$\ can be realized
as a following  direct sum of Fock modules:
\eqn\mxn{{\pi}_Z^{NS}=\oplus_{k\in Z} {\cal F}_{k}\ , }
where NS-vacuum is identified with \ ${\bf f}_0$.
The expression for operators
$Z_a(\zeta)$ in terms of bosonic field \ $\phi$\ is given
by formula \lshh \ again. The main difference here in comparison
with Ramond sector case is
that operator\ $X'$\ does not act in NS sector.
It happens since the integration contour in the
definition
of \ $ X'$ is not closed.

The states in the  Fock module \ ${\cal F}_k$\ can be classified
with respect to
the action of direct product of Virasoro and $\ sl(2)$\ algebras:
\eqn\hf{{\cal F}_{2m}\cong\oplus_{k=0}^{\infty}\
{\cal L}_{2|m|+2k}\otimes {\bf e}^{m}_{2|m|+2k+1},
 \ m\in \frac{1}{2}{\bf Z}\ , }
where we denoted by
$\  {\cal L}_l, \ l>0$\
irreducible modules of Virasoro algebra built upon highest
vector \ ${\bf v}_l$\
with highest weight
\eqn\mvbnv{\Delta_{l-1}=\frac{(l-1)^2}{2}-\frac{1}{8}\ .}
Vectors \ ${\bf e}^{m}_{l}$\ are basic vectors in
the \ $l$-dimensional irreducible
representation of the \ $sl(2)$-algebra.
Again, the correspondence between
representations is given
by the maps of vectors $X^{j-m}\ {\bf f}_{l}\rightarrow
{\bf v}_l\otimes {\bf e}^{m}_{l} $\ and bosonization rules
for generators of Virasoro algebra and \ $sl(2)$\ algebra.
The scalar product in the space \ ${\pi}_Z^{NS}$\ can be introduced
analogously as for R-sector. Notice, that
in this case
the space \ ${\pi}_Z^{NS}$\ can be considered as selfdual representation of
ZF algebra:
$${\pi}_Z^{NS}={\pi}_Z^{NS\star}.$$

\subsec{Vertex operators for \ $Vir_{-2}$\ algebra.}
Now, we wish to find an exact meaning of the
operators \ $Z_{\pm}$\ in terms of \ $Symm$\ algebra.
Before going on, let us explain the situation on which
we will focus.  In   case under consideration
the algebra \ $Symm$\  turns to be a direct product
of infinite-dimensional Virasoro algebra
and finite algebra.
We expect, this statement is general in the sense
that it is possible to
associate for any ZF algebra another algebra which
would be a direct product of two parts, infinite-dimensional
(like Virasoro)
and finite ones.  We will demonstrate by
executing simple example \ki\ - \lik \ that
these operators can be
identified with  vertex operators for
the total symmetry algebra \ $Symm$.
Keeping in the mind the decompositions\ \jshdg\
it is rather natural to
study the action of the operators \ $Z_{\pm}$\
on the Virasoro states
separately from the action on the
finite-dimensional part of the representation space.
The operators \ $Z_{\pm}$\
have the spins \ ${1\over 2}$\ and momentum projections
\ ${\pm}{1\over 2}$\ with respect to \ $sl(2)$ algebra.
Then, if
\ ${\bf e}^m_{l}$\ are basic vectors in irreducible
representations\ ${\cal V}_l$\ and \ ${\bf v}\in
 {\cal L}_{p}$, where\ $p=l-\frac{1}{2}, p=l-1$\
correspondingly for R and NS sectors,
the operators  \ $\ Z_a(\zeta)$\
in the representations \ ${\pi}_Z^R$\ and \ ${\pi}_Z^{NS}$\
\  can be represented
in the form:
\eqn\kdjhf{\eqalign{Z_a(\zeta)\ {\bf v}\otimes {\bf e}^m_l}=
\sum_{b=\pm 1} \pmatrix{&\frac{1}{2}&j&j+\frac{b}{2}\cr
                      &\frac{a}{2}&m&m+\frac{a}{2}\cr}
\Phi^{p+b\ p}_{21}(\zeta)\ {\bf v}\otimes {\bf e}^{m+\frac{a}{2}}_{l+b}
\ ,}
where \ $l=2j+1$.
The numerical coefficients in this formula
coincide with the Clebsh-Gordan coefficients for \ $sl(2)$\ algebra.
At the same time
the  operators
\ $\Phi_{21}$\
act as following:
\eqn\jdhgd{\eqalign{ \Phi^{p\pm 1\ p}_{21}(\zeta)&:\ \
{\cal L}_{\Delta_p}\matrix{\Phi_{21}\cr
\longrightarrow\cr {}}{\cal L}_{\Delta_{p\pm 1}}\otimes\C[\zeta]\
\zeta^{\Delta_{p\pm 1}-\Delta_p}\
 \ ,} }
where\ $\C[\zeta]$\ denotes the Laurent series in\ $ \zeta $.
Using operators  \ $Z^{\star}\in End\ ({\pi}_Z^{R,NS\star})$\
one can define the operators \ $\Phi_{21}^{\star}$\
by similar expressions. It follows from the
formula \mj \ that the operators \ $\Phi_{21}$\  and
\ $\Phi_{21}^{\star}$\
\jdhgd\ satisfy the relation:
\eqn\mcnvnb{\big({\bf v}_1^{\star} \Phi^{\star\  p\  p\pm 1}_{21}(\zeta),
\ {\bf v}_2\big)=
\big({\bf v}_1^{\star},\Phi^{p\  p\pm 1}_{21} (\zeta)\ {\bf v}_2 \big)\ ,}
where\ ${\bf v}_1^{\star}\in {\cal L}^{\star}_{p}$ \ and
${\bf v}_2\in {\cal L}_{p\pm 1}.$

We claim that
the operators\ \jdhgd\   are vertex operators for the
Virasoro algebra. Indeed, using the bosonization \ \lshh
\ and conjugation condition\ \mcnvnb\
one can show that they satisfy   the following
commutation relations with Virasoro algebra generators \ $L_n$\ :
\eqn\gsfddwge{[L_n,\Phi^{\pm}_{21}(\zeta)]=\zeta^{n+1}\p_{\zeta}
\Phi^{\pm}_{21}(\zeta)
+\Delta_{21}\ n\ \zeta^n\ \Phi^{\pm}_{21}(\zeta)\ ,}
where \ $\Delta_{21}=1$. Here and after we will
use the short notation\
$\Phi^{\pm}_{21}$\  for operators\ $\Phi^{p\pm 1\  p}_{21}$.

Thus, ZF algebra appeared to be naturally connected
with chiral primary operators
of Virasoro algebra. The last objects are of a great important
for our next constructions.
It is well-known \refs{\BPZ},\ \refs{\DotsFat} that
chiral primary operators are uniquely determined by the
commutation relations with \ $L_n$, intertwining property and
conjugation condition.
{}From now on in this section let us
assume
that symbol \ ${\cal L}_p$\
denotes irreducible representation of \ $c=-2$\
Virasoro algebra with highest weight
$$\Delta_p=\frac{p^2}{2}-\frac{1}{8}\ ,$$
where \ $p\in {\bf R}$ rather then just integer or half-integer.
This generalization is very useful,
since, considering general case we will
be able separate general features of the bosonic
construction
from the specific properties of \ $SL(2)$-invariant
fermion model \ki -\lik .

Usually, the special attention is paid to
the studying of the properties of two chiral primary
operators  \refs{\BPZ},\ \refs{\DotsFat}.
The first of these operators is given by \ $\Phi_{21}$\ with properties
\jdhgd\ -\gsfddwge \
while the
second can be introduced
by the following conditions:
\par
\noindent
1. Intertwining properties
\eqn\jdhgdii{
\Phi_{12}^{p\mp\frac{1}{2}\ p}(\zeta)\ :\
{\cal L}_{p} \matrix{\Phi_{12}\cr
\longrightarrow\cr {}}
{\cal L}_{p\mp\frac{1}{2}}
\otimes \C[\zeta] \ \zeta^{\Delta_{p\mp\frac{1}{2}}-\Delta_p}\ . }

\noindent
2. Conjugation conditions:
\eqn\jdhr{
({\bf v}_1^{\star}\Phi_{12}^{\star\  p\ p\mp\frac{1}{2}}(\zeta),
{\bf v}_2)=
({\bf v}_1^{\star},\Phi_{12}^{p\ p\mp\frac{1}{2}}(\zeta), {\bf v}_2)\ .}
3. Commutation relations with Virasoro algebra generators:
\eqn\ksjd{[L_n,\Phi^{\pm}_{12}(\zeta)]=\zeta^{n+1}\p_{\zeta}
\Phi^{\pm}_{12}(\zeta)
+\Delta_{12}\ n\ \zeta^n\ \Phi^{\pm}_{12}(\zeta)\ ,}
with
$\Delta_{12}= -{1\over 8} $.
\par\noindent
Again, short notations
$\Phi^{\pm}_{12}$\
stand for operators\ $\Phi^{p\mp\frac{1}{2}\ p}_{12}$\
correspondingly.
The properties \  \jdhgd -\gsfddwge\
and \ \jdhgdii -\ksjd \
describe chiral primary operators \ $\Phi_{21}$\ and  \ $\Phi_{12}$\
uniquely up to multiple constant.
This means that basing only on
these formulae one can reconstruct matrix elements
of any combination of such operators.
Indeed, as it was shown by Belavin, Polyakov and Zamolodchikov \refs{\BPZ},
due to formulae \gsfddwge\ - \ksjd\
matrix elements of operators built from \ $\Phi_{21}$\ and  \ $\Phi_{12}$\
turn to
be solutions of certain linear differential
equations.
For instance, the functions
\eqn\mcn{\eqalign{&G_p^{\pm}
(\zeta_1\zeta^{-1}_2)=({\bf v}^{\star}_{p},\Phi^{\pm}_{21}(\zeta_2)
\ \Phi_{21}^{\mp } (\zeta_1) {\bf v}_p )\ ,\cr
&G'^{\pm}_p
(\zeta_1\zeta^{-1}_2)=({\bf v}^{\star}_{p},\Phi^{\pm}_{12}(\zeta_2)
\ \Phi_{12}^{\mp } (\zeta_1) {\bf v}_p )\ , \cr}}
where \ ${\bf v}_p$\ is highest vector in the
irreducible Virasoro module\ ${\cal L}_p$,
satisfy to the second order ordinary differential equations:
\eqn\skjdgh{L_{\Delta_{21}}\ G^{\pm}_p(\zeta)=0\ ,
\ \ \ L_{\Delta_{12}}\ G'^{\pm}_p(\zeta)=0\ .}
The explicit form of linear
differential operator\ $L_{\Delta}$\ is given by:
\eqn\mdhf{L_{\Delta}= \biggl\{ \frac{3}{2(2\Delta+1)}\p_{\zeta}^2+
\big(\frac{1}{\zeta}+\frac{1}{\zeta-1}\big)
\p_{\zeta}-\frac{\Delta_p}{\zeta^2}-\frac{\Delta}{(1-\zeta)^2}+
\frac{2\Delta}{\zeta(\zeta-1)}
\biggl   \}\zeta^{-\Delta}\ .}
According to this formula
the functions \mcn \
have the following asymptotics under \ $\zeta\to 0$:
$$G^{\pm}_p(\zeta)=O( \zeta^{\frac{1}{2}\mp p})\ ,\ \ \
G'^{\pm}_p(\zeta)=O( \zeta^{\frac{1}{8}\pm \frac{p}{2} })\ .$$
Up to normalization, the solutions of the
equations \skjdgh \ with boundary
conditions above can be easily
expressed through hypergeometric functions \ $F(a,b,c;\zeta)$\
as:
\eqn\jdhf{\eqalign{&G^{\pm}_p(\zeta)=  \zeta^{\frac{1}{2}\mp p}
\ \frac{1\mp 2p+\zeta(1\pm 2 p)}{(1-\zeta)^2}\ ,
\cr
&G'^{\pm}_p(\zeta)=\frac{\Gamma(\pm p+\frac{1}{2})}
{\pi^{\frac{1}{2}}  \ \Gamma(\pm p+1)}\
\zeta^{\frac{1}{8}\pm\frac{p}{2}}(1-\zeta)^{\frac{1}{4}}\
F(\pm p+\frac{1}{2},\frac{1}{2},\pm p+1;\zeta)\ .}}

Let us now take numbers \ $p\neq 0,\frac{1}{2},1,...$\ .
Knowing explicit formulae \jdhf\ one can find
the commutation relations of the
algebra of chiral primary operators. It can be done by
using the relation:
\eqn\fhj{\eqalign{
F(a,b,c;z)&=\frac{\Gamma(c)\ \Gamma(b-a)}{\Gamma(b)\ \Gamma(c-a)}
\ (-z)^{-a}\ F(a,a-c+1,a-b+1;z^{-1})\cr
&\ \ \ \ \ +\frac{\Gamma(c)\ \Gamma(a-b)}{\Gamma(a)\ \Gamma(c-b)}
\ (-z)^{-b}\
F(b,b-c+1,b-a+1;z^{-1})\ .}}
The straightforward calculation ensures that
operators\  $ \Phi_{1,2}$\ generate ZF
algebra of IRF type:
\eqn\hdfa{
\Phi_{12}^a(\zeta_1)\Phi_{12}^b(\zeta_2)|_{{\cal L}_p}
=\sum_{c+d=a+b}{\bf W}
\left[\matrix{p+\frac{a+b}{2}&p+\frac{c}{2}\cr p+\frac{b}{2}  &p}
\biggl|\ \sigma_{12}\ \right]\
\Phi_{12}^d (\zeta_2)
\Phi_{12}^c(\zeta_1)|_{{\cal L}_{p}}\ .}
The
precise form of nontrivial elements of matrix ${\bf W}$ is
the following:
\eqn\ksh{\eqalign{&{\bf W}
\left[\matrix{p\pm 1&p\pm \frac{1}{2}
\cr p\pm \frac{1}{2}&p}\biggl|\sigma\right]=e^{\frac{i\pi \sigma}{4}}
\ ,\cr
&{\bf W}
\left[\matrix{p&p\pm \frac{1}{2}\cr p\pm \frac{1}{2}\ &p}\biggl|\sigma\right]=
\mp \frac{e^{- \frac{i\pi \sigma(1\pm 4p)}{4}}}{\sin\pi p}, \cr
&{\bf W}
\left[\matrix{p&p\mp\frac{1}{2}\cr p\pm \frac{1}{2}\ &p}\biggl|z\right]=
\pm e^{-\frac{\pi i \sigma    }{4}}\ \cot\pi p\  .}}
The commutation relations \hdfa \ should be considered
as a rule of analytical continuation of
functions
from region \ $|\zeta_1|>|\zeta_2|$\ to
\ $|\zeta_1|<|\zeta_2|$\ along the paths \ $C_{\sigma_{12}},
\sigma_{12}=\pm$.
\ifig\fprodoljenie{Two contours of analytical continuations
\ $C_{\pm}$\ of functions \ $G'^{\pm}_p(\zeta)$\ .  }
{\epsfxsize1.5in\epsfbox{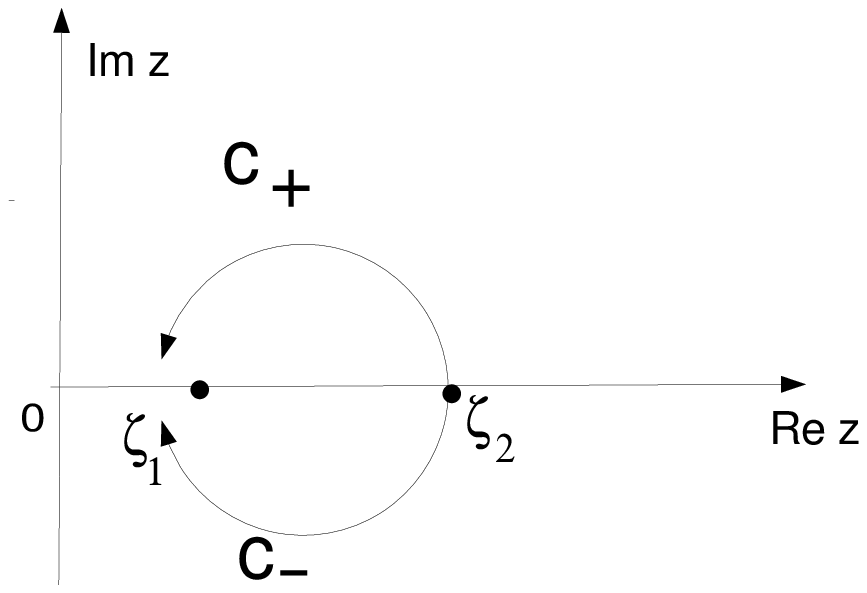}}
\par\noindent
Therefore, the
dependence of the matrix
\ ${\bf W}$\ on the variable\ $z$\ appears
only in the choice of the
contour of the analytical continuation.
The commutation relations of operators \ $\Phi_{21}$\
are more simple. It
hold for any value of \ $p$:
\eqn\hdfa{
\Phi_{21}^a(\zeta_1)\Phi_{21}^b(\zeta_2)
=-\Phi_{21}^b(\zeta_2)\Phi_{21}^a(\zeta_1)\ .}
If number \ $p$\ is integer or half-integer,
then commutation relations of
chiral primaries \ $\Phi_{12}^a$\ become undetermined
and algebra of chiral primaries would be
meaningless. The technical reason for this inconsistency
is in the fact
that under half-integer values of \ $p$\ the hypergeometric
function in \ $G'^{\pm}_p(\zeta)$\ degenerates.
Such
phenomena is well-known. It appears if one
study degenerated modules
of Virasoro algebra with rational
central charge.
\foot{In certain cases (minimal models) it is possible to reduce
the space of states such that chiral primaries would be correctly defined
and generate associative algebra \refs{\ffk} . }
To avoid additional complications we omit such cases,
considering either special value of central charge \ $c$\
and modules in generic position
(further in this section and in section 4),
or degenerated modules but for
generic central charge (as in section 5 below).

In spite of failure with algebra of chiral primaries,
the model
with \ $c=-2$\ and \ $p=0,\frac{1}{2},1,...$\ elaborated above
has several attractive features. First,
the algebra of operators \ $\Phi_{21}$\ is still
the same as for the generic \ $p$. The
fermionic representation of these operators
drastically simplify the analysis of the construction. Second,
in this model we have explicit bosonic realization of
finite-dimensional subalgebra in the symmetry algebra
which is given by operators \ $X,X', P$. And finally,
such model gives us example
how to find the bosonization
of chiral primary operators. One can expect that
form of bosonic  realization of chiral
operators should be the same, no matter whether the action of
intertwining operators \ $X,X'$\ is defined or not.

\subsec{Bosonization of vertex operators}
Definitions of chiral primaries do not assume bosonization procedure.
However,
we will find bosonic realization
of such operators because it admits direct generalization
for less trivial cases.
Since chiral primary operators
do not depend on the realization of \ ${\cal L}_{p}$, then
its matrix elements are the same for any
irreducible Virasoro module with given conformal
dimension \ $\Delta_p$. Let us consider at first
the cases when \ $p$\ is positive integer or half-integer number.
It is convenient to realize the irreducible Virasoro
modules  \ ${\cal L}_p,
({\cal  L}^{\star}_p$ )
as a subspaces (factorspaces)
$Ker_{{\cal F}_{p}}[X^l] ,
{\cal F}_{-p}/Im[X^{l}]$
in the Fock modules where \ $l=p+\frac{1}{2}$\ for R sector and
\ $l=p+1$\ for NS sector. Then we have
\par
\noindent
{\bf Proposition 3.3}

\noindent
\it
The following vertex operators
of representation \ $\pi_Z$\
can be realized as the  bosonic operators
in the Fock submodules
\par
\noindent
i)
\eqn\hsbgdgf{\eqalign{&\Phi^{+}_{21}(\zeta)\ {\bf v}=
e^{i\phi(\zeta)}\ {\bf v}, \cr
&\Phi^{\star\  +}_{21}(\zeta)\ {\bf v}^{\star}=
\int_C\frac{d z}{2\pi i z}\
e^{-2i\phi(z)} \ e^{i\phi(\zeta)}  \ {\bf v}^{\star}\ ,}}

\par
\noindent
ii)
\eqn\nxhc{\eqalign{
&\Phi_{12}^{\star\ -}(\zeta) {\bf v}^{\star}
=e^{-\frac{i}{2}\phi(\zeta)}\ {\bf v}^{\star}\ ,\cr
&\Phi^{-}_{12}(\zeta)\ {\bf v}=
\int_C\frac{d z}{2\pi i z}\
e^{i\phi(z)} \ e^{-\frac{i}{2}\phi(\zeta)}  \ {\bf v}\ ,}}
here\ ${\bf v}\in {\cal L}_{p}$\ and\ ${\bf v}^{\star}\in
{\cal L}^{\star}_{p}\  ;$
\rm
\par\noindent
The integration contour \ $C$\ going counterclockwise is
chosen to have the beginning and end at the origin of
complex \ $z$-plane. It encloses all singularities of which positions
are determined by the vector \ ${\bf v}^\star({\bf v})$.
Notice, that
due to this prescription the
integrals are well defined.
%

\par\noindent
This proposition is rather
evident since
part {\it i)} is a direct
consequence of the equation\  \lshh \ while the
proof of part {\it ii)} is based on the following important
property of operator\ $ e^{-\frac{i}{2}\phi(\zeta)}$\ :
\eqn\nmshs{X|_{\pi^{NS}}\
e^{-\frac{i}{2}\phi(\zeta)}=- e^{-\frac{i}{2}\phi(\zeta)}\
X|_{\pi^R}\ .}
Now we want to check our guess that
formulae \hsbgdgf -\nxhc \ describe the bosonization
of chiral primaries for any \ $p>0$.
Let us note, that
irreducible Virasoro algebra
modules \ ${\cal L}_{p}\ ({\cal L}_{p}^\star)$\
in the case of generic \ $p$\ are isomorphic to
Fock modules \ ${\cal F}_{p}\ ({\cal F}_{-p})$.
Indeed, the action
of Virasoro algebra in \ ${\cal F}_{p}$\ can be determined through the
formulae \kdhfg  \ but
there are no intertwining
operators between Fock modules.
\foot{For instance, the action of the integral operators like  \ $X,X'$\
is not defined in these modules.}
Therefore operators \hsbgdgf -\nxhc \
act from irreducible Virasoro
algebra module into irreducible one.
So, one need to check only the commutation relations
of these operators with Virasoro algebra generators.
But it are rather evident such as
the integrals in \hsbgdgf -\nxhc \       are well-defined in the
general case
too.
Now straightforward computation prove that
proposition 3.3  is still true for the chiral primary operators in the
case of generic value of \ $p$.

Bosonization prescription \ \hsbgdgf -\nxhc  \ and conjugation
condition\ \mcnvnb \ , \jdhr\  allow to
work out the integral representation for any matrix element
of operators \ $\Phi^{\pm}_{12}$\ and \ $\Phi^{\pm}_{21}$.
For instance,
let us write down alternative derivation of
the function
\ $G'^{\pm}_p(\zeta)$.
It would seems at first glance that calculation of
matrix elements using the scalar product \gfor -\hsfdd \ is
very non-trivial problem. The essence of the bosonization
method is in the fundamental fact
that
the scalar product\ \vcsda\   restricted on
the submodules of Fock modules

\par
\noindent
i)\ \ $Ker_{{\cal F}_{p} }[X^l]\ ,\
{\cal F}_{-p}/Im[X^{l}]$\ ,
\ $l=p+\frac{1}{2}$\ for R sector, \ $l=p+1 $\ for NS sector; \ $p$\
is positive integer or half-integer,

\par
\noindent
ii)\ \ ${\cal F}_{p}\ , \ {\cal F}_{-p}$ \ , \ for generic\  $ p$,
\par
\noindent
which are isomorphic to irreducible modules of
Virasoro algebra,
coincides with\ \hsfdd .
Using this fact
and \  \hsbgdgf -\nxhc \ we get the formula:
\eqn\vxccs{({\bf v}^{\star}_{p},\Phi^{+}_{12}(\zeta_2)
\ \Phi_{12}^{- } (\zeta_1) {\bf v}_{p} )=
\int_{C}\frac{d z}{2\pi i z }\ ({\bf f}_{-p},
e^{-\frac{i}{2}\phi(\zeta_2)}\ e^{i\phi(z)}\ e^{-\frac{i}{2}\phi(\zeta_1)}\
{\bf f}_{p}
)_0\ ,}
where the integration contour\ $C$\ is determined by the same prescription as
in \nxhc .
The formula \ \vxccs\ leads to integral representation for functions
\ $G'^{+}_p(\zeta)$\ if we use the well-known rules
for average of the exponent operators
in the bosonic Fock space:
\eqn\vx{ ({\bf f}_{-p},
e^{i \ell_n\phi(\zeta_n)}....e^{i \ell_1\phi(\zeta_1)}{\bf f}_p)_0
=\prod_{k>m}(\zeta_k-\zeta_m)^{\ell_k \ell_m}\
\prod_k \zeta_k^{\frac{l_k^2}{2}+l_k p}
\ \delta_{\ell_1+...+\ell_n,0}\ .}
Then  the function\ \vxccs\
can be rewritten in term of hypergeometric function
via integral representation:
\eqn\ytww{
\int_C
\frac{d z}{2\pi i} z^{c-1}\
(1-z^{-1})^{-a}\ (1-\zeta z)^{-b}
=\
\frac{\Gamma(c+a)}{\Gamma(c+1)\ \Gamma(a)}
\ F(a+c,b,c+1;   \zeta)\ \ \ Re [c]>0.}
To find the function \ $\ G'^-_p$\ we can use
the relation
\eqn\sopr{
[G'^{+}_p(\zeta^{-1})]^{*}=G'^{-}_{-p}(\zeta),\ |\zeta|=1 \ ,}
which follows from the conjugation conditions\  \jdhr  .
In this way we reproduce  formula \  \jdhf .

\ifig\fgiper{Integration contour for hypergeometric function.}
{\epsfxsize2.0in\epsfbox{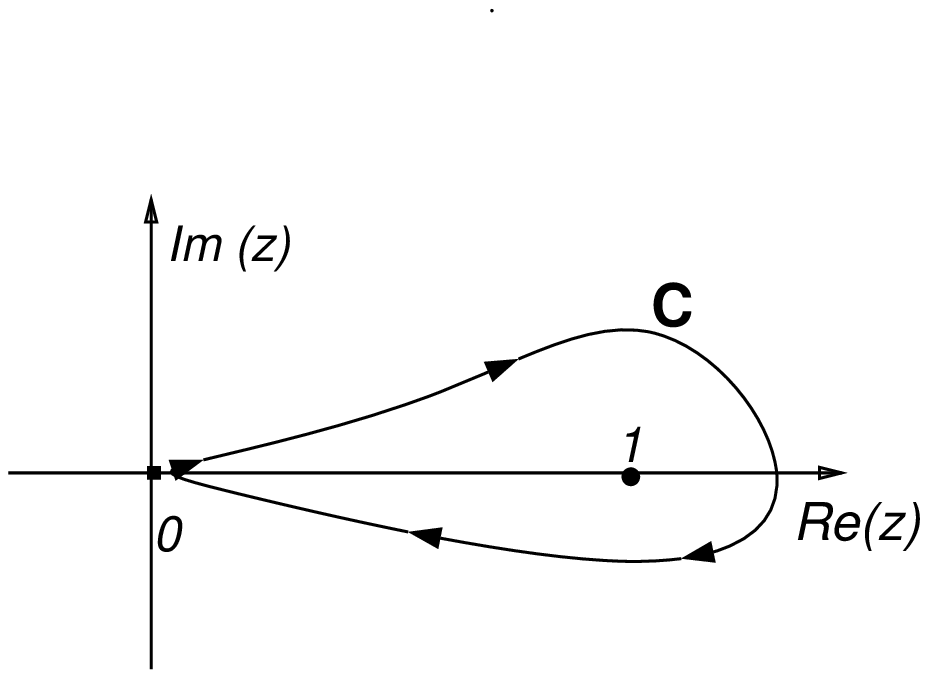}}
%


\newsec{Deformed Virasoro algebra with c=-2.}
In the previous section we have developed
bosonization procedure to study the representations
of ZF algebra. The worth important step there was the
introduction of integral operators \ $ X,X'$\ acting in the
bosonic space \ $\oplus_{k\in {\bf Z}}
{\cal F}_{\frac{k}{2}} $. In this space we introduced
the action of\ $ c=-2$\  Virasoro algebra by defining generators
\ $T(z)=
\sum L_nz^{-n-2}$ as composite operators built from Heisenberg
generators \ $b_n$. The property, that Virasoro algebra
generators commute with  \ $X$\ and \ $X'$\ allowed us
classify the representation space of ZF algebra by the
action of \ $Symm$\ algebra. Moreover, we pointed out that
operators \ $Z_\pm$\ itself can be constructed
as some linear combinations of vertex operators \ $\Phi_{21}$\
of Virasoro algebra. We also determined another elementary chiral
primary
\ $\Phi_{12}$\ the commutation relations of which are defined
by the constant R-matrix of IRF type.

Of course, representations of algebra \ki \ - \lik \
could be investigated directly in the fermionic language. However,
the bosonization procedure can be applied
in constructions where
fermionic description is lacking.
One of the lessons we have studied above is that ZF algebra, Virasoro
algebra, algebras of chiral primary operators and algebras
of screening operators are
mutually related in the bosonic picture.
We would like to examine further these
interrelations considering less trivial situations .

It will be convenient
to start with bosonic space and
deformed integral operators \ $X$\ and \ $X'$\
rather than with ZF algebra. Indeed,
operators \ $X$\ and \ $X'$\ could be treated as
completely
specifying the construction. For instance,
universal enveloping
algebra of Virasoro algebra could be considered
as subalgebra of operators from Heisenberg algebra
commuting with \ $X$\ and \ $X'$, while operators
\ $\Phi_{21},\Phi_{12}$\ were specified with
respect to Virasoro algebra, etc.
{} From this point
of view,  generalizations of the
proposed construction are determined by the appropriate
deformations of screening operators.

The well-known generalization of proposed bosonic
construction is originated in
deformation of operators \ $X$\ and \ $X'$\ in
a way which leads to
Virasoro algebra with central charge \ $-2<c<1$. This
direction will be discussed later.

There exists, however, another class of deformations \refs{\singordon}
relating with generalizations of the Virasoro algebra.
These deformations will be called hereafter by x-deformations.
Again,
x-deformation is determined by the appropriate
continuous
deformation of screening operators.  In distinction
with Virasoro algebra cases, x-deformation
leads
to more general
associative algebras of deformed vertex operators
\ $\Phi_{21},\Phi_{12}$.

In the present section we want to elaborate simplest example
of x-deformation
corresponding to
\ $c=-2$\ Virasoro algebra. In this particular instance
the deformation has remarkably simple form.
Working out this toy model we try to extract information on
essential properties of general
x-deformation. Namely, we argue that
the whole Fock
space can be decomposed
under the action of
operators \ $X$\ and \ $X'$\ into direct sum of
irreducible representations
of deformed Virasoro algebra
proposed in \refs{\kitaitzi}.
This algebra is determined as
subalgebra of operators from the universal enveloping algebra
of Heisenberg algebra which commute with
\ $X$\ and \ $X'$.
It should be clear that such commutants
always generate associative algebra. Indeed,
the linear space of commuting with \ $X$\ and \ $X'$\ operators
is closed under the operation
of multiplication, while
associativity condition follows from the fact that \ $T(\zeta)$\ are
build upon generators of the Heisenberg algebra.
Hence, in general case one can also expect that Fock
space will be divided into direct sum of representations of some
infinite-dimensional algebra
generalizing
Virasoro algebra.

Another purpose of this example is to demonstrate
how one can find the deformation of
chiral primary operators \ $\Phi_{12}$\ and \ $\Phi_{21}$.
More explicitly, how to obtain bosonic realization for such
operators. We will show that commutation relations of
operators \ $\Phi_{12}$\ are determined by elliptic R-matrix
of IRF type. This justify proposed deformation.

Further we will use
the same notations as an earlier, pointing only those
definitions and
formulae which contain significant distinctions.
For instance, symbol \ ${\cal L}$\ will mean the irreducible
representation of deformed \ $c=-2$\  Virasoro algebra, \ $\Phi_{12}$\
will be deformation of correspondent chiral primary etc.

\subsec{Deformation of the Virasoro algebra with \ $c=-2$}
Let us start with consideration
of the bosonic space \ $\pi_Z=\oplus_{k\in {\bf Z}}
{\cal F}_{k-\frac{1}{2}} $\ occurred in the section 3.
The x-deformation of the construction given in the
previous section is determined by the following
redefinition of the
operator \ $X$:
\eqn\efg{
X=\oint\frac{d z}{2\pi i z}\  e^{- i (\phi(zx)+\phi(zx^{-1}))}\ .
}
where deformation parameter
is a real number \ $0<x<1$. If second operator \ $X'$\
is still unchanged, then it is not hard to see that
formulae \gstaqfd \ -  \kdfjfhj \ are valid.
Let us fix notations of new operators, exact sense
of which will become transparent later:
\eqn\hhh{\eqalign{
&Z_+(\zeta)= \ e^{i\phi(\zeta)}\ ,\cr
&Z_-(\zeta)=[X,Z_+(\zeta)]}}
The integral in the definition of \ $Z_-$\ could be
computed explicitly:
\eqn\iii{Z_-(\zeta)=\frac{e^{- i \phi(\zeta x^2)} \
- \  e^{- i \phi(\zeta x^{-2})}}{x-x^{-1}}\ .}
In accordance with the scheme of previous section
let us describe at first the structure of the subspaces
\ $Ker_{{\cal F}^+_{l-\frac{1}{2}}}[X^l]$\ in Fock space. To do this
one need find operators which would generalize \bxvc .
It might be easily checked that the action of the
following operators in the Fock space \ $\pi_Z$
commute with \ $X$\ and \ $X'$\ :
\eqn\rrr{-\zeta^2\ T^{\nu}   (\zeta)=
\frac{Z_+(\zeta x^{-2\nu-1})
Z_-(\zeta x^{2\nu+1})-\ Z_-(\zeta x^{-2\nu-1})\ Z_+(\zeta x^{2\nu+1})}
{2(x+x^{-1})}\ ,}
where \ $\nu\in {\bf Z}$\ . Note that
$$ T^{\nu}(\zeta)\equiv T^{-\nu-1}(\zeta)\ .$$
In the
limit \ $x \rightarrow 1$\ operators \ $T^\nu(\zeta)$\ become
Virasoro algebra
generators expressed by formula \kdhfg .
Straightforward calculation ensure that
modes  of the
Laurent expansion
\ $T^\nu(\zeta)=\sum_{n\in{\bf Z}} \ L_n^\nu \zeta^{-n-2}$\
generate so-called deformed Virasoro algebra \refs{\kitaitzi}:
\eqn\kkk{\eqalign{
2[L_n^\nu,&L_m^\mu]=
[(\mu+1)n-(\nu+1)m]_{x^2}\ L^{\nu+\mu+1}_{n+m}
+[(\mu+1)n+\nu m]_{x^2}\ L_{m+n}^{\mu-\nu}-\cr
&-[\mu n+(\nu+1)m]_{x^2}\ L^{\nu-\mu}_{n+m}-
[\mu n-\nu m]_{x^2}\ L^{\nu+\mu}_{n+m}-
\cr
&\ \ \ \ -\frac{\delta_{m+n,0}}{4(x^2-x^{-2})^2}\ (C_n^{\nu , \mu}+
C_n^{\nu,   -1-\mu}+C_n^{-1-\nu , \mu}+C_n^{-1-\nu ,  -1-\mu})\ ,}}
where
$$C_n^{\nu, \mu}=\frac{[n(\mu+\nu)]_{x^2}}{[\mu+\nu]_{x^2}}-
\frac{[n(\nu-\mu)]_{x^2}}{[\nu-\mu]_{x^2}}\ .$$
We denote this
algebra as \ $Vir_{-2,x}$.
Let us adjoin to the algebra \kkk \
a derivation \ $D$\
with property
 $[D,L_n^\nu]=n L_n^\nu \ $\
for any \ $\nu\in{\bf  Z}$. In the bosonic Fock spaces this operator
can be realized as
\ $D=\sum_{m>0} b_{-m}b_m + \frac{P^2}{2}-\frac{1}{8}$.
It
provides the
universal enveloping algebra of
\ $Vir_{-2,x}$ with structure of \ $Z$-graded algebra.
Then the triangular decomposition of \ $Vir_{-2,x}$\  is
a decomposition into elements of positive
(\ $\ L_n^\nu , n<0$\ ), zero (\ $D, L_0^\nu$\  ) and
negative \ $(\ L_n^\nu , n<0\ )$ degrees for any \ $\nu\in{\bf  Z}$.
By the definition, Verma module of the deformed Virasoro algebra
is a \ $Z$-graded module generated by operators \ $\ L_n^\nu , n<0$\
under the action on the unique highest weight vector \ ${\bf v}$.
Verma module inherits structure of \ $Z$-graded space
from the grading of universal enveloping algebra.
Vector \ ${\bf v}$\ is the highest weight vector of Verma module with
weight
\ $\Delta$\
if it is annihilated by any operator
\ $\ L_n^\nu , n>0$, and the action of elements
with zero grading on it, yields
\ $ L_0^\nu{\bf v}=\Delta^{\nu}{\bf v}$\ ;\ $D{\bf v}=\Delta{\bf v}$.
In general situation
Verma module is degenerated, i.e. it contains invariant subspaces
created by action of operators \ $L_n^\nu ,\ n<0$\ on
null vectors. Null vectors, by definition,
obey the equation
\ $ L_n^\nu{\bf v}^0=0\ , n>0;\  \ D{\bf v}^0=(\Delta+N){\bf v}^0\ $\
with some integer \ $N$\ (degeneration level). In order to
obtain an irreducible representation, we have to put all
null-vectors together with the whole subspace
generated by it equal to zero; that is we have to
factorize Verma module by all
invariant submodules.

In spite of the fact that operator \ $X$\
and its commutants in the Fock
space have been
deformed, the description of structure of Fock space
\ $\oplus_k{\cal F}_{k-1/2}^+$\ in terms of
irreducible \ $Vir_{-2,x}$\ modules is almost
the same as an earlier. Indeed,
let generators \ $L_n^\nu$\ are
realized through \rrr . Then one can
demonstrate that the highest vectors \ ${\bf f}_{l-\frac{1}{2}}$\
of Fock module
turn to be highest vectors of the Verma modules of
\ $Vir_{-2,x}$. It is possible to show that
acting on the highest vectors of Fock space
\ ${\bf f}_{l-\frac{1}{2}}
\in {\cal  F}^{+}_{l-\frac{1}{2}}\ ,  \ l=1,2,...\ $\
generators \ $L_n^\nu$\ create subspaces
\ $Ker_{{\cal F}^{+}_{l-\frac{1}{2}}}[X^l]
$\
which are isomorphic to irreducible \ $Vir_{-2,x}$\  modules
${\cal L}_{l-\frac{1}{2}}$\ correspondingly. Notice, that
the character of irreducible
module of \ $Vir_{-2,x}$\ coincides with
the character of \ $ c=-2$\  Virasoro algebra:
\eqn\char{
Tr_{{\cal L}_{l-\frac{1}{2}}}\ [\  q^{D-\frac{c}{24}}\ ]
=q^{\frac{1}{12}}\ \frac{q^{\frac{l(l-1)}{2}}-q^{\frac{l(l+1)}{2}}}
{\prod_{n=1}^{\infty} (1-q^n)}\ . }
The numbers of the states
on given level will be conserved in a general x-deformation too.
Recall, that similar situation take place in the representation
theory of quantum groups \refs{\dr},\ \refs{\kirre},\ \refs{\Lu}\
where characters of deformed
irreducible modules remain to be the same as undeformed one
if the deformation parameter is not a primitive root of unit.

The spaces\ ${\cal F}_{l-\frac{1}{2}}^{+}$\ and
\ ${\cal F}_{l+1/2}^\star $\ can be
decomposed into the direct sum of \ $Vir_{-2,x}$\ modules
analogously to \hdgff \ and \lakjsk \ where, of course,
we should bear in the mind that
symbol
\ ${\cal L}_{l-\pol}$\ means now irreducible module of
deformed Virasoro algebra.

Now the whole algebra \ $Symm$\ acting on the space \
$\pi_Z^R\oplus\pi_Z^{R\star}$\
is the tensor product of infinite-dimensional
deformed Virasoro algebra with \ $c=-2\ , \ 0<x<1$\
and finite-dimensional part
generated by
operators
\ $X,  X^{\star},X',  X'^{\star}$\ and \ $P$\
with commutation relations \gsfddw \ - \jshdgp .
For this reason, the decomposition of the space\ ${\pi}_Z^R\oplus
\pi_Z^{R\star}$\
into direct sum of irreducible representation of
algebra  \ $ Symm $\  has the form \jshdg .
The scalar product in the space \ $\oplus_{k\in{\bf Z}}
{\cal F}_{k-\frac{1}{2}}$\
is given by \gfor
\ and the conjugation conditions
\ $({\bf v}_1L_n^{\nu},{\bf v}_2)=({\bf v}_1,L_{-n}^{\nu}{\bf v}_2)$.
Moreover,
in the calculations one can use the scalar product \ \vcsda . The
arguments here are practically the same as in subsec.{\it 3.6} .

In the NS sector the finite-dimensional part
of \ $Symm$\ algebra is still given by the \ $sl(2)$\
algebra
and similarly to undeformed situation we have isomorphism
\ $Ker_{{\cal F}_l}X^l \cong \ {\cal L}_{l}\ . $\
In addition, the decomposition of Fock spaces
into direct
sum of irreducible modules of deformed Virasoro algebra
is described
by the formula \hf \ again.
The scalar product in the
space \ $\pi_Z^{NS}$\  is the same as \gfor -\hsfdd \ and
the space \ $\pi_Z^{NS}$\ can be treated as selfdual
\ $\pi_Z^{NS}=\pi_Z^{NS\star}.$

\subsec{Deformation of vertex operators}
Now we want to establish
x-deformation of chiral primary operators
\ $\Phi^{\pm}_{21}$\ and \ $\Phi^{\pm}_{12}$. The construction
of first operator \ $\Phi^{\pm}_{21}$\ is rather evident.
Indeed, it
can be realized in the Fock spaces in the same manner as \hsbgdgf :
 \eqn\phihh{\eqalign{&\Phi^{+}_{21}(\zeta)\ {\bf v}=
e^{i\phi(\zeta)}\ {\bf v}, \ \ {\bf v}\in {\cal L}_p\ ,\cr
&\Phi^{\star\  +}_{21}(\zeta x^2)\ {\bf v}^{\star}=
\int_C\frac{d z}{2\pi i z}\
e^{-i\ [\phi(zx)+\phi(zx^{-1})]} \ e^{i\phi(\zeta)}  \ {\bf v}^{\star},
 \ \ {\bf v}^{\star}\in {\cal L}^{\star}
_p\ ,}}
where we explicitly show the states on which such operators are
well-defined. As an earlier, we take
contours in the integral
\ $\Phi^{\star\  +}_{21}(\zeta)$\
with beginning and ending in zero, and enclose all singularities
which positions are determined by vector\ $v^{\star}$.
Demanding conjugation properties
of \ $\Phi_{21}^{\pm}(\zeta)$\ for\  $|\zeta|=1$\  as
\eqn\conjuga{\big({\bf u}^{\star}
\Phi^{\star\  \pm}_{21}(\zeta x^2),
\ {\bf v}\big)=
\big({\bf u}^{\star},\Phi^{\mp}_{21}
(\zeta)\ {\bf v} \big)\ ,}
where\ ${\bf u}^{\star}\in \ {\cal L}^{\star}_p$ \ and
${\bf v}\in \ {\cal L}_{p\pm 1}$, one can easily
obtain any
matrix element of such operators.
The bosonic realization \phihh \ possesses to find that
commutation relations of these vertex operators with deformed
Virasoro algebra
are given by formula
\eqn\vira{\eqalign{2\ \zeta^{-n}\ [L_n^\nu ,\Phi^{\pm}_{21}]=
 &\frac{x^{2\nu+3}\ \Phi^{\pm}_{21}(\zeta x^{4\nu+4})-
x^{-2\nu-3}\ \Phi^{\pm}_{21}(\zeta x^{-4\nu-4})}{x^2-x^{-2}}-\cr
&\ \ \
\ \ -\ \frac{x^{2\nu-1}\ \Phi^{\pm}_{21}(\zeta x^{4\nu})-
x^{-2\nu+1}\ \Phi^{\pm}_{21}(\zeta x^{-4\nu})}{x^2-x^{-2}}\ .}}

To define deformed chiral primary operators
\ $\Phi_{12}$\
\jdhgdii ,
let us remind that the crucial
property
in the proving of the
proposition 3.3 was the equation \nmshs .
We demand that
this basic relation is preserved in the
deformed theory in the following sense. The second chiral primary
have to be constructed in terms of new field \ $\phi'$\ such that
\ $\lim_{x\rightarrow 1}\phi'(z)=-\frac{1}{2}\phi(z)$\ and
\eqn\nmritors{X_{|\pi_{NS}}\
e^{ i \phi'(\zeta)}=- e^{ i \phi'(\zeta)}\
X_{|\pi_R} .}
Let \ $\phi'$\ is
built from generators \ $P,Q,b_n'$ as:
\eqn\phishtrih{\phi'(\zeta)=-\frac{1}{2}(Q-i\ P\ln \zeta) -
\sum_{\scriptstyle m \in {\bf Z}\scriptstyle\atop
m\not=0}\ \frac{b_m'}{i\  m}\  \zeta^{-m}\ ,}
then condition \nmritors \ is obviously satisfied
if new creation-annihilation operators be
\eqn\dgfhrt{b_n'=(x^n+x^{-n})^{-1}\ b_n\ .}
Notice,
that, due to this definition, the formula expressing
\ $X'$ in terms of field
\ $\phi'(\zeta)$\ has the form:
\eqn\efgsh{
X'=\oint\frac{d z}{2\pi i z}\  e^{- i\ [\phi'(zx)+\phi'(zx^{-1})]}\ .}
Comparing this with equation \efg \  one can observe that integral
operators \ $X,X'$\ are built from \ $\phi$\ and \ $\phi'$\
correspondingly
in the remarkably symmetrical form. Further
we will see that this property allows direct generalization
for the arbitrary x-deformation with \ $0<x<1$.
Now the chiral primary
operators, intertwining R and NS sectors can be realized
as
\eqn\phishtri{\eqalign{&\Phi^{\star\ -}_{12}(x^2\zeta)\ {\bf v}^{\star}=
e^{i\phi'(\zeta)}\ {\bf v}^{\star}, \ \ {\bf v}^{\star}
\in \ {\cal L}_p^{\star}\ ,\cr
&\Phi^{-}_{12}(\zeta)\ {\bf v}=
\int_C\frac{d z}{2\pi i z}\
e^{-i\ [\phi'(zx)+\phi'(zx^{-1})]} \ e^{i\phi'(\zeta)}  \ {\bf v},
 \ \ {\bf v}\in \ {\cal L}_p\ .}}
The integral in \ $\Phi^{-}_{12}(\zeta)$\
would be well-defined if contour is chosen as in \nxhc .
These bosonic
prescriptions together with conjugation condition
\eqn\coa{\big({\bf u}^{\star}
\Phi^{\star\  \pm}_{12}(\zeta x^2),
\ {\bf v}\big)=
\big({\bf u}^{\star},\Phi^{\mp }_{12}
(\zeta)\ {\bf v} \big)\ ,}
where \ $|\zeta|=1$,
completely
fix
the action of
the operators \ $\Phi^{\pm}_{12}(\zeta)$\ in the
Fock
space.
The matrix elements of products of such operators
can be derived by the standard bosonization
technique. We leave the explicit calculations
till the section 6 where it will be worked out
as particular \ $\xi =1$\ case in context of general x-deformation.
So, let us just note, that knowing matrix elements
one can obtain by
standard way the commutation relations of these operators.
The essential difference of this case
in comparison with undeformed one is that
these commutation relations are determined by elliptic R-matrix
of the IRF type rather than constant R-matrix:
\eqn\haaadfa{
\Phi_{12}^a(\zeta_1)\Phi_{12}^b(\zeta_2)|_{{\cal L}_p}
=\sum_{c+d=a+b}{\bf W}'
\left[\matrix{p+\frac{a+b}{2}&p+\frac{c}{2}\cr p+\frac{b}{2}  &p}
\biggl|\ \frac{\zeta_1}{\zeta_2}\ \right]\
\Phi_{12}^d (\zeta_2)
\Phi_{12}^c(\zeta_1)|_{{\cal L}_{p}}\ .}
The nontrivial elements of the matrix \ ${\bf W}'$\ read:
\eqn\kh{\eqalign{&{\bf W}'
\left[\matrix{p\mp 1&p\mp \frac{1}{2}
\cr p\mp \frac{1}{2}&p}\biggl|\zeta\right]=r(\zeta)
\ ,\cr
&{\bf W}'
\left[\matrix{p&p\mp
\frac{1}{2}\cr p\mp \frac{1}{2}\ &p}\biggl|\zeta\right]=r(\zeta)\
\zeta^{\pm p-\frac{1}{2}}\
\frac{\Theta_{x^4}(x^2)\ \Theta_{x^4}(x^{\pm 4p}\zeta )}{\Theta_{x^4}
(x^2 \zeta)\
\Theta_{x^4}(x^{\pm 4 p })}\ ,\cr
&{\bf W}'
\left[\matrix{p&p\pm\frac{1}{2}\cr p\mp
\frac{1}{2}\ &p}\biggl|\zeta\right]=
-r(\zeta) \ x^{\pm 2 p }\zeta^{-\frac{1}{2}}\
\frac{\Theta_{x^4}(x^{2(\pm 2 p +1)})\ \Theta_{x^4}
(\zeta)}{\Theta_{x^4}(x^{\pm 4p})\
\Theta_{x^4}(x^2 \zeta)}\ , }}
where
$$r(\zeta)=\zeta^{\frac{1}{4}}\
\exp\big[\ \sum_{m=1}^{+\infty}\frac{ \zeta^m-\zeta^{-m}}
{ m\ (x^m+x^{-m})^2}\ \big]\ . $$
The associativity of algebra \haaadfa \ follows from
the fact that matrix \ ${\bf W}'$\ is a solution
of Yang-Baxter equation. This justify proposed deformation
of chiral primary operators \ $\Phi_{12}$.

\subsec{The x-deformation of fermions ZF algebra}
So far we were interested in vertex operators
of \ $Vir_{-2,x}$\ algebra. To construct vertex operators of \ $Symm$\
algebra, one need take into account remaining finite-dimensional part
of
symmetry algebra.

Let \ ${\bf e}^m_l$\ be basic vectors of irreducible spin
\ $j=\frac{l-1}{2}$\
representation \ ${\cal V}_l$\ of algebra \ $sl(2)$\ and
\ ${\bf v}\in \ {\cal L}_l$. Define operators \ $Z_{\pm}(\zeta)$\
acting on the space \ $\pi_Z^R$ \ and\ $\pi_Z^{NS}$\
by the formula \kdjhf .
The straightforward computation shows that
operators \ $Z_{\pm}(\zeta)$\
are generators of ZF algebra given by commutation
relations
\eqn\Zam{Z_{a}(\zeta_1)Z_{b}(\zeta_2)=-Z_{b}(\zeta_2)Z_{a}(\zeta_1),
\ \ \zeta_1\not=\zeta_2\ }
and operator product expansion:
\eqn\zetik{\eqalign{&Z_{\pm}(\zeta_2)Z_{\mp}(\zeta_1)=
\pm
\frac{(x+x^{-1})\
 \zeta_1\ \zeta_2}{(\zeta_2-\zeta_1 x^2)(\zeta_2-\zeta_1 x^{-2})
}
+O(1)\ ,
\cr
&Z_{\pm}(\zeta_2)Z_{\pm}(\zeta_1)=O(1)\ . }}
It is rather evident  now that operators \hhh \ introduced in the
beginning of this section are exactly generators of ZF algebra
\Zam -\zetik.

\newsec{Nondeformed Virasoro algebra with \ $c<1$}

In the present section we remind how to construct
representations of ZF algebra (more explicitly,
a pair of ZF algebras) with constant R matrix
corresponding to quantum group \ $U_q(sl(2))$.
As an earlier, the representation space
of ZF algebra can be realized in the direct sum of irreducible
representations of Virasoro algebra with central charge \ $c<1$.
The operators of ZF algebras will be expressed in terms of
vertex operators of Virasoro algebra \ $Vir_{c}$. Therefore,
the main objects of our investigation will be the irreducible
representations of Virasoro algebra and vertex operator algebra.
To describe these objects we will use bosonization method,
examples of which was presented in the previous sections.

In the case under the consideration the explicit realization
of Virasoro algebra generators \ ${L_n}$ in
the bosonic Fock space \ ${\cal F}_p$\ is well-known \refs{\DotsFat}. Then
universal enveloping
algebra \ $U(Vir_c)$\ turn to be subalgebra in the universal
enveloping algebra of Heisenberg algebra. Moreover,
any highest weight
vector in the Fock module will be the highest weight vector
of Verma module of \ $Vir_c$. The original motivation of
bosonization of Virasoro algebra \refs{\fei}
is that in the Fock space
one can explicitly construct the intertwining operators
between Verma modules. In the  physical literature
these operators were historically
called by screening operators \refs{\DotsFat}. According
to the definition,
intertwining operators
commute with any element of \ $U(Vir_c)$,
hence it have to map singular vectors of Verma module into
singular ones. So the analysis of the structure of
the reducible Verma  modules becomes very simple. Indeed,
to determine singular vectors one needs to find such vectors
in the Fock space which are mapped under the action of screening
operators into highest weight vectors, or such vectors which
can be obtained from the highest
weight vectors by the action of screening
operators.
For this reason, knowledge of all possible
intertwining operators
is equivalent to the knowledge of
the invariant subspaces in the Verma modules.
This subject was intensively discussed in the literature.
It is well known,
that structure of Verma modules of \ $Vir_c$\
drastically depends on the arithmetical
properties of real number\ $ c$.
Having in the mind further \ $x$-deformation
we will omit complicated cases of completely reducible
Verma modules just
considering the generic case,
i.e. case 2 in the Feigin-Fuks
classification \refs{\fei}. Our task is to remind the
essential features which occur
in the bosonization
of representations of
Virasoro algebra and ZF algebra. (The analysis of more complicated cases
can be found in the refs. \refs{\felders},\ \refs\bernard). In particular we
would like to clear up the idea that knowledge of
the intertwining operators between Verma modules of
\ $Vir_c$\ is sufficient to describe the
irreducible representations of
Virasoro algebra and algebra of vertex operators without
appealing to the commutation relations of generators \ $L_n$.
Another very important point we recall and
constantly use in the
construction is the
existence of remarkable discrete symmetry which in our
notations is just a change \ $\xi \leftrightarrow -1-\xi$\
(\ $\alpha_-\leftrightarrow \alpha_+$\ in the Dotsenko-Fateev
work \refs{\DotsFat}). This symmetry, origin of which is still hardly
understood, seems to be essential not only in CFT \refs{\ffk}
but also in general theory of integrable models \
\refs{\singordon},\
\refs{\japellip}.

Our present consideration is
based on slightly different
point of view on the bosonization (see also section 3,4). Namely,
we consider screening operators as basic
objects which completely determine the whole construction.
In particular,
Virasoro algebra can be treated as an algebra
of generators in the Fock space which commute with
correspondent
screening operators. In this approach
we can investigate the irreducible representations
of Virasoro algebra independently on the
facts what basis in \ $U(Vir_c)$\
or what commutation relations are
really involved. The similar idea was applied in the
development of the theory of W-algebra \refs{\LF}.
So, the most important step is
the introduction of correctly defined screening operators.
These operators have remarkably simple form.
They are given by some powers of operators
\ $X$\ and \ $X'$\ which are deformations of
\idreg . We explicitly enlist the subspaces in the Fock space
where these integral operators are well-defined.
As soon as we determine screening operators, we
are able to describe the irreducible representations
of Virasoro algebra as some submodules (or factor modules)
of Fock modules.
Our next step is the definition of the
chiral vertex operators of Virasoro algebra. To do this,
we need fix basis \ $L_n$\ in the \ $U(Vir_c)$\ since,
in general, the knowledge of intertwining operators
is not enough to uniquely determine
vertex operators. Our task is to extract
necessary properties of these operators
which are determined by screening operators
rather than by choice of basis \ $L_n$. Defining
the bosonization of
chiral primaries,
we demonstrate that matrix elements of these operators
can be easily computed using Wick theorem. The studying
the analytical properties of four point functions
shows that commutation
relations of chiral primary operators \ $\Phi_{12}$, \ $\Phi_{21}$\
are determined by two different constant
\ $\bf W$-matrices which are solutions of Yang-Baxter equation
of IRF type. To construct the ZF algebras
of "vertex type"  \refs{\sau} one needs take into
account multiplicities of irreducible representations
of Virasoro algebra. We argue that there are two ZF algebras associated
to
the algebras of chiral primaries \ $\Phi_{12}$\ and \ $\Phi_{21}$\
correspondingly. The R-matrices of these ZF algebras
correspond to the R-matrices of quantum groups
\ $U_p(sl(2))$\ with different deformation parameters \ $p$ \refs{\ffk} .
These R-matrices are connected by the transformation
\ $\xi\leftrightarrow -1-\xi $\
which is originated from the symmetry between
\ $\Phi_{12}$\ and \ $\Phi_{21}$\ (or \ $\phi'\leftrightarrow \phi$).
The irreducible representation of a pair of
ZF algebras
coincides with direct sum of Fock modules, hence
it admit classification by
the representations of the Virasoro algebra.

The bosonization is just a useful method
to study the representations of Virasoro
and ZF algebras. Of course, the results do not
depend on it. In particular, without any
bosonization the irreducible representations of
a pair of
\ $[\ \xi\leftrightarrow -1-\xi\ ]$-symmetrical
ZF algebras are isomorphic to direct sum
of irreducible representations of Virasoro algebra.

This construction can be considered as
case \ $x=1$\  of general two parametric deformation with
parameters \ $x$\ and \ $\xi$. In the following
section we will generalize
main statements for \ $x\neq 1$\ deformation.

\par
\noindent
{\it 3.1}\
Let us introduce the free bosonic field
\eqn\hsgdffw{\phi(z)=\sqrt{\frac{\xi+1}{2\xi}}\ (Q-i\ P\ln z)
+\sum_{\scriptstyle m \in {\bf Z}\scriptstyle\atop
m\not=0}\ \frac{b_m}{i\  m} z^{-m}\ ,}
where the commutation relation of the zero modes\ $P, Q$\
is defined by \ $[Q,P]= i,$\ while
determining relation for modes \ $b_n$\ is:
\eqn\hsfgdf{[b_m,b_n]=\frac{\xi+1}{2\xi}\ m\ \delta_{m+n,0}\ .}
Further we will consider cases with deformation parameter \ $\xi>1$\ .
Moreover, in order to avoid additional complications we assume that
this number is irrational. Note that the case \ $\xi=1$\ corresponds
to the example discussed in the section 3.
Together with the field \ $\phi$\ , it is convenient to define
another one:
\eqn\ksjh{\phi'(z)=-\sqrt{\frac{\xi}{2(\xi+1)}}\ (Q-i\ P\ln z)
-\sum_{\scriptstyle m \in {\bf Z}\scriptstyle\atop
m\not=0}\ \frac{b'_m}{i\  m} z^{-m}\ ,}
where
$$\xi\  b_m= (\xi+1)\  b'_m \ .$$
As we will see below,
there is a
remarkable symmetry with respect the transformation\
$\xi\leftrightarrow -1-\xi$. This is the
reason why we introduce independent notation for field\ $\phi'(z)$\ .
In the
case \ $x=1$\ fields \ $\phi'(z)$\ and \ $\phi(z)$\
are just proportional. However, in the x-deformed construction
their
connection turns to be more complicate.

In the set of Fock modules
\eqn\jdhfg{{\cal F}_{k,k'}\equiv
{\cal F}_{\frac{(\xi+1) k-\xi k'}{\sqrt{ 2\xi(\xi+1)}}},\
k,k'\in {\bf Z}\ }
one might introduce the following formal operators:
\eqn\hdgswqr{\eqalign{&X^l
=\int_{C_1}...\int_{C_l}
\frac{d z_1}{2\pi i z_1}...\frac{d z_l}{2\pi i z_l}\  e^{-2 i \phi(z_1 )}
.. e^{-2 i \phi(z_l )},\cr
&X'^{\ l'}
=\int_{C_1}...\int_{C_{l'}}
\frac{d z_1}{2\pi i z_1}...\frac{d z_l'}{2\pi i z_{l'}}\
e^{-2 i \phi'(z_1)}... e^{-2 i \phi'(z_l' )}\ .}}
Here we used Felder's prescription \refs{\felders}
for
integration contours (see fig.4.). Namely, any contour
\ $C_i,\ 1\leq i\leq l-1 $\  has
beginning and ending in the point \ $z_l$\ chosen
on the unit circle. It encloses the origin and all singularities
depending on the variables \ $z_k(k<i)$. The last integration on the
variable \ $z_l$\ is provided along the circle \ $C_l$.
\ifig\fcftcontura{The integration contours. }
{\epsfxsize2.5in\epsfbox{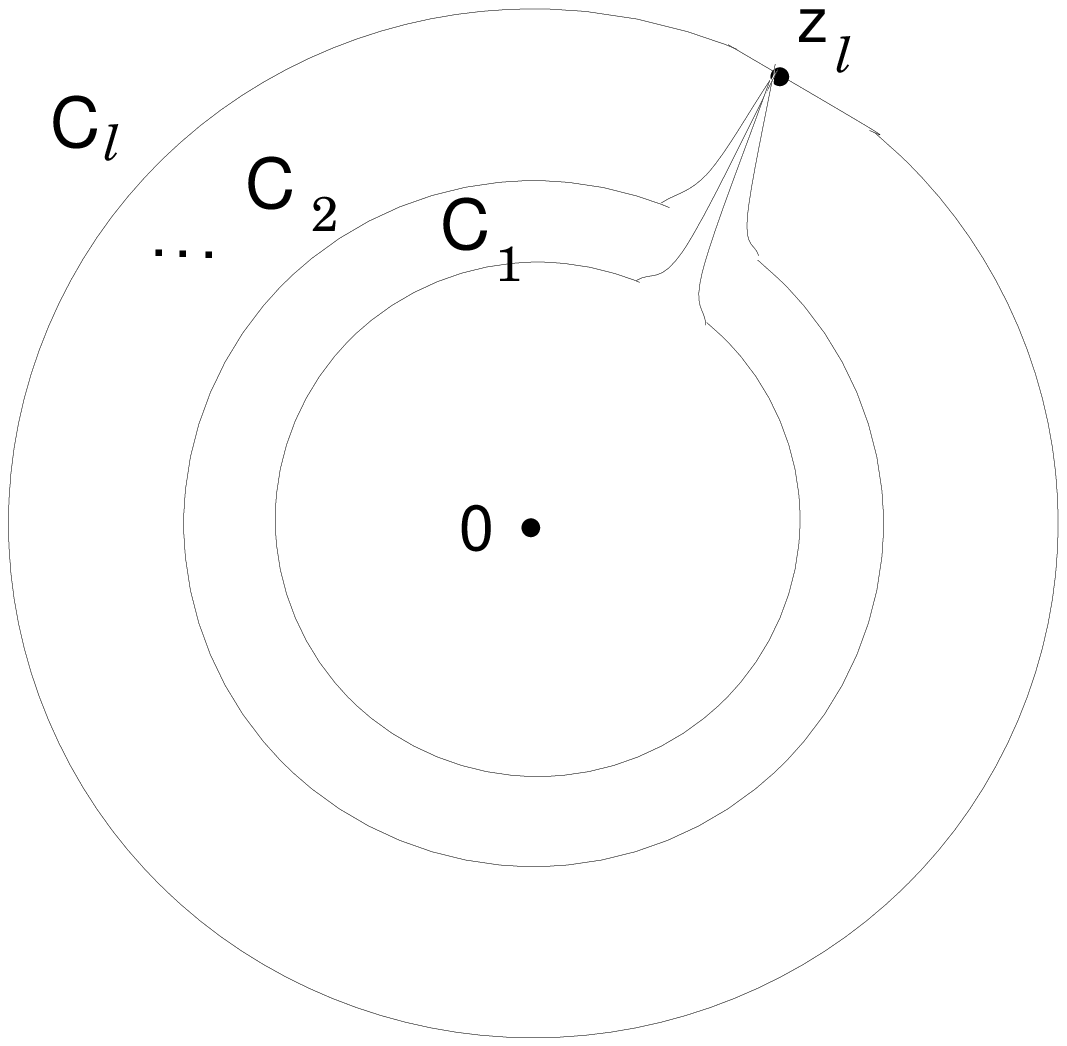}}

\par\noindent
In essential distungish with the case \ $c=-2$\ the action of these
operators is ill defined in the whole
set of Fock modules since the last
integration contours \ $C_l\ (C_{l'})$\
in the definitions \ \hdgswqr\ is not closed.
Hence, first
of all, we must formulate where the operators \hdgswqr \ act.

\noindent
{\bf Proposition 5.1}

\noindent
\it
The action of operator \ $X^l\ (X'^{\ l'})$ \ is defined
only on the Fock modules\ ${\cal F}_{l,k}\ ({\cal F}_{k,l'}), k\in{\bf Z}\ $
and\ $l, l'>0$. Then
\eqn\mdbnfb{\eqalign{
&Ker_{{\cal F}_{l,l'}}[X^l]=Ker_{{\cal F}_{l,l'}}[X'^{\ l'}]\ ,\cr
&Ker_{{\cal F}_{l,-l'}}[X^l]=
Ker_{{\cal F}_{l,-l'}}[X'^{\ l'}]=0\ ,\cr
&Im_{{\cal F}_{l,l'}}[X^l]={\cal F}_{-l,l'}\ ,\cr
&Im_{{\cal F}_{l,l'}}[X'^{\ l'}]={\cal F}_{l, -l'}\ ,\cr
&Im_{{\cal F}_{l,-l'}}[X^l]=
Im_{{\cal F}_{-l,l'}}[X'^{\ l'}]
\subset {\cal F}_{-l,-l'}\ .}}
\noindent
\rm
We will call integral operators \hdgswqr \
defined on the correspondent spaces by screening operators.
Considering the screening operators as basic objects,
one can define Virasoro algebra as follows:
\par
\noindent
{\bf Definition.} \it Let operators \ ${P, Q, b_n, n\in {\bf Z}}$\
satisfy the commutation relations  \hsfgdf \
and screening operators are given by \hdgswqr .
Virasoro algebra
is a subalgebra in the universal enveloping
algebra of Heisenberg algebra of operators \ $b_n$\ and \ $P$.
An element from universal enveloping of Heisenberg
algebra belongs to Virasoro algebra if it
commutes with an action of the screening operators.
\rm
\par
\noindent
In present case one can simply
write down explicit expressions for basic generators
of the space of invariants in the universal enveloping of Heisenberg
algebra:
\eqn\jsgdfgo{\eqalign{
&L_n=\sum_{k+m=n} \ b_m b'_k -(\sqrt{2\xi(\xi+1)}P+n)
(b_n-b'_n)\ ,\ \  n\not = 0\ ;\cr
&L_0=\sum_{m>0} ( b_{-m} b'_{m}+b'_{-m} b_{m})+
\frac{P^2}{2}-\frac{1}{4\xi(\xi+1)}\ .}}
The \ $L_n$\ obey Virasoro commutation relations with the
central charge \ $c=1-\frac{6}{\xi(\xi+1)} $.
So, the Fock space is given
the structure of Virasoro
module. The highest Fock
state\ ${\bf f}_{k,k'}\equiv
{\bf f}_{\frac{(\xi+1)k-\xi\ k'}{\sqrt{ 2\xi(\xi+1)}}}$\
will also be the highest weight vector of Verma module of Vir
with the conformal
dimension
\eqn\jshdg{\Delta_{k,k'}=\frac{((\xi+1) k-\xi k')^2-1}
{4\xi(\xi+1)}\ .}
The most important property of
operators \hdgswqr \ is that they commute with generators of the
form \jsgdfgo \ , i.e. are intertwining operators between Verma modules
of the Virasoro algebra. Knowing intertwining operators one might
study the structure of Verma modules \ ${\cal M}_{k,k'}$\ of
\ $Vir_c$\ and construct the irreducible representations
\ ${\cal L}_{k,k'}$\  as some subspaces or factor spaces of
Fock spaces. Namely, introduce the following notations:
\eqn\ksi{\eqalign{{\cal L}_{k,k'}=\cases{Ker_{{\cal F}_{k,k'}}[X^k],
&if $k,k'\geq 1$;\cr
{\cal F}_{k,k'}/Im_{{\cal F}_{-k,k'}}[X^{-k}],
&if  $k,k'\leq -1$;\cr
{\cal F}_{k,k'},& otherwise\ ,}}}
We claim that the following proposition is hold:

\noindent
{\bf Proposition 5.2}

\noindent
\it
Let generators of Virasoro algebra are given by formulae \jsgdfgo\
and parameter\ $\xi >1$\ be irrational number. Then

\par\noindent
1. The space\ ${\cal L}_{k, k'}$\ for any integer number\ $k, k'$\ is
irreducible representation of Virasoro algebra with central
charge \ $c=1-{6\over \xi(\xi+1)}$\ . The highest weight vector of \ ${\cal
L}_{k, k'}$\
coincides with the vector
\ ${\bf f}_{k,k'}$\ and has the conformal dimensions \jshdg .

\par\noindent
2. For \ $k\cdot k'\leq 0$\ or\ $ k,k'<0$\ ,
the Verma module \ ${\cal M}_{k,k'}$\
built upon highest weight vector\ ${\bf f}_{k,k'}$ \
coincides with the total Fock module \ ${\cal F}_{k,k'}$.

\noindent
\rm
We would like now to comment propositions 5.1,\ 5.2 \refs{\bernard}.
As we have noted, any highest weight vector
\ ${\bf f}_{k,k'}$\
of Fock module \  ${\cal F}_{k,k'}$\ turns to be highest weight vector of
Verma module \  ${\cal M}_{k,k'}$\ of Virasoro algebra.
Generically speaking, Fock space \ ${\cal F}_{k,k'}$\ does
not coincide neither with  Verma module
\  ${\cal M}_{k,k'}$, nor with
irreducible module \  ${\cal L}_{k,k'}$\ of \ $Vir_c$.
It might contain some subspaces which are
invariant with respect to the action of generators
\jsgdfgo  . Such as screening operators
\hdgswqr \ commute with any generator
from universal enveloping of \ $Vir_c$\ then they
have to map invariant subspace of the Vir into invariant one.
In the case under the consideration the structure of embedding
of Verma modules given by the actions of screening
operators is rather simple:
\par
\noindent
{\it (i)}\ Consider at first Fock module
\ ${\cal F}_{l,l'}, \ l,l'>0$. According to
proposition 5.1 this module contains a vector
\ ${\bf f}^0_{l,l'}$\
such that \ $X^l\ {\bf f}^0_{l,l'}={\bf f}_{-l,l'}$
(\ $X'^{\ l'}\ {\bf f}^0_{l,l'}=
{\bf f}_{l,-l'}$). This vector can not be obtained by the
action of the Virasoro algebra generators on the highest weight
vector \ ${\bf f}_{l,l'}$\
and Fock submodule built upon state \ $ {\bf f}^0_{l,l'}$\
turns to be
invariant space with respect to the action of generators of
Virasoro algebra. Indeed,
if \ ${\bf f}^0_{l,l'}$\ is produced by the action of any element from
universal enveloping of \ $Vir_c$\ on the vector \ ${\bf f}_{l,l'}$\ , then
operator \ $X^l\ (X'^{\ l'})$\ would map it into zero rather than into
\ ${\bf f}_{-l,l'}\ ({\bf f}_{l,-l'})$\ , since \ $X^l\ (X'^{\ l'})$\
commutes with any generator of \ $Vir_c$\ and \ $X^l
\ {\bf f}_{l,l'}=X'^{\ l'}\ {\bf f}_{l,l'}=0$\ .
The proposition 5.2 means that
subspace \ $Ker_{{\cal F}_{l,l'}}[X^l]=Ker_{{\cal F}_{l,l'}}[X'^{\ l'}]$\
is irreducible representation \ ${\cal L}_{l, l'}$\
of Virasoro algebra. The modules
\ ${\cal F}_{-l,l'}=Im_{{\cal F}_{l,l'}}X^l$\ and
\ ${\cal F}_{l,-l'}=Im_{{\cal F}_{l,l'}}{X'^{\ l'}}$\ coincide with
Verma modules of \ $Vir_c$\ and do not have any invariant
subspaces. Therefore it can be identified with \ ${\cal L}_{l, -l'}$\
and \ ${\cal L}_{-l, l'}$\ correspondingly.
\par
\noindent
{\it (ii)}
The Fock space \ ${\cal F}_{-l,-l'}$\ is a
reducible Verma module
of \ $Vir_c$\ with unique
null-vector \ ${\bf f}_{-l, -l'}^0$\ which is the image
of the highest weight vector
\ ${\bf f}_{l, -l'}\ ({\bf f}_{l, -l'})$\ under
the action
of the operator \ $X^l\ (X'^{\ l'})$\ :
$${\bf f}_{-l, -l'}^0=X^{l}\ {\bf f}_{l,-l'}=
X'^{\ l'} {\bf f}_{-l,l'} \ \in {\cal F}_{-l,-l'}\ .$$
The
irreducible representation \ ${\cal L}_{-l, -l'}$\
of Vir turn to be isomorphic to
factor space
\ ${\cal F}_{-l,-l'}/Im_{{\cal F}_{l,-l'}}[X^{l}]=
{\cal F}_{-l,-l'}/Im_{{\cal F}_{-l,l'}}[X'^{\ l'}]$ of Fock space.
\ifig\fobschee{The structure of spaces \ ${\cal F}_{l,l'}$\
and \ ${\cal F}_{-l,-l'}$\ }
{\epsfxsize3.5in\epsfbox{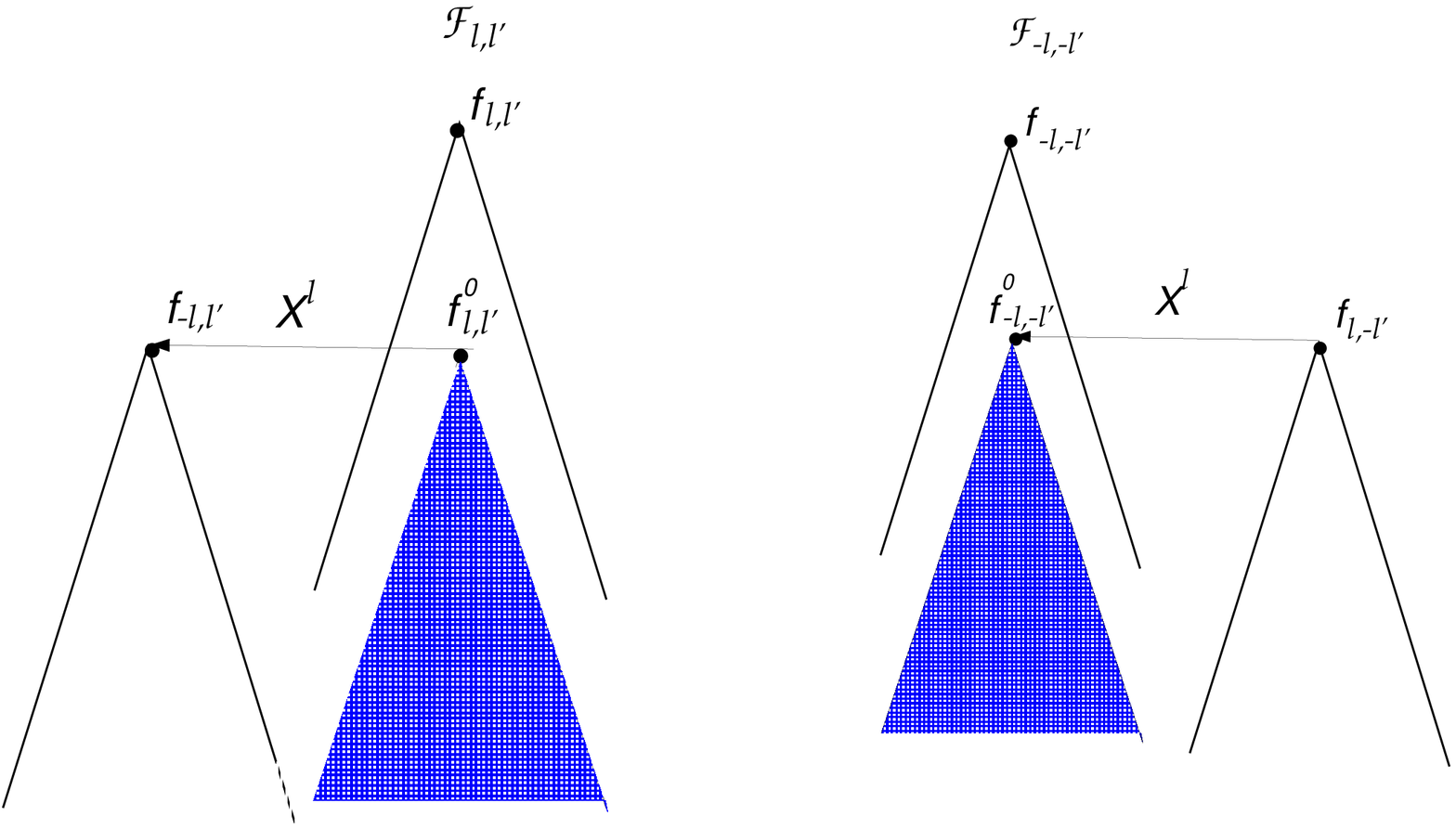}}
The most important objects for us are the sets of irreducible
representations\ $\{{\cal L}_{l,l'}|l,l'>0\}$\ and
$\{{\cal L}^{\star}_{l,l'}|l,l'>0\}$\ where,
by definition, \ ${\cal L}^{\star}_{l,l'}=
{\cal L}_{-l,-l'}$. These spaces
can be endowed by the
scalar product through the procedure which
was explained in the section 3 on the
example of Virasoro algebra with central charge\ $c=-2$. Namely,
we introduce dual field \ $\phi^\star$\ and repeat
analysis above for dual representations. Then we identify
dual modules of Virasoro algebra with subspaces (or factor spaces)
of Fock space
by demanding the condition \ $T^\star(\zeta)=T(\zeta)$.
Again, choosing the field \ $\phi$\ as basic field
and determining dual field via Riccati type equation,
we destroy
the symmetry between \ $\phi$\ and  \ $\phi^\star$.
For instance, Fock submodule
\ ${\cal L}^\star_{l,l'}=Ker_{{\cal F}_{l,l'}^\star}[{X^\star}^l]$
will be identified with factor module
\ ${\cal L}_{-l,-l'}={\cal F}_{-l,-l'}/Im_{{\cal F}_{l,-l'}}[X^l]$ etc.
Thus, we have

\noindent
{\bf Proposition 5.3}

\noindent
\it
Let irreducible representation \ ${\cal L}_{l,l'}$\
and dual Verma module \ ${\cal M}_{l,l'}^\star$\ of Virasoro
algebra are realized as
bosonic modules \ $Ker_{{\cal F}_{l,l'}}[X^l]$\ and
\ ${\cal F}_{-l,-l'}$\ correspondingly, where \ $l=1,2,..$.
Then the scalar product \ $( , )_0$\ \vcsda \
restricted on the vectors
from \ ${\cal L}_{l,l'}$\ and \ ${\cal M}_{l,l'}^\star$\ coincides
with the following:
\eqn\lsjdhg{
({\bf u}L_n,{\bf v})=({\bf u}, L_{-n}{\bf v})\ .}

\noindent
\rm
Let us explain the meaning of this statement. Consider at first
vector \ ${\bf u}\in Ker_{{\cal F}_{l,l'}}[X^l]\cong {\cal L}_{l,l'}$.
Then due to both scalar products \ $ ( , )_0$\ and \lsjdhg\ \ ${\bf u}$\
has to be orthogonal to any vector \ ${\bf v}\not\in {\cal F}_{-l,-l'}$.
Let now \ ${\bf v}\in {\cal F}_{-l,-l'}$. We know from the proposition 5.2
that \ ${\cal F}_{-l,-l'}\cong {\cal M}_{l,l'}^\star$. It is clear
that taking \ ${\bf v}\in Im_{{\cal F}_{l,-l'}}[X^l]$ we would find that
this vector is orthogonal to any vector from
\ $Ker_{{\cal F}_{l,l'}}[X^l]\cong{\cal L}_{l,l'} $. The reason
for this is that
subspace \ $Im_{{\cal F}_{l,-l'}}[X^l]$\ is generated by null vector
in the \ ${\cal M}_{l,l'}^\star$\ which is orthogonal
to \ ${\cal L}_{l,l'}$. So, the only vectors which have
non-trivial scalar
product are \ ${\bf u}\in Ker_{{\cal F}_{l,l'}}[X^l]$\ and
\ ${\bf v}\in {\cal F}_{-l,-l'}/Im{\cal F}_{l,-l'}[X^l]\cong
{\cal L}_{l,l'}^\star $. But scalar products
\vcsda \ and \lsjdhg\ are the same
in these spaces. It should be clear from explicit
bosonic realization of generators \ $L_n$\ \jsgdfgo .

Now let us turn to consideration of the
vertex operator algebra acting in the set of
irreducible representations of \ $Vir_c$. It is easy to see that
such an algebra is generated by the following operators
\eqn\hgd{\eqalign{ \Phi^{\pm}_{21}(\zeta)&:\ \
{\cal L}_{l,l'}\matrix{\Phi^{\pm}_{21}\cr
\longrightarrow\cr {}}{\cal L}_{l\pm 1,l'}  \otimes\C[\zeta]\
\zeta^{\Delta_{l\pm 1,l'}-\Delta_{l,l'}}\
 \ ,} }
\eqn\hgdaa{\eqalign{ \Phi^{\pm}_{12}(\zeta)&:\ \
{\cal L}_{l,l'}\matrix{\Phi^{\pm}_{12}\cr
\longrightarrow\cr {}}{\cal L}_{l,l'\pm 1}  \otimes\C[\zeta]\
\zeta^{\Delta_{l,l'\pm 1}-\Delta_{l,l'}  }\
 \ .} }
The commutation relations of these operators
with the Virasoro generators are given by
formulas\  \gsfddwge\ , \ksjd \
where \ $\Delta_{1,2}$\ and\ $\Delta_{2,1}$ are
defined as in \jshdg .
In addition, the following conjugation conditions take
place
\eqn\mcnvnb{\eqalign{&\big({\bf u}^{\star} \Phi^{\star\  \pm}_{21}(\zeta),
\ {\bf v}\big)=
\big({\bf u}^{\star},\Phi^{\mp}_{21} (\zeta)\ {\bf v} \big)\ ,\cr
&\big({\bf u}^{\star} \Phi^{\star\  \pm}_{12}(\zeta),
\ {\bf v}\big)=
\big({\bf u}^{\star},\Phi^{\mp}_{12} (\zeta)\ {\bf v} \big)\ ,}}
where\ ${\bf u}^{\star}\in {\cal L}_{l,l'}^{\star}$ \ and
${\bf v}\in {\cal L}_{ l\pm 1,l'}\ ({\cal L}_{l,l'\pm 1}).$

Now we wish to explain on the example of
the operators
\ $\Phi_{21}$\ how one can find the
bosonic realization of such vertex operators.
Let us introduce formal operator
\eqn\hsgdf{V_-(\zeta)=\int_C\frac{d z}{2\pi i z}\
e^{-2i\phi(z)} \ e^{i\phi(\zeta)}\ ,}
where the integration contour \ $C$\ is chosen as in \hsbgdgf .
The action of this operator is defined only in the
Fock modules \ ${\cal F}_{k,k'}$\ with \ $(\xi+1)\ k-\xi\  k'<0$\ and
$$
V_-(\zeta):\  {\cal F}_{k,k'}\matrix{V_-\cr\longrightarrow\cr
{}} {\cal F}_{k-1,k'}\ .$$
The chiral primary
operator which increases value
\ $k$\ by unit can be obviously
constructed as  \ $V_+(\zeta)=e^{i\phi(\zeta)}$\ :
\eqn\jshgg{V_+(\zeta):
\ {\cal F}_{k,k'}\matrix{V_+\cr\longrightarrow\cr
{}} {\cal F}_{k+1,k'} \ .}
It is easy to see that operators \hsgdf -\jshgg \ ,
satisfy the same commutation relations with Virasoro algebra
as \ $\Phi_{21}$\ .
Moreover, the following proposition holds:

\noindent
{\bf Proposition 5.4}

\it
\par\noindent
Let $ l, l'$ be positive integer numbers such that
\ $ (\xi+1)\  l-\xi\  l'<0$. Then
the  following diagrams are commutative:

\noindent
i.
\eqn\nsgdfwt{\eqalign{\matrix{&{\ } & {\ }    &V_{\pm}&{\ } &{\ }\cr
&{\ } &{\cal F}_{l, l'}&\longrightarrow &{\cal F}_{l\pm 1,l'} &{\ }\cr
& X'^{\ l'}&\biggl\downarrow & V_{\pm}&\biggl\downarrow &(-1)^{l'}X'^{\ l'}\cr
&{\ } &{\cal F}_{l, -l'}&\longrightarrow &{\cal F}_{l\pm 1,-l'}&{\ }\cr}\ ,}}
ii.
\eqn\nsgdf{\eqalign{\matrix{&{\ } & {\ }    &V_{\pm}&{\ } &{\ }\cr
&{\ } &{\cal F}_{-l, -l'}&\longrightarrow &{\cal F}_{-l\pm 1,-l'} &{\ }\cr
& X'^{\ l'}&\biggl\uparrow & V_{\pm}&\biggl\uparrow & (-1)^{l'}X'^{\ l'}\cr
&{\ } &{\cal F}_{-l, l'}&\longrightarrow &{\cal F}_{-l\pm 1,l'}&{\ }\cr}\ .}}
\rm
The propositions 5.1-5.4 ensure that \ $V_{\pm}$\ maps one highest weight
representation of \ $Vir_c$\ into another and provide
all properties of vertex operators. However, we are capable
to identify such bosonic operators with vertex operators \ $\Phi_{21}$\
only on the states, where \ $V_{\pm}$\ are defined but not in the
whole Fock space. Let \ ${\bf v}$\ and \ ${\bf v^\star}$\
be vectors from \ ${\cal L}_{l,l'}$\ and \ ${\cal L}_{l,l'}^\star$\
correspondingly, then
\eqn\ksdgh{\eqalign{
&V_+(\zeta){\bf v}=\Phi_{21}^+(\zeta){\bf v} \
\ \ \  (l>0,l'>0,\ (\xi+1)\  l-\xi\  l'>0)\ , \cr
&V_+(\zeta){\bf v}^{\star}=\Phi_{21}^{\star -}(\zeta){\bf v}^{\star} \
\ (l>1,l'>0,\ (\xi+1)\  l-\xi\  l'<0)\ , \cr
&V_-(\zeta){\bf v}=\Phi_{21}^-(\zeta){\bf v} \
\ \ \ (l>1,l'>0, \ (\xi+1)\  l-\xi\  l'<0)\ , \cr
&V_-(\zeta){\bf v}^{\star}=\Phi_{21}^{\star \ +}(\zeta){\bf v}^{\star} \
\ (l>0,l'>0,\ (\xi+1)\  l-\xi\  l'>0)\ ,}}
here we explicitly write down in the parenthesis the restrictions
on the numbers \ $l,l'$\
of the
spaces \ ${\cal L}_{l,l'}$\ where the bosonic
realization is defined correctly.
The bosonization for vertex operators of another type\
($\Phi_{12}$\ and\   $\Phi^{\star}_{12}$)
can be provided at the similar fashion.
The result will be the following:
\eqn\kshdgh{\eqalign{
&V'_+(\zeta){\bf v}=\Phi_{12}^+(\zeta){\bf v} \
\ \ \  (l>0,l'>0,\ (\xi+1)\  l-\xi\  l'<0)\ , \cr
&V'_+(\zeta){\bf v}^{\star}=\Phi_{12}^{\star -}(\zeta){\bf v}^{\star} \
\ (l>0,l'>1,\ (\xi+1)\  l-\xi\  l'>0)\ , \cr
&V'_-(\zeta){\bf v}=\Phi_{12}^-(\zeta){\bf v} \
\ \ \ (l>0,l'>1\ (\xi+1)\  l-\xi\  l'>0)\ , \cr
&V'_-(\zeta){\bf v}^{\star}=\Phi_{12}^{\star +}(\zeta){\bf v}^{\star} \
\ (l>0 ,l'>0,\ (\xi+1)\  l-\xi\  l'<0)\ ,}}
where \ ${\bf v}\in {\cal L}_{l,l'}$\
and \ ${\bf v}^{\star}\in {\cal L}^{\star}_{l,l'}$.
The operators\ $V'_{\pm}(\zeta)$\ reads
\eqn\gdfsr{\eqalign{&V'_+(\zeta)=e^{i\phi'(z)}\ ,\cr
&V'_-(\zeta)=\int_C\frac{d z}{2\pi i z}\
e^{-2i\phi'(z)} \ e^{i\phi'(\zeta)}\ }}
and its properties are quite similar.

The bosonization prescriptions \ksdgh-\kshdgh \
together with the conjugation conditions \mcnvnb\  allow us to
calculate any matrix element of vertex operators
\ $\Phi_{12},\ \Phi_{21}$\ .
Using the technique described in the previous sections,
one can calculate the functions \ $G^{\pm}_p,\ G'^{\pm}_p$\ \mcn \
for any number \ $\xi$\ and then find the
commutation relations in the algebra of vertex operators.
We will arrive to the following result \refs{\ffk}, \refs{\mor}, \refs{\alg}:
\eqn\hdfasa{\eqalign{&\Phi_{21}^a(\zeta_1)\Phi_{21}^b(\zeta_2)
|_{{\cal L}_{l,l'}}
=\sum_{c+d=a+b}{\bf W}
\left[\matrix{l+a+b&l+c\cr l+b&l}\biggl| \sigma_{12}, p
\right]\
\Phi_{21}^d(\zeta_2)
\Phi_{21}^c(\zeta_1)|_{{\cal L}_{l,l'}}\ ,\cr
&\Phi_{12}^a(\zeta_1)\Phi_{12}^b(\zeta_2)
|_{{\cal L}_{l,l'}}
=\sum_{c+d=a+b}{\bf W}
\left[\matrix{l'+a+b&l'+c\cr l'+b  &l'}
\biggl|\sigma_{12}, p'  \right]\
\Phi_{12}^d (\zeta_2)
\Phi_{12}^c(\zeta_1)|_{{\cal L}_{l,l'}}\ ,}}
where \ $p=e^{i\pi\frac{\xi+1}{\xi}}$ \ and
\ $p'=e^{i\pi\frac{\xi}{\xi+1}}$.
As before, these formulae define the rule of analytical
continuation of matrix elements of the vertex operators
from the region \ $|\zeta_1|>|\zeta_2|$\ to \ $|\zeta_2|>|\zeta_1|$
along the contours\ $C_{\sigma_{12}},\ \sigma_{12}=\pm$\
depicted in fig.2.
The nontrivial elements of \ ${\bf W}$\ read explicitly:
\eqn\ksh{\eqalign{&{\bf W}
\left[\matrix{l\pm 2&l\pm 1\cr l\pm 1&l}\biggl|\sigma
\right]=p^{\frac{1}{2}
\sigma }\ ,\cr
&{\bf W}
\left[\matrix{l&l\pm 1\cr l\pm 1\ &l}\biggl|\sigma\right]=
\mp \frac{p^{-(\frac{1}{2}\pm l)\sigma}}{[l]_p}, \cr
&{\bf W}
\left[\matrix{l&l\pm 1\cr l\mp 1\ &l}\biggl|\sigma\right]=
-p^{-\frac{1}{2}\sigma}\ \frac{[l\pm1]_p}{[l]_p}\ ,}}
where\ $[l]_p=\frac{p^l-p^{-l}}{p-p^{-1}}$.
It is easy also to find the commutation relation
\eqn\hsgsfgq{\Phi_{21}^a(\zeta_1)\Phi_{12}^b(\zeta_2)
=a\ b\ d(\sigma_{12})
\ \Phi_{12}^b(\zeta_2)\Phi_{21}^a(\zeta_1)\ ,}
with
\eqn\kdshfd{d(\sigma)=e^{-i\frac{\pi}{2}\sigma}\ .}
Let us discuss the structure of commutation relations.
The algebra \hdfasa \ , \hsgsfgq \ is self-consistent since
\ $d(\sigma_{12})\  d(\sigma_{21})=1\
(\sigma_{12}=-\sigma_{21})$\ and matrices \ ${\bf W},{\bf W}'$\
satisfy to the  so-called unitarity condition:
\eqn\kdjhh{\sum_{l}{\bf W}
\left[\matrix{l_4&l\cr l_1&l_2}\biggl| \sigma_{21}
\right]\ {\bf W}
\left[\matrix{l_4&l_3\cr l&l_2}\biggl|\sigma_{12}
\right]=\delta_{l_1,l_3}\ .}
The
associativity condition for this algebra is provided by
Yang-Baxter equation in IRF form \refs{\mor}, \refs{\alg}.
\eqn\kh{\eqalign{&\sum_{l}{\bf W}
\left[\matrix{l_4&l_3\cr l&l_2}\biggl|\sigma_{23} \right]\
{\bf W}
\left[\matrix{l_5&l_4\cr l_6&l}\biggl|\sigma_{13} \right]\
{\bf W}
\left[\matrix{l_6&l\cr l_1&l_2}\biggl|\sigma_{12} \right]\
\cr
&=\sum_l
{\bf W}
\left[\matrix{l_5&l_4\cr l&l_3}\biggl|\sigma_{12} \right]\
{\bf W}
\left[\matrix{l&l_3\cr l_1&l_2}\biggl|\sigma_{13} \right]\
{\bf W}
\left[\matrix{l_5&l\cr l_6&l_1}\biggl|\sigma_{23}\right]\ ,}}
and analogously for\ \ ${\bf W}'$.

\par
\noindent
{\it 3.3}\
We wish now to construct the representations of ZF algebras
associated with algebras of chiral primaries
\ $\Phi_{21}$\  \refs{\sau}.
It was explained on the simple example \ $\xi=1$\ (section 3),
that to describe representations of ZF algebras one needs
involve in the consideration some finite-dimensional algebra
\ $U$\
together with Virasoro algebra. Then irreducible representations of two
ZF algebras are decomposed into direct sum of irreducible
representations
of whole symmetry algebra
\eqn\hdgffd{Symm=Vir_c
\otimes U \ . }
It will be shown that this proposal leads to ZF algebras
in general situation
(\ $\xi \neq 1$\ ) too.
The finite dimensional
subalgebra \ $U$\ of whole symmetry algebra for \ $\xi \neq 1$\  case
was considered
in the work \refs{\ffk}. It turns to be
a direct product of two quantum algebras
of the form \kdjh .
We refer to the notions of quantum algebras
and their representations
to the section {\it 2.2}.

Let us consider the direct product
\ $U_p(sl(2))\otimes U_{p'}(sl(2))$\ of two
quantum algebras with generators (\ $X^{\pm},T$) and
(\ $X'^{\pm},T'$) correspondingly. The commutations
relations of both algebras are determined by \kdjh \
where the
deformation parameters \ $p,p'$\ are given by
\eqn\hdfgf{p=e^{i\pi\frac{\xi+1}{\xi}}\ , \
p'=e^{i\pi\frac{\xi}{\xi+1}}\ .}
We will concentrate on the
situation when \ $\xi$\ is a real irrational number more then 1.
In this case, as it was noted in section 2, the representations
of quantum algebra \kdjh\
are similar to representations of ordinary
\ $sl(2)$\ algebra \refs{\dr},\ \refs{\kirre}
For this reason, algebra \ $U_p(sl(2))\otimes U_{p'}(sl(2))$\
admits a set of finite dimensional irreducible
representations parameterized by a pair of natural numbers
\ $l,l'$. We will denote these representations
as\ ${\cal V}_{l,l'}$.
The basis vectors
of space \ ${\cal V}_{l,l'}$\ is given by
\ ${\bf e}_{l,l'}^{m,m'}={\bf e}_{l}^{m}\otimes{\bf e}_{l'}^{m'} $,
where \ ${\bf e}_{l}^{m}$\ were described in \hsgd .

Let us consider now the whole symmetry algebra
\ $Symm=Vir_c\otimes U_p(sl(2))\otimes U_{p'}(sl(2))$.
The irreducible
representation of this algebra is given by tensor product
\eqn\alsk{\pi_Z=\oplus_{l,l'}{\cal L}_{l,l'}\otimes {\cal V}_{l,l'}\ .}
The scalar product in this space is induced by scalar product
in \ ${\cal L}_{l,l'}$\ and in  \ ${\cal V}_{l,l'}$.
We claim that \ $\pi_Z$\ has the structure of irreducible
representation of two ZF algebras. Indeed, let us define the action
of generators of ZF algebras on the space \alsk \ by the
formulae \refs{\sau}\ :
\eqn\Zamo{\eqalign{Z_a(\zeta)\ {\bf v}&\otimes {\bf e}^{m,m'}_{l,l'}=\cr
=&(-1)^{j'-m'}\sum_{b=\pm 1} \left(\matrix{\frac{1}{2}&j&j+\frac{b}{2}\cr
\frac{a}{2}&m&m+\frac{a}{2}\cr}\right)_p
\Phi^{b}_{21}(\zeta)\ {\bf v}\otimes {\bf e}^{m+\frac{a}{2},m'}_{l+b,l'}
\ ,\cr}}
\eqn\sonf{\eqalign{Z_a'(\zeta)\ {\bf v}&\otimes {\bf e}^{m,m'}_{l,l'}=\cr
=&(-1)^{j-m}\sum_{b=\pm 1}
\left(\matrix{\frac{1}{2}&j'&j'+\frac{b}{2}\cr
\frac{a}{2}&m'&m'+\frac{a}{2}\cr}\right)_{p'}
\Phi^{b}_{12}(\zeta)\ {\bf v}\otimes {\bf e'}^{m,m'+\frac{a}{2}}_{l,l'+b}
\ .\cr}}
Here, as usual, \ $l=2j+1, l'=2j'+1$.
Using precise expressions \mdsj \ and \ksdgh-\kshdgh \
one can
find that operators \Zamo -\sonf \ satisfy for
$\zeta_1\not = \zeta_2$\  the commutation relations:
\eqn\alZam{\eqalign{&Z_{a}(\zeta_1)Z_{b}(\zeta_2)=
R_{ab}^{cd}(\sigma_{12},p)\  Z_{d}(\zeta_2)Z_{c}(\zeta_1)\ , \cr
&Z_{a}'(\zeta_1)Z_{b}'(\zeta_2)=
{R}_{ab}^{cd}(\sigma_{12},p')\ Z_{d}'(\zeta_2)Z_{c}'(\zeta_1)\ ,
\cr
&Z_{a}(\zeta_1)Z'_{b}(\zeta_2)=\ a \ b \ d(\sigma_{12})\
Z'_{b}(\zeta_2)Z_{a}(\zeta_1)\ .}}
The meaning of the parameter \ $\sigma_{12}=\pm$\ is the
same as in the formula \ \hdfa.
The explicit form of the matrix \ $R_{ab}^{cd}(\sigma_{12},p)$\
is defined by the relations \fsdq -\osnm \ .

\newsec{Deformed Virasoro algebra and elliptic ZF algebras.}
\par
\noindent
Now we want to consider a general case of the
deformed Virasoro algebra
\ $Vir_{c,x}$\ with  \ $ 0<x<1 $ and the central charge
\eqn\virasl{c=1-\frac{6}{\xi(\xi+1)}\ ,}
where \ $\xi>1$\ is again
an irrational number. Our method, various aspects of
which were demonstrated above, is based on the very simple idea.
{\it
The structure of representations of deformed Virasoro
algebra is determined by the deformed intertwining operators.}
The problem of proper deformation of screening operators
is not quite evident.
Here we will use the deformation suggested in the work
 \refs{\singordon}.
Having the deformed screening
operators depending on two continuous parameters
\ $x$\ and \ $\xi$,
we claim that there is a correspondent
two parametric algebra \ $Vir_{c,x}$\
which coincides with the Virasoro algebra
with central charge \virasl\ at the limit \ $x\to 1$.
Indeed, for given parameters
\ $\xi ,x$\ any operator constructed from annihilation-creation
operators of Fock space belong, by definition, to
universal enveloping algebra of
deformed Virasoro algebra if it
commutes with screening operators. Unfortunately,
at present time, we do not know explicit
expressions for proper basis of generators of the deformed
Virasoro algebra.
Therefore   the generalization of
\ $\Phi_{21}^\pm, \Phi_{12}^\pm$\ will be based only
on properties of the deformed screening operators.
In such a way, we obtain the bosonic
representation of the  elliptic ZF algebra of
the deformed vertex operators.
We will demonstrate on examples how to provide
real computations of matrix elements of the vertex operators.
The main point we wish to emphasize
in this section is that
{\it the ideas and technique developed in
CFT can be generalized for other integrable models}.

\subsec{Screening operators for \ $Vir_{c,x}$\ algebra}
Let bosonic fields
\ $\phi(z)$\ and \ $\phi'(z)$\  are given by the formulas
\hsgdffw ,\ \ksjh .
We will assume that modes \ $b_m$\ satisfy to
deformed commutation relations \refs{\singordon}:
\eqn\nsbdf{[b_m,b_n]=m\ \frac{[m]_x\ [(\xi+1)m]_x}{[2 m]_x\ [\xi m]_x}
\ \delta_{m+n,0}\ ,}
and
\eqn\cbv{ b'_m\ [m(\xi+1)]_x=
b_m\ [m\xi]_x\ .}
Here we use the notation\ $[a]_x=\frac{x^a-x^{-a}}{x-x^{-1}}.$
One can consider these commutation
relations as a two-parametric deformation
of Heisenberg algebra \lnsgd \ with real
parameters \ $\xi >1$\ and \ $0<x<1$\ . The meaning of
deformation with parameter
\ $\xi$\ was discussed in section 5, while the simple example of
x-deformation was represented in section 4.
The Fock module\ ${\cal F}_p$\
for algebra \nsbdf \ can be constructed
as an ordinary one. As a vector space
it is isomorphic to undeformed Fock
space considered in the previous sections because \ ${\cal F}_p$\
is still covered by vectors
\ $\oplus_{k=1}^{\infty}\C b_{-n_1}...b_{-n_k}{\bf f}_p$\ with \ $n_i>0$.
The module\ ${\cal F}_p$\ can be endowed by the
structure of \ $Z$-graded module if we introduce grading
operator as
\eqn\jsghg{D=\frac{1}{2}\sum_{m>0} \frac{[2m]_x}{[m]_x}
( b_{-m} b'_m+b'_{-m} b_m)+
\frac{P^2}{2}-\frac{1}{4\xi(\xi+1)} \ .}
The x-deformation \nsbdf \ does not change
the number of states on every level, and character of
the Fock module will be the same as an earlier. Let us note here, that
choosing deformation parameter x as real positive number,
we are considering x-deformation in generic point.
As before the main object of our consideration will
be the set of the Fock space\
$\{{\cal F}_{k,k'}\equiv{\cal F}_{\frac{(\xi+1) k-\xi k'}
{\sqrt{ 2\xi(\xi+1)}}}\ | \
k,k'\in{\bf Z}\}. $
Introduce two-parameters family of formal operators
\eqn\swqr{\eqalign{&X^l
=\int_{C_1}...\int_{C_l}
\frac{d z_1}{2\pi i z_1}...\frac{d z_l}{2\pi i z_l}\  e^{- i\ [ \phi(z_1 x )+
\phi(z_1 x^{-1} )]}
..
e^{- i\  [\phi(z_l x )+
\phi(z_l x^{-1} )]}
,\cr
&X'^{\ l'}
=\int_{C_1}...\int_{C_{l'}}
\frac{d z_1}{2\pi i z_1}...\frac{d z_{l'}}{2\pi i z_{l'}}
\  e^{- i\ [ \phi'(z_1 x )+
\phi'(z_1 x^{-1} )]}
..
e^{- i\ [ \phi'(z_{l'} x )+
\phi'(z_{l'} x^{-1} )]}\ .}}
Evidently, these integral operators are
deformations of operators \hdgswqr . We have seen example
of such deformation \efg \ in section 4.
Let us note that \ $X^l$\ and \ $X'^{\ l'}$\
are related by transformation
\ $\xi\leftrightarrow -1-\xi $ (or equivalently, \ $\phi\leftrightarrow \phi'\
,\
l\leftrightarrow l',$).
The worth important fact is that,
as in the case\ $x=1$,
operators  \ $X^l$\ and \ $X^{'l'}$\
obey the proposition:
\par
\noindent
{\bf Proposition 6.1}
\noindent
\it
The action of operator \ $X^l\ (X'^{\ l'})$ \ is defined
only on the Fock modules\ ${\cal F}_{l,k}\ ({\cal F}_{k,l'}), k\in{\bf Z}\ $
and\ $l, l'>0$. Then
\eqn\mdbhjob{\eqalign{
&Ker_{{\cal F}_{l,l'}}[X^l]=Ker_{{\cal F}_{l,l'}}[X'^{\ l'}]\ ,\cr
&Ker_{{\cal F}_{l,-l'}}[X^l]=
Ker_{{\cal F}_{l,-l'}}[X'^{\ l'}]=0\ ,\cr
&Im_{{\cal F}_{l,l'}}[X^l]={\cal F}_{-l,l'}\ ,\cr
&Im_{{\cal F}_{l,l'}}[X'^{\ l'}]={\cal F}_{l, -l'}\ ,\cr
&Im_{{\cal F}_{l,-l'}}[X^l]=
Im_{{\cal F}_{-l,l'}}[X'^{\ l'}]
\subset {\cal F}_{-l,-l'}\ .}}
\noindent
\rm
Now we would like to give the
\par
\noindent
{\bf Definition.}\
\it
Let operators \ ${P, Q, b_n, n\not = 0}$\
satisfy the commutation relations \nsbdf \
and screening operators are given by \swqr .
The  x-deformed Virasoro algebra \ $Vir_{c,x}$\
is a subalgebra in the universal enveloping
algebra of deformed Heisenberg algebra of operators \ $b_n$\ and\ $P$.
An element from universal enveloping of Heisenberg
algebra belongs to deformed Virasoro algebra if it
commutes with screening operators.
\rm

The example of the deformed Virasoro algebra was given
in section 4. This algebra inherits \ ${\bf Z}$\ grading from
the algebra \nsbdf \ of operators \ $b_n$.
Due to definition, screening operators
are intertwining operators for representations
of deformed Virasoro algebra. Using proposition 6.1 one can
investigate
the irreducible representations of \ $Vir_{c,x}$.
Let spaces \ ${\cal L}_{l,l'}$\ are determined as
in \ksi . We assume that in the generic point of x-deformation
the analogue of proposition 5.2 is also correct.
It can be rewritten now as:

\noindent
{\bf Conjecture 6.1}

\noindent
\it
Let parameter\ $\xi$\ be irrational number and \ $\xi>1$, then

\noindent
1. The space\ ${\cal L}_{l, l'}$\ for any integer number\ $l, l'$\ is
irreducible representation of
deformed  Virasoro algebra with central
charge defined by \gsfd .
The highest weight vector of \ ${\cal L}_{l, l'}$\
coincide with the vector
\ ${\bf f}_{l,l'}$\ and its weight is given by the formula \jshdg .

\noindent

2. For \ $l\cdot l'\geq 0$\ or\ $ l,l'<0$\ ,
Verma modules \ ${\cal M}_{l,l'}$\
built upon highest vector\ ${\bf f}_{l,l'}$ \
coincide with Fock modules\ ${\cal F}_{l,l'}$\ .

\noindent
\rm
Now we need make one more assumption concerning to
conjugation condition for \ $Vir_{c,x}$. Such as in the Fock space
there exists a natural scalar product then we expect
that
the following statement is true:
\par
\noindent
{\bf Conjecture 6.2}

\noindent
\it
Let irreducible representation \ ${\cal L}_{l,l'}$\
and dual Verma module \ ${\cal M}_{l,l'}^\star$\ of deformed
Virasoro
algebra \ $Vir_{c,x}$\ are realized as
bosonic modules \ $Ker_{{\cal F}_{l,l'}}[X^l]$\ and
\ ${\cal F}_{-l,-l'}$\ correspondingly, where \ $l=1,2,...$\ .
Then the scalar product \ $( , )_0$\ \vcsda \
restricted on the vectors
from \ ${\cal L}_{l,l'}$\ and \ ${\cal M}_{l,l'}^\star$\
confirms with
inner conjugation in the deformed Virasoro algebra.
\par
\noindent
\rm
Unfortunately, at present we can not construct
proper basis of generators in "quantum" Virasoro algebra
and prove assumptions above. Nevertheless, let us try to
work out the consequences of these statements. We will
see below that such conjugation condition leads to
the proper anti-involution in the
algebras of chiral primaries.

Let us consider now the generalization of the
operators\ $V_{\pm}(\zeta)$\ and $V_{\pm}'(\zeta)$\
introduced in the section 5. In the \ $x$-deformed
case we will define it as follows:
\eqn\hsgdaaf{\eqalign{&V_+(\zeta)=\ e^{i\phi(\zeta)}\ ,\cr
&V_-(\zeta)=\eta^{-1}\ \int_C\frac{d z}{2\pi i z}\
e^{-i\ [(\phi(zx)+\phi(zx^{-1}]} \ e^{i\phi(\zeta)}\
\cr
&V'_+(\zeta)=\ e^{i\phi'(\zeta)}\ ,\cr
&V'_-(\zeta)=\eta'^{-1}\ \int_C\frac{d z}{2\pi i z}\
e^{-i\ [(\phi'(zx)+\phi'(zx^{-1}]} \ e^{i\phi'(\zeta)}\
,}}
where the prescription for the integration contours is the same
as an earlier.
The constants \ $\eta,\  \eta'$\ we will be specified later to
provide the convenient normalization of operators\ \hsgdaaf .
Note that action of operator \ $V_- \ (V'_-)$\ is  well defined
only on the Fock modules\ ${\cal F}_{k,k'}$\
with \ $(\xi+1)k-\xi k'<0$\ $((\xi+1)k-\xi k'>0)$.
One can prove the
\par
\noindent
{\bf Proposition 6.2}

\noindent
\it
Let
\ $ (\xi+1)\  l-\xi\  l'<0$  and \ $l,l'>0$. Then

$\bullet$ the action
of the operators \ $V_{\pm}(\zeta)$\ is
described
by the
commutative diagrams \nsgdfwt\ \nsgdf .

$\bullet\bullet$\
the action of the operators \ $V'_{\pm}(\zeta)$\ is
defined by the following
commutative diagrams

\noindent
i.
\eqn\nsgdfwt{\eqalign{\matrix{&{\ } & {\ }    &V'_{\pm}&{\ } &{\ }\cr
&{\ } &{\cal F}_{l, l'}&\longrightarrow &{\cal F}_{l,l'\pm 1} &{\ }\cr
& X^{l}&\biggl\downarrow & V'_{\pm}&\biggl\downarrow & (-1)^l X^{l}\cr
&{\ } &{\cal F}_{-l, l'}&\longrightarrow &{\cal F}_{-l,l'\pm 1}&{\ }\cr}\ ,}}
ii.
\eqn\nsgdf{\eqalign{\matrix{&{\ } & {\ }    &V'_{\pm}&{\ } &{\ }\cr
&{\ } &{\cal F}_{-l, -l'}&\longrightarrow &{\cal F}_{-l,-l'\pm 1}
&{\ }\cr
& X^{l}&\biggl\uparrow & V_{\pm}&\biggl\uparrow &(-1)^l X^{l}\cr
&{\ } &{\cal F}_{l, -l'}&\longrightarrow &{\cal F}_{l,-l'\pm 1}&{\ }\cr}\ .}}

\rm
\par
\noindent
Let us
illustrate this proposition on the simple example.
\subsec{Example of calculations}
\par
\noindent
{}From the proposition 6.1 follows that if Fock module \ $F_{-1,-1}$\
is considered as \ $Vir_{c,x}$\ module then it
contains singular vector \ $X'{\bf f}_{-1,1}$\ .
It is easy to show that this vector turns to be proportional to
the state\ $\p_t\ \big\{t e^{-i[\phi'(tx)+\phi'(tx^{-1})]}
\big\}|_{t=0}{\bf f}_{-1,1}\sim
b_{-1}'{\bf f}_{-1,-1}$.
According to proposition 6.2, operator \ $V_-(\zeta_1)$\
acting on this state set it into null vector
in the  Verma module \ ${\cal M}_{-2,-1}\cong {\cal F}_{-2,-1}$.
Due to our conjecture on the scalar product
such null vector has to be orthogonal to all states
from irreducible module \ ${\cal L}_{2,1}$ and, in particular,
to the state \ $V_+(\zeta_2) {\bf f}_{1,1}$:
\eqn\fsda{\zeta_1^{-1} f(\zeta_1 \zeta_2^{-1})=
({\bf f}_{1,1} V_{+}(\zeta_2),\  V_-(\zeta_1)X'{\bf f}_{-1,1})=0\ .}
Let us check this formula.
As a consequence of bosonic representation
it can be represented in the form:
\eqn\jsgdg{\zeta^{-1}_1 f(\zeta_1 \zeta_2^{-1})=\p_t\big\{t
\int_C\frac{d z }{2\pi i z}\
({\bf f}_{1,1}, e^{i\phi(\zeta_2)}
e^{-i \bar\phi( z )}
e^{i\phi(\zeta_1)} e^{-i \bar\phi'(  t)}{\bf f}_{-1,1})_0\big\}|
_{t=0}\ ,}
where  short notations \ $\bar\phi(z)$  and
\ $\bar\phi'(z)$\ are introduced for \ $\phi( z x)+\phi(z x^{-1})$\ and
\ $ \phi'( t x)+\phi'(t x^{-1})$  correspondingly.

The technique of calculation of similar vacuum expectation values
in the bosonic space was developed by Dotsenko and Fateev \refs{\DotsFat}.
Let us recall how it works in the case of x-deformed operators.
First of all, it is convenient to extract the contribution
coming from the zero modes of operators in \ \jsgdg .
Using the commutation
relation for \ $P,Q$\ it is easy to get
that ordering of these operators is resulted in the
multiple of the form
\ $\zeta^{\frac{\xi-1}{4\xi}}_1\
\zeta_2^{-\frac{\xi+3}{4\xi}} \ z^{\frac{1}{\xi}}\ t^{-1}.$
The ordering of oscillator modes \ $b_n$,
as an ordinary, means that annihilation operators have to stand
at right hand side while creation operators should be at left hand side.
It is convenient to distinguish here two
cases. In the first case one needs provide ordering of
oscillators from different exponents while in the second
from the same exponent.

First of all, we want to remind the procedure in the
first case.
Let us elaborate for instance the ordering of the
expression \ $e^{i \phi( \zeta_2 )}
e^{i\phi(\zeta_1)}$. Using Campbell-Baker-Hausdorff  formula\
$e^A\ e^{B}=e^{A+B} e^{\frac{[A,B]}{2}}\ $ one can find
that contributions appearing from coupling of these exponents
are equal to
\eqn\gsfqr{g(\zeta_1 \zeta_2^{-1})=e^{-[\phi_+(\zeta_2 ),
\phi_-(\zeta_1)]}\ ,}
where we denote by $\phi_\pm$\
positive and
negative frequency parts of the field\ $\phi$ correspondingly.
Straightforward computation with using of commutation relations \nsbdf\
leads to
the following representation of
function \ $g(\zeta_1 \zeta_2^{-1})$:
\eqn\hdsgd{g(z)=\exp\big[-\sum_{m=1}^{+\infty}\frac{[m]_x\ [m(\xi+1)]_x}
{ m\ [2m]_x\ [m\xi]_x}
\  z^m\big]\ .}
The sum in this expression converges only for\ $|z|<1$.
Its analytical continuation
on the whole complex plane is given by the following
infinite product:
\eqn\js{g(z)=\frac{(z;x^{2\xi})_{\infty}
\ (x^4 z;x^{2\xi},x^4)_{\infty} \
(x^{4+2\xi} z;x^{2\xi},x^4)_{\infty}}
{(x^2 z;x^{2\xi})_{\infty}
\ (x^6 z;x^{2\xi},x^4)_{\infty} \
(x^{2+2\xi} z;x^{2\xi},x^4)_{\infty}}\ .}
The contributions coming
from the averaging of others pairs of exponents
can be obtained by the
same fashion. Slightly different procedure have to be provided
in the ordering of the oscillators in the
same exponent. Proceeding as above, one formally get that the
ordering of \ $e^{i\phi(\zeta_1)}$\ gives
constant\ $\rho$\ such that
$\rho^2=g(1)$.
However, as it is seen from \js , \ $g(z)$\ has a simple zero at \ $z=1$.
Let us adopt the following conventions:
\eqn\ghdfgl{\rho^2=\lim_{z\to 1}\ \frac{1-x^2}{1-z}\  g(z)
\ .}
Ordinary, one use exponent operators in the
normal ordered form
from  the very beginning.
Then there is no need to take care on
ordering of oscillators belonging to the same exponent. Making
this ordering in the final step is more convenient
for us. Indeed, in deformed case
this step is resulted in non-trivial constants depending on
the deformation parameter that is unlike
to the conformal case where renormalization has the similar
form for every exponent.
We introduce constants like\ $\rho$\ since they
provide the natural normalization of operators.
Now let us present the final expression for averaging of the
integrand in \jsgdg \ :
\eqn\jshdg{\eqalign{({\bf f}_{1,1}, e^{i\phi(\zeta_2)}
e^{-i \bar\phi(z)}
&e^{i\phi(\zeta_1)} e^{-i \bar\phi'( t )}{\bf f}_{-1,1})_0
=\rho^2\bar\rho\bar\rho'\
\zeta^{\frac{\xi-1}{4\xi}}_1\
\zeta_2^{-\frac{\xi+3}{4\xi}}\  z^{\frac{1}{\xi}}\ t^{-1}\times\cr
&g(\zeta_1 \zeta_2^{-1})\
w(\zeta_1 z^{-1})\ w(z \zeta_2^{-1})\ u(t \zeta_1^{-1})\
u(t \zeta_2^{-1})\ \bar h(t z^{-1})\ .}}
Here \ $g(z)$\ is defined by \js , while other functions
yield
\eqn\jshfgf{\eqalign{
&w(z)=\frac{(x^{1+2\xi}z;x^{2\xi})_{\infty}}{(x^{-1}z;x^{2\xi})_{\infty}},\cr
&u(z)=1-z\ ,\cr
&\bar h(z)=\frac{1}{(1-zx)(1-zx^{-1})}\ .}}
As we mentioned above, the constants\ $\rho , \bar\rho ,\bar\rho'$\
are non-trivial functions of deformation parameter \ $x$. They are given by:
\eqn\jdhfgyt{\eqalign{&\rho^2=(1-x^2)\
\frac{(x^{2\xi};x^{2\xi})_{\infty}}{(x^{2+2\xi};x^{2\xi})_{\infty}}\
g^{-1}(x^2)\ ,\cr
&\bar\rho^2=(1-x^2)\
\frac{(x^{-2};x^{2\xi})_{\infty}}{(x^{2+2\xi};x^{2\xi})_{\infty}}\ ,\cr
&\bar\rho'^2=(1-x^2)\
\frac{(x^{2};x^{2+2\xi})_{\infty}}{(x^{2\xi };x^{2+2\xi})_{\infty}}\ .}}
The integral from expression \jshdg \ can be expressed in terms of
q-special functions.
Let us introduce necessary notations.
We will need definitions of
q-gamma function \ $\Gamma_q(a)$\
and q-hypergeometric function \ $F_q(a,b,c;z)$.
They are usually defined as following
\eqn\nbvb{\Gamma_q(a)=(1-q)^{1-a}
\frac{(q;q)_{\infty}}
{(q^a;q)_{\infty}}\ , }
\eqn\hsfgf{F_q(a,b,c;z)=1+\sum_{n=1}^{\infty}\frac{(q^a;q)_n(q^b;q)_n}
{(q^c;q)_n (q;q)_n}\  z^n\ ,}
here\ $(q^a;q)_n=\prod_{p=0}^{n-1}(1-q^{a+p})$.
There is the following integral representation for
q-hypergeometric function generalizing formula \ytww :
\eqn\jdhgfytww{\eqalign{
\int_C&
\frac{d z}{2\pi i} z^{c-1}
\frac{(q^{\frac{1+a}{2}} z^{-1};q)_{\infty}}
{(q^{\frac{1-a}{2}} z^{-1};q)_{\infty}}\
\frac{(q^{\frac{1+b}{2}}\zeta z;q)_{\infty}}
{(q^{\frac{1-b}{2}}\zeta z;q)_{\infty}}=
\cr
&=\
q^{\frac{c(1-a)}{2}}\  \frac{\Gamma_q(c+a)}{\Gamma_q(c+1)\ \Gamma_q(a)}
\ F_q(a+c,b,c+1;  q^{1-\frac{a+b}{2}} \zeta)\ .}}
Notice, that in this remarkable formula we have
ordinary contour integral rather than Jackson's one.
With using of this integral representation the calculation becomes trivial
and leads to the formula:
\eqn\hsgdf{\eqalign{f(\zeta)=const\  g(\zeta)
\big[\  &F_{x^{2\xi}}(2\xi^{-1},
1+\xi^{-1},\xi^{-1};\zeta x^{-2})-\cr
&(1+\zeta)
F_{x^{2\xi}}(1+2\xi^{-1},
1+\xi^{-1},1+\xi^{-1};\zeta x^{-2})\ \big]\ . }}
In the special case under the consideration we need the following
particular expressions for hypergeometric functions:
\eqn\gdfsfdr{\eqalign{&F_q(a,b,b;z)=
\frac{(q^a z;q)_{\infty}}
{(z;q)_{\infty}}\ ,\cr
&F_q(2b-2,b,b-1;z)=(1+ q^{b-1}\ z)\ \frac{(q^{2b-1} z;q)_{\infty}}
{(z;q)_{\infty}}\ .}}
Substituting these expressions into formula \hsgdf \
one can obtain that \ $f(\zeta)$\
is identical zero. That proves
the orthogonality
of the vectors \ $V_-(\zeta_1)X'{\bf f}_{-1,1} $\
and \ $V_+(\zeta_2) {\bf f}_{1,1}$.
Analogously, one can prove that the following matrix element
is zero:
\eqn\fsdas{
({\bf f}_{1,1} V'_{+}(\zeta_2),\  V'_-(\zeta_1)X{\bf f}_{1,-1})=0\ .}
The procedure here is similar to those done before.
The only difference is that averaging of the integrand
leads to the expression:
\eqn\shdg{\eqalign{({\bf f}_{1,1}, e^{i\phi'(\zeta_2)}
e^{-i \bar\phi'(z)}
&e^{i\phi'(\zeta_1)} e^{-i \bar\phi( t )}{\bf f}_{1,-1})_0
=\rho'^2 \bar\rho' \bar\rho\
\zeta^{\frac{\xi+2}{4(\xi+1)}}_1\
\zeta_2 ^{-\frac{\xi-2}{4(\xi+1)}}\  z^{-\frac{1}{\xi+1}}\ t^{-1}\times\cr
&g'(\zeta_1 \zeta_2^{-1})\
w'(\zeta_1 z^{-1})\ w'(z \zeta_2^{-1})\ u(t \zeta_1^{-1})\
u(t \zeta_2^{-1})\ \bar h(t z^{-1})\ .}}
Let us emphasize here that the function \
$u(\zeta)=e^{[\phi'_{+}(\zeta_2),\bar{\phi}_{-}(\zeta_1)]}$\
in this expression is given by the same formula as before,
although it is appeared from the ordering of other
exponents. This is consequence of the very special
form of the proposed x-deformation. Functions
\ ${\bar h},\ {\bar \rho},{\bar \rho}'^2$\
have been written in \jshfgf \ and  \jdhfgyt\
while others
have the form
\eqn\jshfgfg{\eqalign{&g'(z)=\frac{
(x^2 z;x^{2+2\xi},x^4)_{\infty} \
(x^{4+2\xi} z;x^{2+2\xi},x^4)_{\infty}}
{(x^4 z;x^{2+2\xi},x^4)_{\infty} \
(x^{2+2\xi} z;x^{2+2\xi},x^4)_{\infty}}\ ,\cr
&w'(z)=
\frac{(x^{1+2\xi}z;x^{2+2\xi})_{\infty}}{(xz,x^{2+2\xi})_{\infty}}\ ,\cr
&\rho'^2=\frac{(x^2 ;x^{2+2\xi})_{\infty}}
{(x^{2\xi+2} ;x^{2+2\xi})_{\infty}}\ g'^{-1}(x^2)\ .}}

\subsec{Vertex operators for deformed Virasoro algebra}
The proposition 6.2 means that operators \hsgdaaf \ are
vertex operators which interpolate between irreducible representations
of deformed Virasoro algebra. We assume that
bosonic operators \ $V_{\pm}$\ can be identified with
"quantum" analogues of operators  \ $\Phi_{21}^{\pm}$\
by the formulae
\eqn\kh{\eqalign{
&V_+(\zeta){\bf v}=\Phi_{21}^+(\zeta){\bf v} \
\ \ \  (l>0,l'>0, \ (\xi+1)\  l-\xi\  l'>0)\ , \cr
&V_+(\zeta){\bf v}^{\star}=\Phi_{21}^{\star -}(\zeta x^2){\bf v}^{\star} \
\ (l>1,l'>0,\ (\xi+1)\  l-\xi\  l'<0)\  , \cr
&V_-(\zeta){\bf v}=\Phi_{21}^-(\zeta){\bf v} \
\ \ \ (l>1,l'>0, \ (\xi+1)\  l-\xi\  l'<0)\ , \cr
&V_-(\zeta){\bf v}^{\star}=\Phi_{21}^{\star \ +}(\zeta x^2){\bf v}^{\star} \
\ (l>0,l'>0,\ (\xi+1)\  l-\xi\  l'>0)\ .}}
Bosonic realization of operator \ $\Phi_{12}^\pm$\
is determined by similar expressions:
\eqn\kih{\eqalign{
&V'_+(\zeta){\bf v}=\Phi_{12}^+(\zeta){\bf v} \
\ \ \  (l>0,l'>0,\ (\xi+1)\  l-\xi\  l'<0)\ , \cr
&V'_+(\zeta){\bf v}^{\star}=\Phi_{12}^{\star -}(\zeta x^2){\bf v}^{\star} \
\ (l>0,l'>1, \ (\xi+1)\  l-\xi\  l'>0)\ , \cr
&V'_-(\zeta){\bf v}=\Phi_{12}^-(\zeta){\bf v} \
\ \ \ (l>0,l'>1\ (\xi+1)\  l-\xi\  l'>0)\ , \cr
&V'_-(\zeta){\bf v}^{\star}=\Phi_{12}^{\star +}(\zeta x^2){\bf v}^{\star} \
\ (l>0 ,l'>0,\ (\xi+1)\  l-\xi\  l'<0)\ ,}}
If we
adjoin the conjugation
conditions:
\eqn\mcnvnb{\eqalign{&\big({\bf v}_1^{\star} \Phi^{\star\  \pm}_{21}(\zeta
x^2),
\ {\bf v}_2\big)=
\big({\bf v}_1^{\star},\Phi^{\mp}_{21} (\zeta )\ {\bf v}_2 \big)\ ,\cr
&\big({\bf v}_1^{\star} \Phi^{\star\  \pm}_{12}(\zeta x^2),
\ {\bf v}_2\big)=
\big({\bf v}_1^{\star},\Phi^{\mp}_{12} (\zeta)\ {\bf v}_2 \big)\ ,}}
where\ ${\bf v}_1^{\star}\in {\cal L}_{l,l'}^{\star}$\ ,
${\bf v}_2\in {\cal L}_{l,l'}$\ and \ $|\zeta|=1$\ ,
then formulae \ \kh ,\ \kih \
uniquely specify an action of x-deformed vertex operators
in irreducible representation of \ $Vir_{c,x}$.
Having the bosonic representation one
can get all information on the vertex operators. In principle,
our construction is very similar to those known in CFT.
The difference appears only in the explicit form of vacuum averaging
of exponents. For instance, let us
write down the matrix elements of the product of two
x-deformed vertex operators. It can be obtained
by using the formula \jdhgfytww \ and knowing coupling of
corresponding exponents.
If we fix the constants
\ $\eta$\  and\ $\eta'$\ in definition \hsgdaaf \
as following:
\eqn\jnb{\eta=\ \eta_{x^{2\xi}}(\frac{\xi+1}{\xi}),\ \ \
\eta'=\ \eta_{x^{2+2\xi}}(\frac{\xi}{\xi+1})\ ,}
where
$$\eta^2_q(a)=(1-x^2)\
\frac{q^{\frac{a(a-1)}{2}}(1-q)}{\Gamma_q(a)\Gamma_q(1-a)}\ ,$$
then we easily find that x-deformation of the functions
\ $G^{\pm}_p$\ \mcn\ has the form:
\eqn\jdgg{\eqalign{&({\bf v}_{l,l'}^{\star},
\Phi_{2,1}^{\pm}(\zeta_2)\Phi_{2,1}^{\mp}(\zeta_1)\ {\bf v}_{l,l'})=
C\
\frac{g(\zeta_1\zeta^{-1}_2 )}
{g(x^2)}\ G_{x^{2\xi}}\big(\mp\frac{\xi+1}{\xi}l\pm l',\frac{\xi+1}{\xi};
\zeta_1\zeta^{-1}_2\big)\ ,\cr
&({\bf v}_{l,l'}^{\star},
\Phi_{1,2}^{\pm}(\zeta_2)\Phi_{1,2}^{\mp}(\zeta_1)\ {\bf v}_{l,l'})=
C'\ \frac{g'(\zeta_1\zeta_2^{-1})}{g'(x^2)}\ G_{x^{2+2\xi}}
\big(\mp\frac{\xi}{\xi+1}l'\pm l,\frac{\xi}{\xi+1};\zeta_1\zeta_2^{-1}\big).}}
Both formulae here are described by the same function
\ $G_q(c,a;z)$\ taken with different parameters.
It is given by the expression:
\eqn\mncx{G_q(c,a;z)=\ q^{\frac{(1-a)(2c+a)}{4}}\ (1-q)^{2a-2}
\frac{\Gamma_q(
a+c)}
{\Gamma_q(c+1)}\ z^{\frac{a}{4}+\frac{ c}{2}}
F_q(a+c,
a,c+1; q^{1-a}z)\ .}
If we choose the constants \ $\eta,\ \eta'$\ as in \jnb  , then
the constants \ $C, C'$\ will have the form
$$C=(1-x^2)\ \Gamma_{x^{2\xi}}(\frac{\xi+1}{\xi}),\ \  C'=
\frac{1-x^{2+2\xi}}{\Gamma_{x^{2+2\xi}}(\frac{1 }{\xi+1})}.$$
This normalization of vertex operators is convenient since
it provides the following  normalization of functions \ \jdgg\ :
\eqn\dhgdfgtw{\eqalign{
&({\bf v}_{l,l'}^{\star},
\Phi_{21}^{\pm}(\zeta_2)\Phi_{21}^{\mp}(\zeta_1)\ {\bf v}_{l,l'})=
\frac{1-x^2}
{1-x^{-2}\zeta_1\zeta^{-1}_2}+...\ ,\ \ \ \zeta_1\to x^2\zeta_2\ ,\cr
&({\bf v}_{l,l'}^{\star},
\Phi_{12}^{\pm}(x^2\zeta)\Phi_{12}^{\mp}(\zeta)\ {\bf v}_{l,l'})= 1\ .}}
Further we will see that expressions \jdgg \ determine the commutation
relations of deformed operators likewise to conformal case.
For this reason, in order to describe
arbitrary matrix element of operators
\ $\Phi_{12}^\pm, \ \Phi_{21}^\pm$\ it is enough to
present the explicit formulae
for functions
\eqn\gsfddqr{\eqalign{\Im_{nm}(\zeta_1,...\zeta_{2n},\zeta_1',...,\zeta_{2m}')=
&({\bf v^{\star}}_{11},
\Phi_{12}^{-}(\zeta'_{2m})...
\Phi_{12}^{- }(\zeta'_{m+1})\Phi_{12}^{+}(\zeta'_m)...
\Phi_{1,2}^{+}(\zeta'_1)\times\cr
&\Phi_{21}^{-}(\zeta_{2n})...
\Phi_{21}^{-}(\zeta_{n+1})\Phi_{21}^{+}(\zeta_n)...
\Phi_{21}^{+}(\zeta_{1})
\ {\bf v}_{11})\ .}}
To calculate these matrix elements of vertex operators
we need know
together with the functions \js\ ,\jshfgf\ ,\jshfgfg\ also explicit
form of the following averaging:
\eqn\jdgf{\eqalign{&
\bar g(\zeta_1 \zeta_2^{-1})=e^{-[\bar\phi_+(\zeta_2 ),
\bar\phi_-(\zeta_1)]}\ ,\cr
&\bar g'(\zeta_1 \zeta_2^{-1})=e^{-[\bar\phi'_+(\zeta_2 ),
\bar\phi'_-(\zeta_1)]}\ ,\cr
&h(\zeta_1 \zeta_2^{-1})=e^{-[\phi_+(\zeta_2 ),
\phi'_-(\zeta_1)]}\ .}}
Providing standard procedure which was explained above
it is not hard to obtain the formulae
\eqn\jsghdf{\eqalign
{&\bar g(z)=
(1-z)\ \frac{(x^{-2}z; x^{2\xi})_{\infty}}
{(x^{2+2\xi}z; x^{2\xi})_{\infty}}\ ,\cr
&\bar g'(z)=
(1-z)\ \frac{(x^{2}z; x^{2+2\xi})_{\infty}}
{(x^{2\xi}z; x^{2+2\xi})_{\infty}}\ ,\cr
&h(z)=
\frac{(x^3 z;x^4)_{\infty}}{(xz;x^4)_{\infty}}\ .}}
Bosonization technique allows one
to represent
the functions\ \gsfddqr \ in the form of contour
integral from meromorphic functions:
\foot{For convenience, we collect all necessary
averaging of the exponents in the Appendix.}
\eqn\vcf{\eqalign{&\Im_{nm}(\zeta_1,...\zeta_{2n},\zeta_1',...,\zeta_{2m}')=
\rho^{2n}(\bar\rho\eta^{-1})^n\rho'^{2m}(\bar\rho'\eta'^{-1})^m\
\int_{C_1}...\int_{C_n}
\prod_{k}
\frac{d z_k}{2\pi i z_k }\times\cr
&\int_{S_1}...\int_{S_m}
\prod_k
\frac{d z'_k}{2\pi i z'_k }
\  f^n_{\frac{\xi+1}{\xi}}(\zeta_1,...,\zeta_{2n}|z_1,...z_n)
\ f^m_{\frac{\xi}{\xi+1}}(\zeta'_1,...,\zeta'_{2m}|z'_1,...z'_m)\
\cr
&\prod_{  i<j} g(\zeta_{i}\zeta_{j}^{-1})
\prod_{  i<j} {\bar g}(z_i z_j^{-1}) \prod_{i<n+ j}
w(\zeta_i z^{-1}_j)\prod_{n+i\leq j}
w(z_i\zeta^{-1}_j)
\prod_{i<j} g'(\zeta'_{i}\zeta'^{-1}_{j})
\prod_{  i<j} {\bar g}'(z'_i z'^{-1}_j)
\times\cr
& \prod_{i<m+ j}
w'(\zeta'_i z'^{-1}_j)\prod_{m+i\leq j}
w'(z'_i \zeta'^{-1}_j)
\prod_{  i,j} h(\zeta_{i}\zeta'^{-1}_{j})
\prod_{  i,j} {\bar h}(z_i z'^{-1}_j) \prod_{i, j}
u(\zeta_i z'^{-1}_j)\prod_{i,j}
u(z_i \zeta'^{-1}_j)\ .}}
Here we denoted by symbols
\  $f^n_{\frac{\xi+1}{\xi}}, f^m_{\frac{\xi}{\xi+1}}$\
functions representing the contributions of zero modes
in averaging of corresponding exponents. Their explicit form
is defined by the relations:
\eqn\jdhfg{f^n_a(\zeta_1,...\zeta_{2n}|z_1,...,z_n)
=\prod_{k=1}^n\zeta_k^{\frac{a}{4}(1+2k)-\frac{1}{2}}\
\ \zeta_{n+k}^{\frac{a}{4}(1+ 2n-2k)-\frac{1}{2}}\
z_k^{a(k-n-1)+1}\ .}
The integration contours \ $\{C_i\}_{i=1}^n$\ and\ $\{S_i\}_{i=1}^m$\
are depicted in the fig.6.
\ifig\fqcontur{Integration contours.}
{\epsfxsize4.5
in\epsfbox{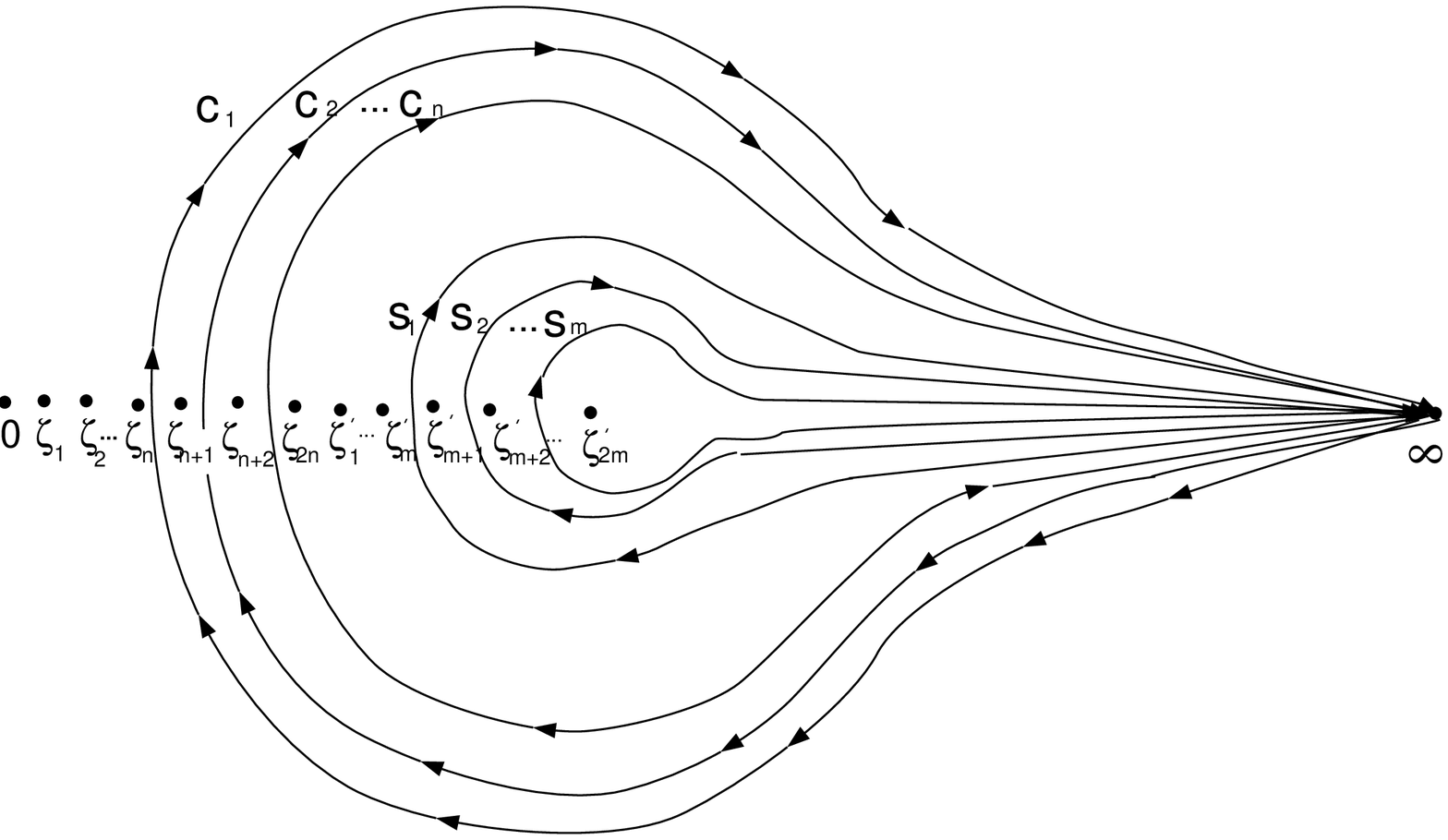}}
%
\par\noindent
Note that any contour \ $C_i$\  encloses those singularities of integrand
which
depend on \ $\zeta_k\ ( 2n\geq k\geq n+i),\ z_k\ (  k>i),\
\zeta_k'\ (  2n\geq k\geq 1)\ , z'_k\ ( n\geq k\geq 1)$.
At the same time a contour \ $S_i$\ encloses all
singularities
determining by \ $\zeta'_k\ ( 2m\geq k\geq m+i),\ z'_k\ ( k>i)$.

\subsec{Elliptic ZF algebra of IRF type.}
\par\noindent
Now let us turn
to consideration of commutation relations
of vertex operators.
They can be derived from the rules of analytical
continuation of the functions \jdgg\
by using the following relations for q-hypergeometric function \refs{\exton}:
\eqn\ldjfhj{\eqalign{
F_q(a,b,c;z)&=\frac{\Gamma_q(c)\ \Gamma_q(b-a)}{\Gamma_q(b)\ \Gamma_q(c-a)}
\frac{\Theta_q(q^a z)}{\Theta_q(z)}F_q(a,a-c+1,a-b+1;q^{c+1-a-b}z^{-1})\cr
&+\frac{\Gamma_q(c)\ \Gamma_q(a-b)}{\Gamma_q(a)\ \Gamma_q(c-b)}
\frac{\Theta_q(q^b z)}{\Theta_q(z)}
F_q(b,b-c+1,b-a+1;q^{c+1-a-b}z^{-1})\ .}}
We have
\par
\noindent
{\bf Proposition 6.3}

\noindent
\it
Vertex operators \ $\Phi_{12},\ \Phi_{21}$\
obey the following commutation relations:
\eqn\hdf{\eqalign{&\Phi_{21}^a(\zeta_1)\Phi_{21}^b(\zeta_2)
|_{{\cal L}_{l,l'}}
=\sum_{c+d=a+b}{\bf W}
\left[\matrix{l+a+b&l+c\cr l+b&l}\biggl|\ \frac{\zeta_1}{\zeta_2}\
\right]\
\Phi_{21}^d(\zeta_2)
\Phi_{21}^c(\zeta_1)|_{{\cal L}_{l,l'}}\ ,\cr
&\Phi_{12}^a(\zeta_1)\Phi_{12}^b(\zeta_2)
|_{{\cal L}_{l,l'}}
=\sum_{c+d=a+b}{\bf W}'
\left[\matrix{l'+a+b&l'+c\cr l'+b  &l'}
\biggl|\ \frac{\zeta_1}{\zeta_2}\ \right]\
\Phi_{12}^d (\zeta_2)
\Phi_{12}^c(\zeta_1)|_{{\cal L}_{l,l'}}\ ,\cr
&\Phi_{21}^a(\zeta_1)\Phi_{12}^b(\zeta_2)=
\ a\ b\ d\big(\frac{\zeta_1}{\zeta_2}\big)\
\Phi_{12}^b(\zeta_2)\Phi_{21}^a(\zeta_2)\ .}}

\rm
\par\noindent
The function \ $d(z)$\ appeared at the formula above has the form:
\eqn\jdghdsaq{d(z)=z^{-\frac{1}{2}}\frac{
\Theta_{x^4}(x z)}{\Theta_{x^4}(x^3 z)}\ ,}
while matrices  \ ${\bf W}$,  \ ${\bf W'}$\ can be represented
in the form:
\eqn\kdshjghf{\eqalign{&{\bf W}\left[\matrix{l_4&l_3\cr l_1&l_2}
\biggl|\ \zeta\ \right]=\zeta^{\frac{\xi+1}{2\xi}}\ \frac{g(\zeta^{-1})}
{g(\zeta)}\
\hat{\bf W}
\left[\matrix{l_4&l_3\cr l_1&l_2}\biggl|\ \zeta\
,\frac{\xi+1}{\xi},x^{2\xi}\
\right]\ ,\cr
&{\bf W}'
\left[\matrix{l'_4&l'_3\cr l'_1&l'_2}\biggl|\ \zeta
\right]=\zeta^{\frac{\xi}{2(\xi+1)  }}\
\frac{g'(\zeta^{-1})}{g'(\zeta)}\
\hat{\bf W}
\left[\matrix{l'_4&l'_3\cr l'_1&l'_2}\biggl|\ \zeta
,\frac{\xi}{\xi+1},x^{2+2\xi}
\ \right]\ .}}
\ $\hat{\bf W}$\ denote
the following matrices:
\eqn\ksjdgh{\eqalign{&\hat{\bf W}
\left[\matrix{l\pm 2&l\pm 1\cr l\pm 1&l}\biggl|\ \zeta,a,q\right]=1\ ,\cr
&\hat{\bf W}
\left[\matrix{l&l\pm 1\cr l\pm1\ &l}\biggl|\ \zeta,a,q \right]=
\zeta^{a(\mp l-1)}\
\frac{\Theta_q(q^a)\ \Theta_q(q^{\mp a l}\zeta )}{\Theta_q(q^a \zeta)\
\Theta_q(q^{\mp al})}\ ,\cr
&\hat{\bf W}
\left[\matrix{l&l\mp 1\cr l\pm 1\ &l}\biggl|\ \zeta,a,q\right]=-
q^{\mp a^2 l}\zeta^{-a}\
\frac{\Theta_q(q^{a(\mp l+1)})\ \Theta_q(\zeta)}{\Theta_q(q^{\mp al})\
\Theta_q(q^a \zeta)}\ . }}
The matrices\ ${\bf W}$,\
${\bf W}'$ \ satisfy unitarity condition and Yang-Baxter
equation in IRF form \refs{\bax},\ \refs{\abf}.
These properties provide correspondingly
self-consistency and associativity conditions for
algebra of vertex operators\ \hdf .
In addition, matrices \ ${\bf W}$\
and \ ${\bf W}'$\ also satisfy to
the
crossing symmetry equation:
\eqn\jhsg{
{\bf W}'
\left[\matrix{l'_4&l'_3\cr l'_2&l'_1}\biggl|\ x^2 z^{-1}\ \right]=
-\frac{\kappa_{l'_3}}{\kappa_{l'_4}}\
{\bf W}'
\left[\matrix{l'_2&l'_4\cr l'_1&l'_3}\biggl|\  z\ \right]\ ,}
where
$$\kappa_{l'}=-x^{\frac{\xi( l'-1)}{\xi+1}(\xi l'-1) }\
\frac{\Theta_{x^{2+2\xi}}(x^{2\xi l'})}{\Theta_{x^{2+2\xi}}
(x^{2\xi})}\ .$$
Analogous equation holds for the matrix \ ${\bf W}$.
Using the commutation relation \hdf \ and the property
\jhsg\ one can show that
the quadratic combination
\ $\kappa_{l'-1 }\ \Phi_{12}^{+}(\zeta)\Phi_{12}^{-}(x^2\zeta)+
\kappa_{l'+1}\ \Phi_{12}^{-}(\zeta)\Phi_{12}^{+}(x^2\zeta)$\
is a central element in the algebra of vertex operators.
In the chosen normalization \dhgdfgtw \
we will have:
\eqn\jfg{\kappa_{l'-1 }\ \Phi_{12}^{+}(\zeta)\Phi_{12}^{-}(x^2\zeta)+
\kappa_{l'+1}\ \Phi_{12}^{-}(\zeta)\Phi_{12}^{+}(x^2\zeta)=1\ .}
Note that the congjugation \mcnvnb \ confirms with the commutation
relations \ \hdf \ and condition\ \jfg . This justifies
proposed choice
of the scalar product in the conjecture 4.2. and bosonization rules
\kh ,\ \kih .

\subsec{Trigonometric ZF algebra}
Let us shortly discuss now limiting cases of x-deformed construction.
The first evident limit is conformal one . It can be obtained
when
parameter\ $x$\ tends to \ $1$\
while variable \ $\ln \zeta $ remains to be finite and non-zero.
The bosonization formulae of matrix elements of vertex operators
in this limit
are obviously equivalent to Dotsenko-Fateev integral
representation for conformal blocks in CFT \refs{\DotsFat}.
In particular, one can find
that the complicated function\ $g(z), g'(z), w(z), w'(z)$\ at
the limit\ $x\to 1,\ i \ln z\sim 1$\ become
correspondingly \ $(1-z)^{\frac{\xi+1}{2\xi}},
(1-z)^{\frac{\xi}{2(\xi+1)}}, (1-z)^{\frac{\xi+1}{\xi}},
(1-z)^{\frac{\xi}{\xi+1}}$.

Another limiting case corresponds to the case when
elliptic matrices W transforms into trigonometric one  \refs{\Zkvadrat},\
\refs{\z}.
It can be
obtained as following.
Let us write down the variables\ $x,\zeta$\ in
the form\ $z=e^{-i\epsilon\beta}$\ and then look at
the limit\ $\epsilon\to 0$, assuming that \ $\beta$\ is finite.
Notice, that it is convenient
to carry out the modular transformation
\ $q=e^{i \pi\tau}\leftrightarrow e^{-i \frac{\pi}{\tau}}$\
of theta functions \ $\Theta_q(\zeta)$\ at first. Then
the limits can be found straightforwardly.
It is not hard to show
that
functions \vcf  , as well as the commutation relations  \hdf  \ are
well-defined in this limit.
These expressions can be naturally treated as vacuum averages of
certain operators
\eqn\gsdfd{\eqalign{\Phi_{21}^\pm(\beta)=\lim_{\epsilon\to 0}
\Phi_{21}^\pm(e^{-i\epsilon\beta})\ ,\cr
\Phi_{12}^\pm(\alpha)=\lim_{\epsilon\to 0}
\Phi_{12}^\pm
(e^{-i\epsilon\alpha)}\ ,}}
which act in the set of spaces
$\lim_{\epsilon\to 0}{\cal L}_{l,l'}.$
\foot{To avoid introducing of an additional notation we will denote
in this  subsection such
spaces as \ ${\cal L}_{l,l'}$.}
Using bosonic realization
of the ZF algebra of IRF type,
one can construct in this limit representations of
ZF algebra of vertex type \refs{\pasqu}.
Indeed,
the finite-dimensional part of total symmetry algebra \ $Symm$\
will coincide with those
considered in the conformal case, i.e. it is given by the
direct product
of two quantum groups \ $U_p(sl(2))\otimes U_{p'}(sl(2)) $.
Let us introduce now space
$\pi_Z$\ as
\eqn\jdhgfg{\pi_Z=\oplus_{l,l'=1}^{\infty} {\cal L}_{l,l'}
\otimes {\cal V}_{l,l'}\ ,}
where \ ${\cal V}_{l,l'}$\ is \ $l\cdot l'$-dimensional
irreducible representation
of the algebra \ $U_p(sl(2))\otimes U_{p'}(sl(2)) $\ and
define an action of the operators \ $Z_{\pm}(\beta),\ Z_{\pm}'(\alpha)$\
by the formula analogous \Zamo , \ \sonf .
It is convenient to consider the
following simple redefinition of these operators
\eqn\sinforg{\eqalign{&Z_\pm(\beta)\to
e^{\mp\frac{\beta}{2\xi }} \ Z_\pm(\beta)\ , \cr
&Z'_\pm(\alpha)\to
e^{\mp\frac{\alpha}{2(\xi+1)}} \ Z'_\pm(\alpha)\ .}}
Then one can show that operators \ $Z_a(\beta),\ Z'_a(\alpha)$\
generate the
ZF algebra of vertex type:
\eqn\vbcj{\eqalign{&Z_a(\beta_1)Z_b(\beta_2)=S_{ab}^{cd}(\beta_1-\beta_2)
Z_d(\beta_2) Z_c(\beta_1),\cr
&Z'_a(\alpha_1)Z'_b(\alpha_2)=R_{ab}^{cd}(\alpha_1-\alpha_2)
Z'_d(\alpha_2) Z'_c(\alpha_1),\cr
&Z_a(\beta)Z'_b(\alpha)=a\ b\ d(\beta-\alpha)
Z'_b(\alpha) Z_a(\beta).}}
The function \ $d(\beta)$ is the limiting value of the
function \ \jdghdsaq.
Its explicit form is just
\ $d(\beta)=\tan(\frac{\pi}{4}+i\frac{\beta}{2})$. The matrices
\ ${\bf S}_{ab}^{cd}(\beta)\ , {\bf R}_{ab}^{cd}(\alpha)$\
have the following form:
\eqn\jdghdg{\eqalign{&{\bf S}_{ab}^{cd}(\beta)=
s(\beta)\  R_{ab}^{cd}(e^{-\frac{2\beta}{\xi}},e^{i\pi\frac{\xi+1}
{\xi}})\ ,\cr
&{\bf R}_{ab}^{cd}(\alpha)=
r(\alpha)\
R_{ab}^{cd}(e^{-\frac{2\alpha}{\xi+1}},e^{i\pi\frac{\xi}
{\xi+1}})\ ,}}
where nontrivial elements of matrix \ $R_{ab}^{cd}(t,p)$\
given by \fsdq .
The explicit forms
of the functions \ $r(\alpha),\ s(\beta)$\
is complicate enough and it can be found in the work
\refs{\singordon}. ZF algebra \vbcj\ was introduced in the context
of Sin-Gordon model  and its physical
meaning was discussed in the works
\refs{\singordon},\ \refs{\josti},\  \refs{\shat}\ .

\newsec{Open problems}
Now we run into the problem how to construct the representations
of the elliptic ZF algebra of vertex type  \refs{\newjap}.
Let us give an abstract definition of this object.
It is a quadratic algebra of the form \vbcj\
where
matrices \ ${\bf R}_{ab}^{cd}, {\bf S}_{ab}^{cd}$\ have the following form:
\eqn\jdghdge{\eqalign{&{\bf S}_{ab}^{cd}(\beta)=
e^{- i \frac{\xi+1}{2\xi}\epsilon \beta}\ \frac{g(e^{i \epsilon \beta})}
{g(e^{-i \epsilon \beta})}\
R_{ab}^{cd}(e^{-\frac{2\beta}{\xi}},e^{i\pi\frac{\xi+1}
{\xi}},e^{-\frac{4\pi}{\epsilon\xi}})\ ,\cr
&{\bf R}_{ab}^{cd}(\alpha)=
e^{- i\frac{\xi}{2(\xi+1)  }\epsilon\alpha}\
\frac{g'(e^{i \epsilon \alpha})}{g'(e^{-i \epsilon \alpha})}\
R_{ab}^{cd}(e^{-\frac{2\alpha}{\xi+1}},e^{i\pi\frac{\xi}
{\xi+1}},e^{-\frac{4\pi}{\epsilon(\xi+1)}})\ .}}
The nontrivial elements of matrix \ $R_{ab}^{cd}(t,p)$\
are defined by the formula \ \ldk .
The function \ $ d(\beta)$\  coincides with
\jdghdsaq\ where \ $z$\ is equal to \ $e^{-i \epsilon\beta}$.
Note that matrices \jdghdge \ satisfy the unitarity, crossing symmetry,
Yang-Baxter equation and at the limit \ $\epsilon\to 0$\
they transform to \ \jdghdg.
As we have studied on the example above,
to construct ZF algebra of vertex type from IRF algebra
of vertex operators, one need know the
comultiplication in the
finite-dimensional subalgebra of symmetry algebra.
This problem might be very non-trivial
since this subalgebra seems to be related
with Sklyanin algebra \refs{\sklyanin}.
At the same time, we want to emphasize that
the Sklyanin
algebra itself is a deformation of
\ $U_p(sl(2))$,
while we expect that this new algebra have to be deformation of
a tensor product of quantum groups
 \ $U_p(sl(2))\otimes U_{p'}(sl(2))$, to give the limits which
are consistent with our constructions
in the trigonometric and conformal cases.

\newsec{Acknoledgments}
We would like to thank V. Brajnikov for discussion.
We are very gratefull to A.B. Zamolodchikov for helpfull
comments and his interest in work.

This work was supported by grant DE-FG05-90ER40559.
Ya.P. was also supported in part by grant ISF M6N000.

\newsec{Appendix.}
In this Appendix we collect the explicit expressions for the
functions and constants, which are necessary to compute
the matrix elements of vertex operators in formula \vcf .
The functions\ $g,w,\bar g,g',w',\bar g',h,u,\bar h$\  are
defined as
\eqn\findu{
\eqalign{
&g(\zeta_1 \zeta_2^{-1})=e^{-[\phi_+(\zeta_2 ),
\phi_-(\zeta_1)]}\ , \cr
&w(\zeta_1 \zeta_2^{-1})=e^{[\phi_{+}(\zeta_2),\bar{\phi}_{-}(\zeta_1)]}\ , \cr
&\bar g(\zeta_1 \zeta_2^{-1})=e^{-[\bar\phi_+(\zeta_2 ),
\bar\phi_-(\zeta_1)]}\ ,\cr
&g'(\zeta_1 \zeta_2^{-1})=e^{-[\phi'_{+}(\zeta_2),\phi'_{-}(\zeta_1)]}\ , \cr
&w'(\zeta_1 \zeta_2^{-1})=e^{[\phi'_{+}(\zeta_2),{\bar \phi}'_{-}
(\zeta_1)]}\ , \ \cr
&\bar g'(\zeta_1 \zeta_2^{-1})=e^{-[\bar\phi'_+(\zeta_2 ),
\bar\phi'_-(\zeta_1)]}\ ,\cr
&h(\zeta_1 \zeta_2^{-1})=e^{-[\phi_+(\zeta_2 ),
\phi'_-(\zeta_1)]}\ , \cr
&u(\zeta_1 \zeta_2^{-1})=e^{[\phi_{+}(\zeta_2),{\bar\phi}_{-}'(\zeta_1)]}\ ,
\cr
&{\bar h}(\zeta_1\zeta_2^{-1})
=e^{-[\bar\phi_{+}(\zeta_2),{\bar\phi}_{-}'(\zeta_1)]} \ . \cr
}
}

They are given by:
\eqn\ytoisa{\eqalign{&g(z)=\frac{(z;x^{2\xi})_{\infty}
\ (x^4 z;x^{2\xi},x^4)_{\infty} \
(x^{4+2\xi} z;x^{2\xi},x^4)_{\infty}}
{(x^2 z;x^{2\xi})_{\infty}
\ (x^6 z;x^{2\xi},x^4)_{\infty} \
(x^{2+2\xi} z;x^{2\xi},x^4)_{\infty}}\ , \
\cr
&w(z)=\frac{(x^{1+2\xi}z;x^{2\xi})_{\infty}}{(x^{-1}z;x^{2\xi})_{\infty}}
,
\cr
&\bar g(z)=(1-z)\ \frac{(x^{-2}z; x^{2\xi})_{\infty}}
{(x^{2+2\xi}z; x^{2\xi})_{\infty}}\ ,
\cr
&g'(z)=\frac{
(x^2 z;x^{2+2\xi},x^4)_{\infty} \
(x^{4+2\xi} z;x^{2+2\xi},x^4)_{\infty}}
{(x^4 z;x^{2+2\xi},x^4)_{\infty} \
(x^{2+2\xi} z;x^{2+2\xi},x^4)_{\infty}}\ ,
\cr
&w'(z)=
\frac{(x^{1+2\xi}z;x^{2+2\xi})_{\infty}}{(xz,x^{2+2\xi})_{\infty}}\ ,
\cr
&\bar g'(z)=(1-z)\ \frac{(x^{2}z; x^{2+2\xi})_{\infty}}
{(x^{2\xi}z; x^{2+2\xi})_{\infty}}\ ,
\cr
&h(z)=\frac{(x^3 z;x^4)_{\infty}}{(xz;x^4)_{\infty}}\ ,
\cr
&u(z)=1-z\ ,\cr
&\bar h(z)=\frac{1}{(1-zx)(1-zx^{-1})}\
.}}
The constants \ $\rho,\bar \rho, \rho',\bar \rho' $\ determined as
\eqn\const{\eqalign{
&\rho^2=\lim_{z\to 1}\ \frac{1-x^2}{1-z}\  g(z)\ , \cr
&{\bar\rho}^2=\lim_{z\to 1}\ \frac{1-x^2}{1-z}\  {\bar g}(z)\ , \cr
&\rho'^2=\lim_{z\to 1}\ g'(z)\ , \cr
&{\bar\rho}'^2=\lim_{z\to 1}\ \frac{1-x^2}{1-z}\  {\bar g}'(z)\ . \cr
}}
have the following  values:
\eqn\bsare{\eqalign{&\rho^2=(1-x^2)\
\frac{(x^{2\xi};x^{2\xi})_{\infty}}{(x^{2+2\xi};x^{2\xi})_{\infty}}\
g^{-1}(x^2)\ ,
\cr
&\bar \rho^2=(1-x^2)\
\frac{(x^{-2};x^{2\xi})_{\infty}}{(x^{2+2\xi};x^{2\xi})_{\infty}}\ ,
\cr
&\rho^{'2}= \frac{(x^2 ;x^{2+2\xi})_{\infty}}
{(x^{2\xi+2} ;x^{2+2\xi})_{\infty}}\ g'^{-1}(x^2)\ ,\cr
&\bar \rho^{'2}=(1-x^2)\
\frac{(x^{2};x^{2+2\xi})_{\infty}}{(x^{2\xi };x^{2+2\xi})_{\infty}}\ .}}

\listrefs
\end